\documentclass[prl,twocolumn,superscriptaddress]{revtex4-1}
\usepackage[colorlinks=true,allcolors=blue]{hyperref} 
\usepackage{graphicx}
\usepackage{amsmath}

\begin{document}

\title{Short-range dynamics in the solid and liquid phases}

\author{E. I. Andritsos}
\email{lefteri.andritsos@kcl.ac.uk}
\affiliation{School of Physics and Astronomy, Queen Mary University of London, Mile End Road, London, E1 4NS, UK}
\affiliation{Department of Physics, King's College London, Strand, London, WC2R 2LS, UK}
\author{M. T. Dove}
\affiliation{School of Physics and Astronomy, Queen Mary University of London, Mile End Road, London, E1 4NS, UK}
\author{F. Demmel}
\affiliation{ISIS Facility, Rutherford Appleton Laboratory, Didcot OX11 0QX, UK}
\author{K. Trachenko}
\affiliation{School of Physics and Astronomy, Queen Mary University of London, Mile End Road, London, E1 4NS, UK}

\begin{abstract}
The existence of the phonon-roton minimum has been widely observed for both the solid and liquid phases but so far there is no sufficient theoretical explanation of its origin.
In this paper we use a range of techniques to study the dynamics and short-range order for a range of simple materials in their crystalline, amorphous and liquid phases.
We perform inelastic neutron scattering (INS) experiments of polycrystalline and liquid barium to study the high-frequency dynamics and understand the mechanisms underlying the atomic motion.
Moreover we perform INS simulations for crystals and supercooled liquids,  compare the collective excitation spectra and identify similarities.
We perform molecular dynamics (MD) simulations for the same materials and present results of population and bond angle distribution showing a short-range order dependence of the different phases, expanding the current knowledge in literature.
Finally, we support our findings with a theoretical explanation of the origin of the phonon-roton minima which is observed in both solids and liquids. 
We study this as a classical phenomenon and we base our explanation on short range interatomic interactions.
\end{abstract}

\maketitle

\section{I. Introduction}

It is known that theories on dynamics of materials describe well the crystalline and gas states but fail to describe the intermediate structures such as glasses and liquids.
Solids are dominated by structural processes and gasses by kinetic processes while the intermediate structures are dominated by a combination of them. 
Due to absence of specific theory for the amorphous and liquid phases, atomic dynamics are usually discussed by either using the theory for crystals or gases, or they are based on statistical solutions \cite{Copley1975, Egelstaff1994,Wallace1997,Scopigno2005}. 
Therefore it is very important to understand the connection between these states and the well defined crystalline state, not only regarding atomic arrangements but also atomic motion.
In this paper we provide evidence from experiments and simulations of connection in dynamics between the crystalline and the amorphous and liquid phases and we discuss the basic mechanism of it.\par

Atomic dynamics is mainly vibrational and it is characterised by normal modes due to the harmonic nature of the interatomic forces.
Atoms in a solid constantly oscillate around their equilibrium positions with an energy governed by temperature.
In liquids the atoms have been observed to initially vibrate around their equilibrium position, with larger vibrational amplitude than those of the corresponding solid \cite{Elliot1984} and then jump to another position where they repeat the same procedure. 
Therefore atoms in liquids have a mixed motion of oscillatory and diffusional motion. 
These processes of hopping and diffusion are time dependent and affect the collective dynamical properties.
This time scale ($\tau_D$) is of the order of $\tau_D = 1/ \omega_D$, where $\omega_D$ is the Debye frequency which corresponds to the frequency of a crystal with similar density.
Thus, at large wavelengths the sound velocity of the crystal and the corresponding liquid are similar \cite{Scopigno2005}.
It is interesting to examine the atomic vibrations in liquids at shorter wavelengths and compare them with the crystalline vibrations. \par 

Despite the experimental and computational advances and the increasing interest on atomic dynamics of liquids over the past years, a direct comparison with crystal dynamics has not been achieved.
Modern experimental techniques such as inelastic neutron scattering (INS) allow us to study microscopic dynamics and determine with accuracy the relative positions and motion of atoms. 
Only lately, scientists working on alkali metals identified similarities on atomic dynamics of the two states and they comment on the solid-like features of the liquids \cite{Pilgrim2006, Monaco2010}.
More specifically, following previous observations Giordano and Monaco \cite{Monaco2010} perform inelastic X-ray scattering (IXS) experiments on sodium and compare collective excitation spectra of polycrystalline and liquid sodium showing intriguing similarities between the two spectra.
They analyse specific modes of vibration along various directions showing a connection of the two phases, something that they relate to short-range order similarities of the two phases.
The similarity on the collective modes is attributed to interatomic interactions specific to alkali metals \cite{Pilgrim2006}, easy to compare due to their distinct collective modes and clear inelastic features, although similarities on collective modes of most liquids are generally observed.\par

The phonon-roton minima which is observed in both solid and liquid collective excitation spectra of many materials indicates a similarity in the dynamics.
The phonon-roton minimum, or roton as it was initially mentioned by Landau for the case of superfluid helium, is a local minimum observed after the initial linear increase in the phonon frequencies spectra at low \textbf{Q}.
The formation of this minimum is a known effect for most liquids but it is also observed in solids (e.g. bcc Ba and fcc Sr \cite{Buchenau1984}, bcc Fe \cite{Minkiewicz1967}, bcc and liquid Na \cite{Woods1962, Monaco2010}, bcc and liquid K \cite{Cowley1966, Ferconi1991}, bcc V \cite{Colella1970}, rare gas solids (fcc) \cite{Raregassolidsch15}, fcc Yb \cite{Stassis1982}, fcc Ca \cite{Stassis1983}, review on some bcc alkali metals \cite{Gajjar1995}, extended review on liquid metals \cite{Scopigno2005}).
This makes clear that the existence of the phonon-roton minimum is not only a universal property of the liquid state but also of the solid state.
Despite the fact this minimum is a great discussion point since its prediction in 1947, there is not a sufficient explanation of its origin taking into consideration both states.
From the initial theory of a quantum phenomenon strictly attributed to superfluid helium we explain how this is a classical phenomenon universal for all phases, directly related to short-range interatomic interactions.\par

\section{II. Inelastic Neutron Scattering}
\subsection{A. INS Experiments}

We have investigated the high-frequency dynamics of polycrystalline barium and analyse the wave vector  and frequency dependence of the dynamic structure factor $S(\it{Q},\omega)$.
The INS experiments were performed on the time-of-flight MARI spectrometer at the ISIS Pulsed Neutron and Muon Facility, UK.
The barium sample (99.0 \% purity, in beads form with an average bead size of 0.5 -- 2.0 mm, purchased from Aldrich) was loaded in a cylindrical aluminium can in an Ar-filled glove box.
The initial measurements performed at 300 K.
For measurements at higher temperatures a barium rod was stored in a niobium can, sealed by electron beam welding.
A furnace was used to heat up the sample at 963 K, 1,033 K, 1,173 K, 903 K and 603 K (following this sequence of temperatures) with a 3 K maximum temperature fluctuation and low heating and cooling rate.
To minimise any background scattering the entire flight path is kept under vacuum.
For all measurements the incident energy was set at 14 meV, the frequency of Fermi chopper at 150 Hz and the frequency of disc chopper at 50 Hz.
The data were placed on an absolute intensity scale by normalisation to a vanadium standard.\par

\begin{figure}[t]
\begin{center}
\includegraphics[width=4.25cm, height=4cm]{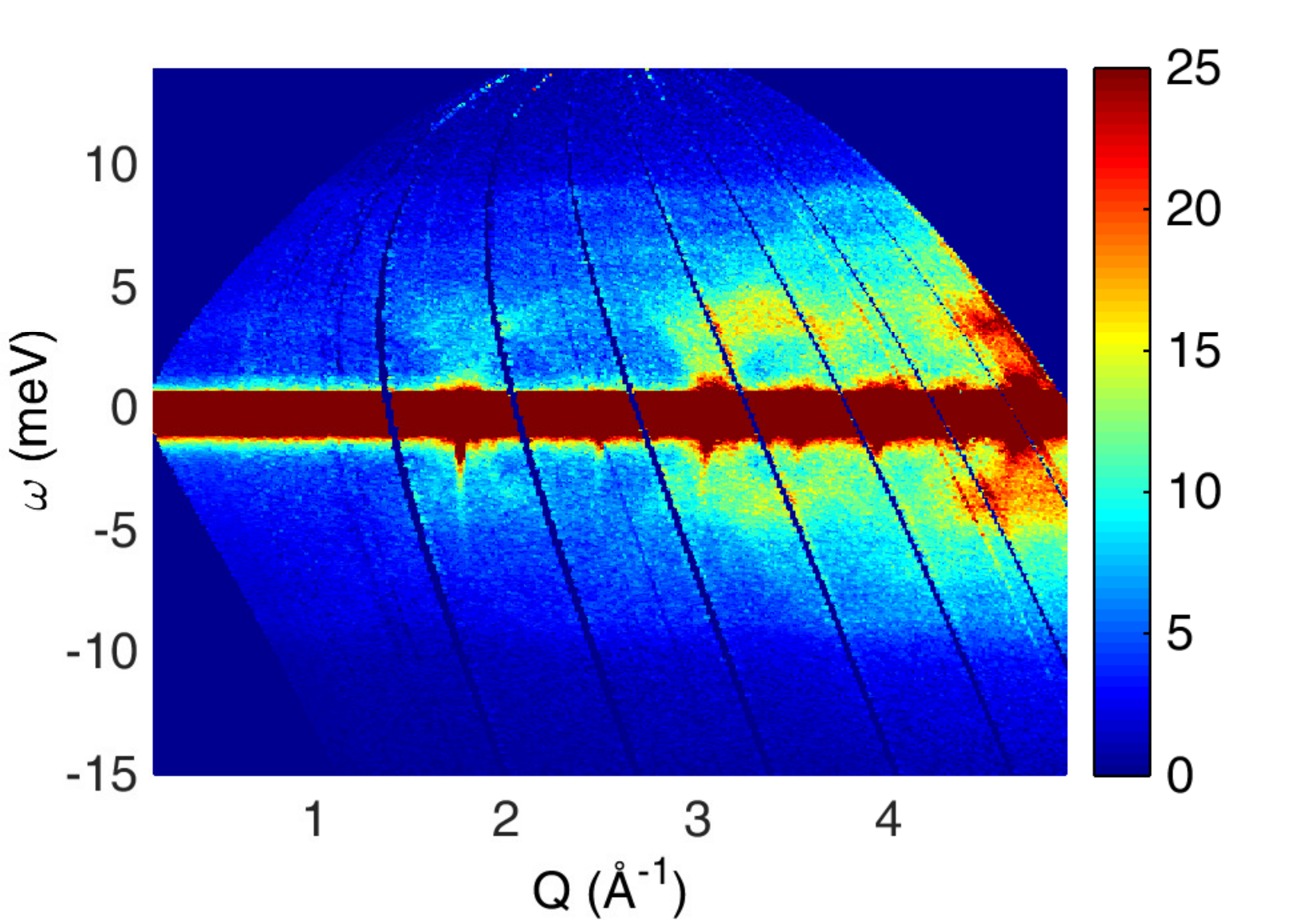}
\includegraphics[width=4.25cm, height=4cm]{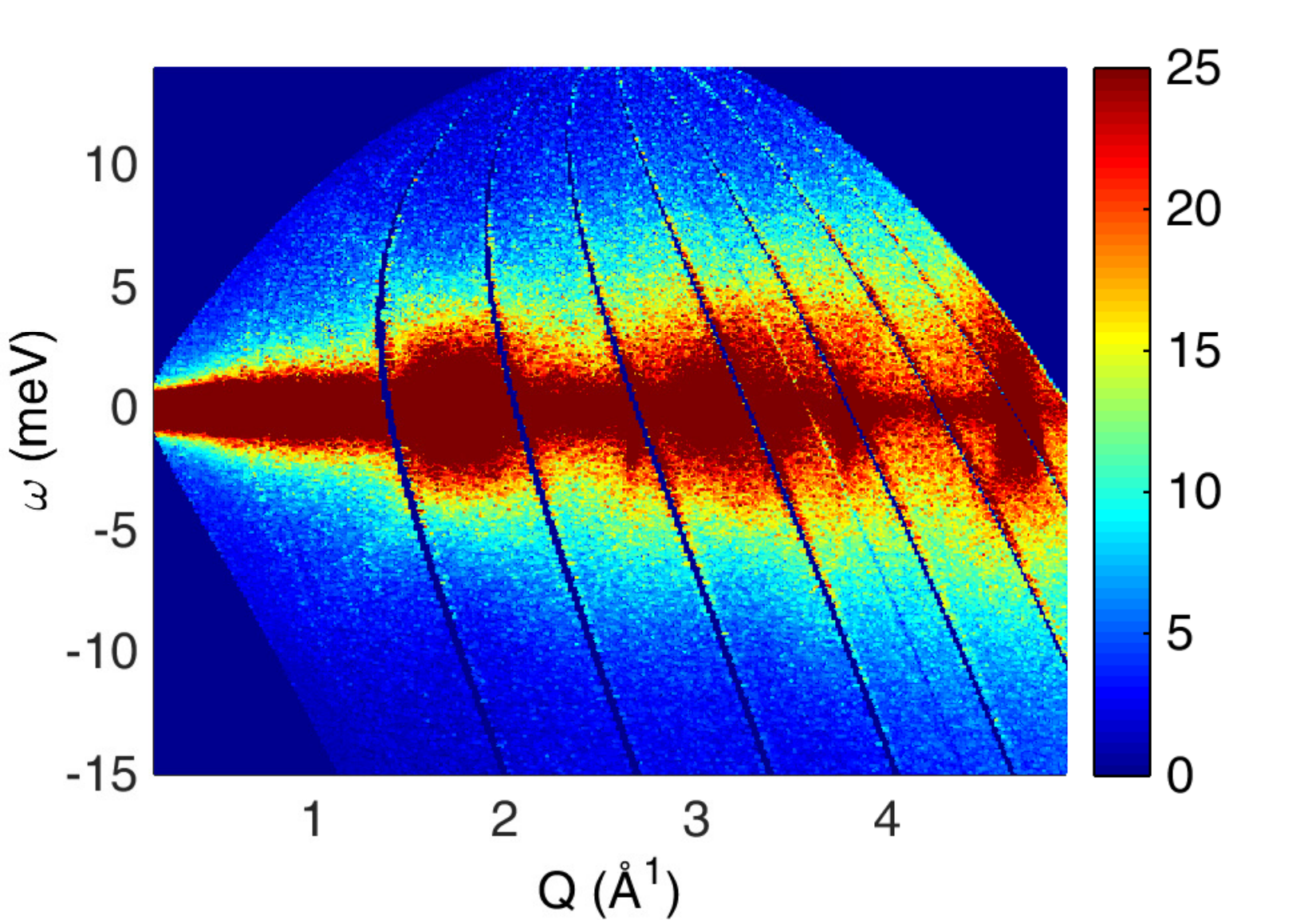}
\includegraphics[width=4.25cm, height=4cm]{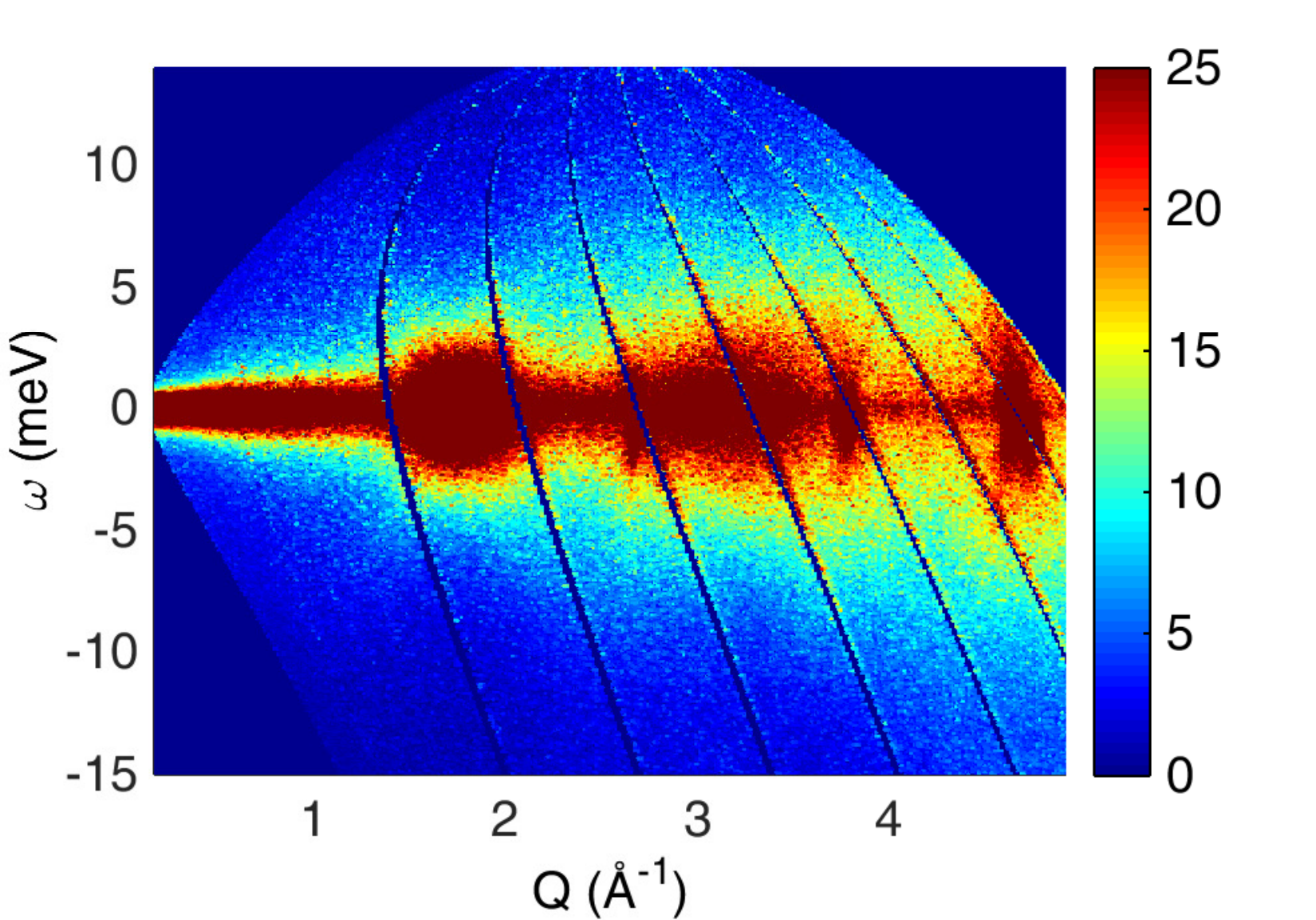}
\includegraphics[width=4.25cm, height=4cm]{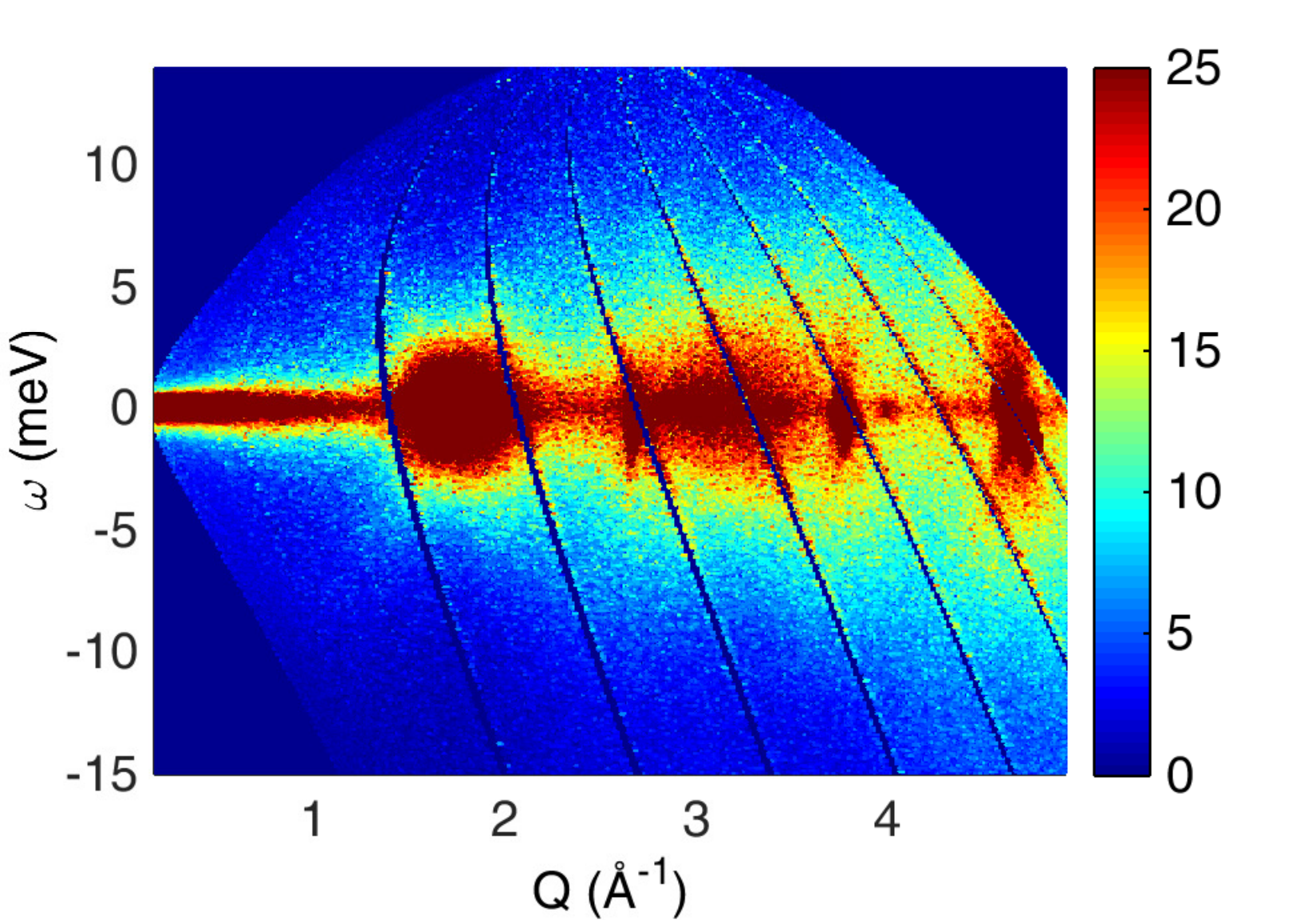}
\includegraphics[width=4.25cm, height=4cm]{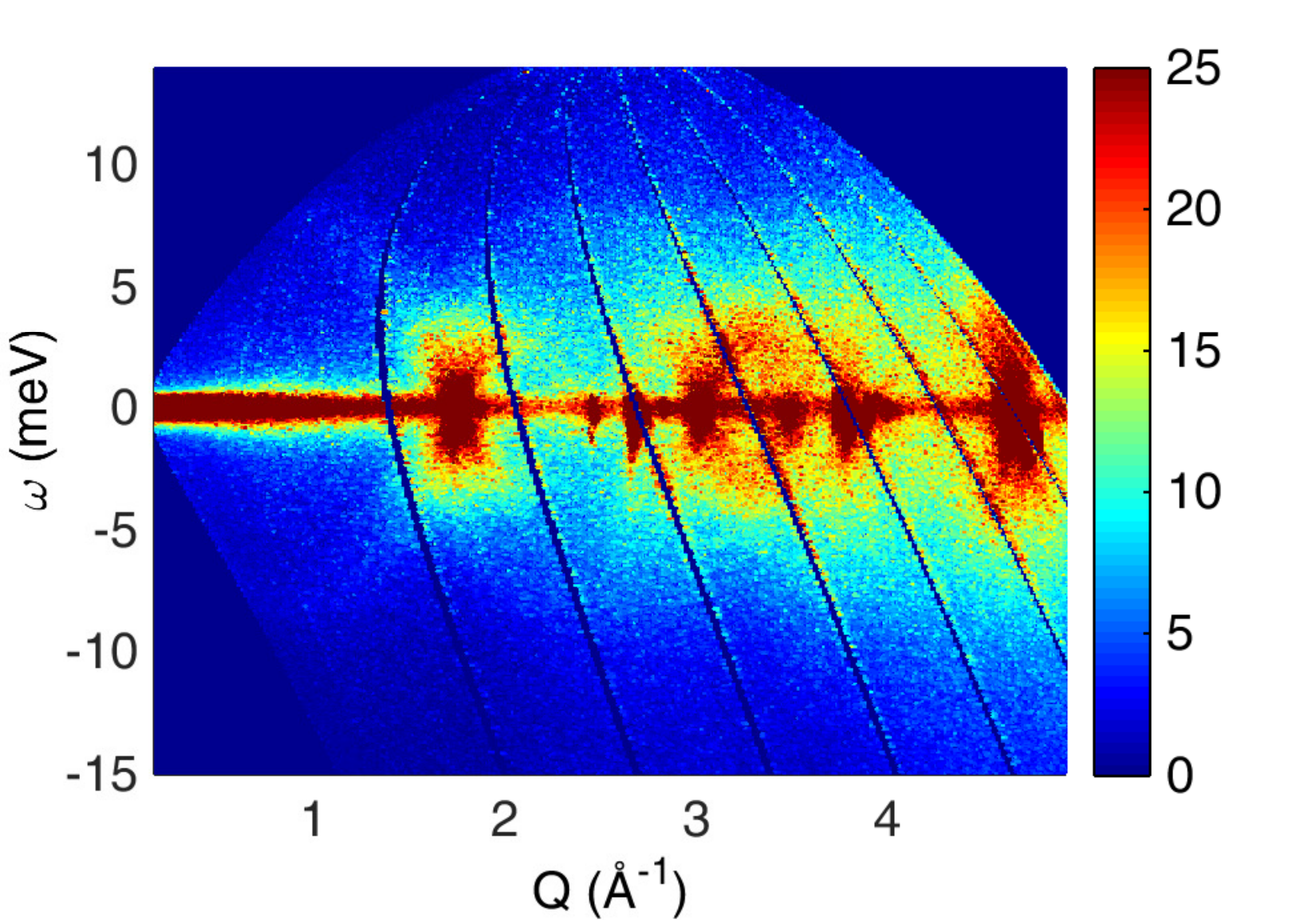}
\includegraphics[width=4.25cm, height=4cm]{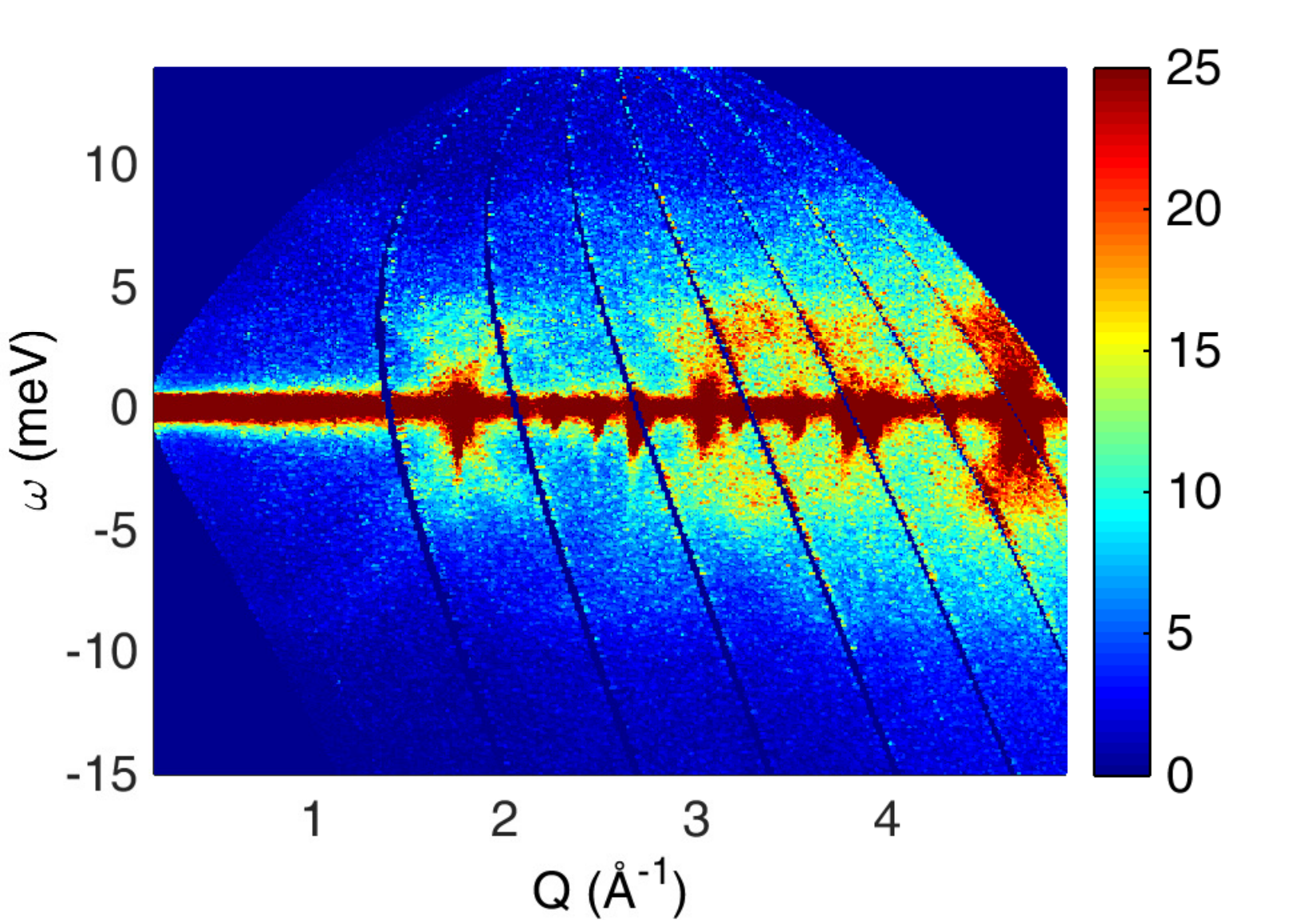}
\caption{Experimental INS spectra of Ba at 300 K, 963 K, 1,033 K, 1,173 K, 903 K and 603 K (left to right, top to bottom).}
\label{fig:INSexp}
\end{center}
\end{figure}
\par

In Fig. \ref{fig:INSexp} we present results from experimental INS spectra of barium.
The plots show the intensity of frequency with respect of the scattering vector in a colour-scale map at 300 K, 963 K, 1,033 K, 1,173 K, 903 K and 603 K (left to right, top to bottom).
Experimental melting point for Ba is at 998 - 1000 K, hence Ba is in liquid state at 1033 K and 1173 K, and in solid state at 300 K, 963 K, 903 K and 603 K.
With blue colour is the area with no detected neutron scattering.
The positive and negative energies represent the energy gain and loss at the acoustic spectra.
Phonon excitations are more intense around the Bragg peaks and on the elastic line, where the acoustic modes go to zero,  and less intense at higher frequencies.
At the low \textbf{Q} regime the intensity is very low and direct study of phonon excitations is challenging without a numerical correction.\par

To emphasise at the lower and intermediate $\bf{Q}$ regime we have multiplied the $S(\it{Q},\omega)$ by $\omega^2/\mathit{Q}^2$.
By using this factor it is easier to visualise and study the collective excitation spectrum.
The scattering intensity S($\mathit{Q},\omega$) multiplied by $\omega^2/\mathit{Q}^2$ is equal to the spectra of the current correlation function J($\mathit{Q},\omega$) \cite{Bosse1978}.
\begin{equation} \label{eq:ccf}
J(\mathit{Q},\omega) = \frac{\omega^2 \, S(\mathit{Q},\omega)}{\mathit{Q}^2}
\end{equation}
The current correlation function describes the correlation of particle velocities rather than particle positions, as is the case for the $S(\mathit{Q},\omega)$.
By using the $J(\mathit{Q},\omega)$, the high-frequency part of the spectra is enhanced, allowing the study of weak high-frequency excitations.
The $\omega^2$ factor has a minor effect on the peak frequencies and thus they do not strictly coincide with the frequencies given by $S(\mathit{Q},\omega)$.
Therefore the longitudinal current correlation function demonstrates the main features of the dispersion \cite{Hensel2014}.

The dispersion curves can be described qualitatively with the damped harmonic oscillator (DHO) line-shape modelling of the spectrum.
In amorphous and liquid structures the collective excitations are strongly damped \cite{Hansen1986, Lovesey1986}, something that agrees with many experimental observations, therefore the DHO fitting describes well the collective excitation spectra. 
The equation that describes the fitted $J(\mathit{Q},\omega)$ is \cite{Bafile2006}
\begin{equation} \label{eq:DHO}
J(\mathit{Q},\omega) =\frac{1}{\pi} \frac{2 \Gamma_{\mathit{Q}} \omega_{\mathit{Q}}^2 \omega^2  }{(\omega^2 - \omega^2_{\mathit{Q}})^2 + 4 \Gamma_{\mathit{Q}}^2 \omega^2}
\end{equation}
where $\omega$ is the energy transfer, $\omega_{\mathbf{Q}}$ is the energy and $\Gamma_{\mathbf{Q}}$ is the width of the phonon excitation or damping factor.
The DHO fitting is a generally accepted line-shape fitting for IXS and INS spectra of solids and liquids (e.g. Refs. \onlinecite{Shapiro1972, Sette1995, Anderson1999, Hippert2006, Ichitsubo2007}) in order to evaluate the appropriate transformation, explained in detail for crystals in Ref. \onlinecite{Lovesey1986} and for liquids in Ref. \onlinecite{Bafile2006}. \par

\begin{figure}[t]
\begin{center}
\includegraphics[scale=0.36]{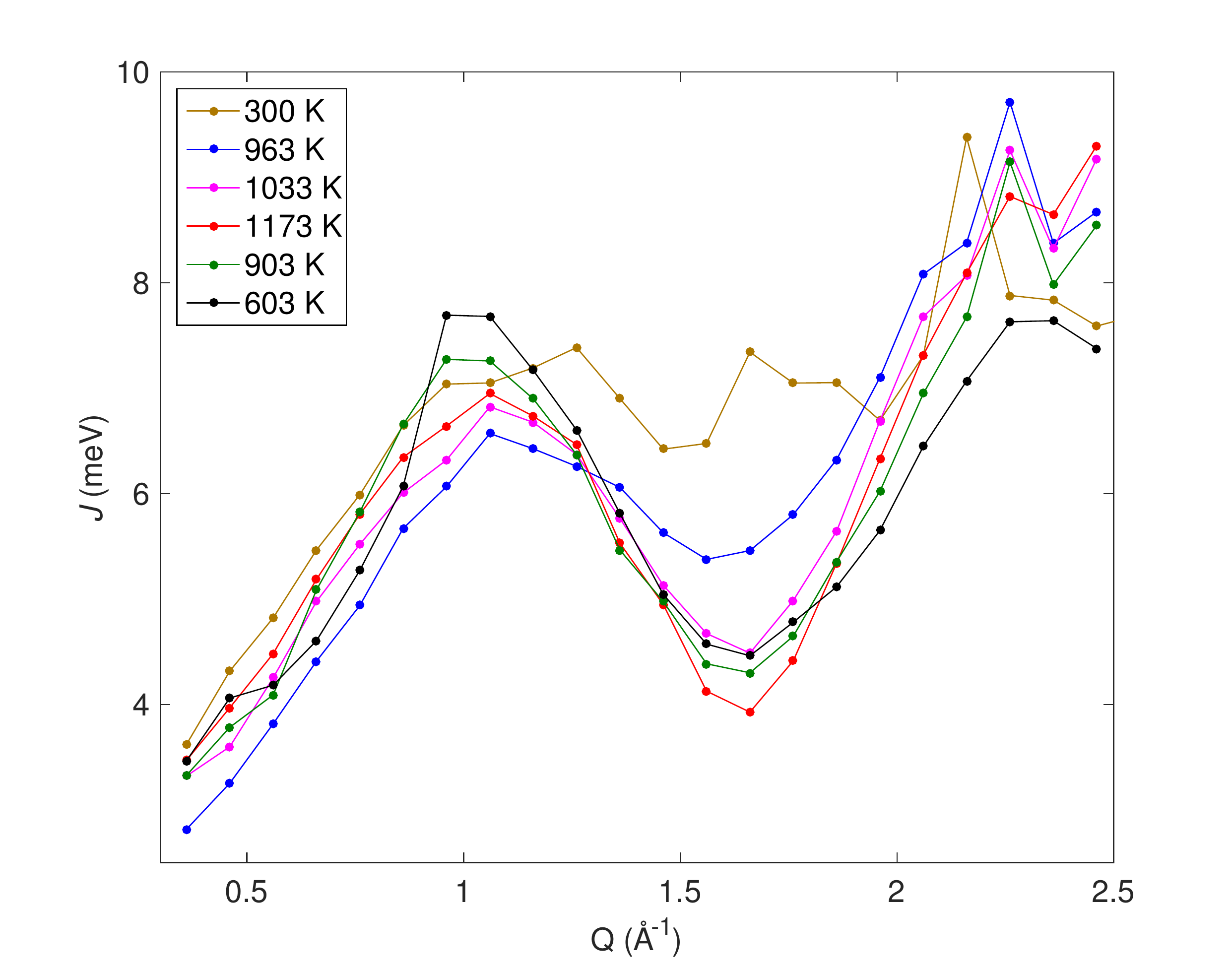} 
\caption{$J(\mathit{Q},\omega)$ of Ba from INS experiments at various temperatures.
At 300 K and 963 K the sample is in beads form, at 1033 K and 1173 K is liquid and at 903 K and 603 K is a solid rod.
The different polycrystalline form of the sample affects the angular averaging of excitations.} 
\label{fig:DHOexpall}
\end{center}
\end{figure}

In Fig. \ref{fig:DHOexpall} we present the $J(\mathit{Q},\omega)$ of the collective excitation spectra of Ba  at low and intermediate $\mathbf{Q}$ regimes from INS experiments at all different temperatures.
As we can see, all curves exhibit an initial almost linear increase up to about $\mathit{Q} = 1.06 \: \mathrm{\AA^{-1}}$, a local minimum at $\mathit{Q} = 1.66 \: \mathrm{\AA^{-1}}$ and then another increase.
Therefore the minimum is formed at all temperatures regardless of the structural state.
The phonon collective excitations are almost degenerate despite the large temperature difference. 
Exceptions are the curves at 300 K and 963 K, where we observe the phonon collective excitations exhibiting higher energy values at the minimum \textbf{Q} point.
Although there is no clear indication, this anomalous behaviour at those temperatures can be attributed to the different polycrystalline average orientation of the sample in beads and rod form.
The different angular averaging and heat treatment of the sample are factors that affect the dispersion. 
At 903 K and 603 K the initial part of the curve at low \textbf{Q} appears not to be linear. 
The non-linearity is justified because barium has already recrystallise at those temperatures.
The fitting of phonon dispersion curves at the crystalline phase has higher complexity because it is highly affected from the distinct dispersion curves, while for liquids the complexity is smaller and results in a smoother fitting curve.\par

\subsection{B. INS Simulations}

We study a range of simple materials and perform classical molecular dynamics (MD) simulations and computational INS.
The materials we present in this study are Fe,  K, Ta, and W with bcc crystalline structure and Ar with fcc crystalline structure.
All materials are examined in their crystalline, amorphous and liquid phase and they have been prepared similarly.
Initially all systems in their crystalline phase undergo a relaxation under constant pressure and temperature at room conditions for 30 ps.
Then we perform a 30 ps simulation of the relaxed structure under constant volume and energy at 10 K for the crystalline phase and at temperature above the melting point for the liquid phase.
We use the liquid configuration for the preparation of the supercooled liquid and we perform a simulation for 30 ps under constant volume and energy at 10 K.
We treat all systems as bulk and we use periodic boundary conditions during the simulations.
All MD simulations are performed with the DL POLY software.\par

The phonon frequencies for each material for the crystalline and supercooled liquid structures are calculated over a wide $\bf{Q}$ range. 
We use a special module of the GULP software \cite{Cope} for the calculation of the collective excitation spectra.
To calculate the phonon frequencies of a system the starting point is the force constant matrix given by the second derivative of the energy with respect to the atomic displacements in space. 
Atomic positions and energies were extracted from the MD simulations. \par

Whilst the different phonon modes for single crystals are possible to be identified, the phonon modes from polycrystalline, amorphous and liquid samples are considered to be very difficult or impossible unless during the refinement procedure there is a good starting model \cite{Fischer2009, Bosak2009}. 
The critical factor for phonon density of states plots apart from the resolution of the integration grid is the system size. 
Because of the periodic boundary conditions, for the crystalline structure only the basis atoms need to be considered while for the supercooled liquid structure of Fe, K, Ta and W we use 432 atoms and for Ar 256 atoms.
The continuous density of states curve has to be approximated by a series of finite regions of frequencies. 
Each phonon mode at each point in $\bf{k}$ space is assigned to a specific frequency region, therefore the more the $\bf{k}$ points the more precise the frequency calculations.
At the crystalline phase we have calculated a set of 100,000 different random $\bf{k}$ points and a set of 40,000 different random $\bf{k}$ points at the amorphous phase.
The interatomic potential parameters for simulations of Fe, Ta and W can be found in Ref. \onlinecite{Dai2006}, for K in Ref. \onlinecite{Girifalco1959} and for Ar in Ref. \onlinecite{DoveBlue}. 
For the calculation of phonon frequencies we use the elastic constants as input parameter to produce accurate results. 
The elastic constants for Fe, Ta and W can be found in Ref. \onlinecite{Dai2006}, for K in Ref. \onlinecite{Girifalco1959} and for Ar in Ref. \onlinecite{Gewurtz1972}. \par

A model for Ba is useful for direct comparison with experiments.
Due to lack of an accurate interatomic potential for Ba compatible with our simulations we consider Fe as the material for comparing with experiments.
Ba and Fe have qualitatevely very similar phonon spectra \cite{Buchenau1984, Minkiewicz1967} and thus it is expected to show similar behaviour in dynamics.
Therefore Fe seems the most appropriate bcc structure to use for scaling to Ba.
Qualitatively scaling of $\bf{Q}$ can be achieved by multiplying with the ration of the two lattice constants, $\mathbf{Q_{\mathrm{rscFe}}} = \mathbf{Q_{\mathrm{Fe}}}\alpha_{\mathrm{Fe}} / \alpha_{\mathrm{rscFE}}$.
For $\alpha_{\mathrm{Fe}} = 2.87$ \AA \ and $\alpha_{\mathrm{Ba}} = 5.02$ \AA \ the rescaled Fe structure has $\mathbf{Q_{\mathrm{rscFe}}} = 0.57 \mathbf{Q_{\mathrm{Fe}}}$.
Despite this being a rough model it can useful for theoretical predictions in cases as this where computational limitations exist. \par

\begin{figure}[t]
\begin{center}
\includegraphics[width=4.25cm, height=4.0cm]{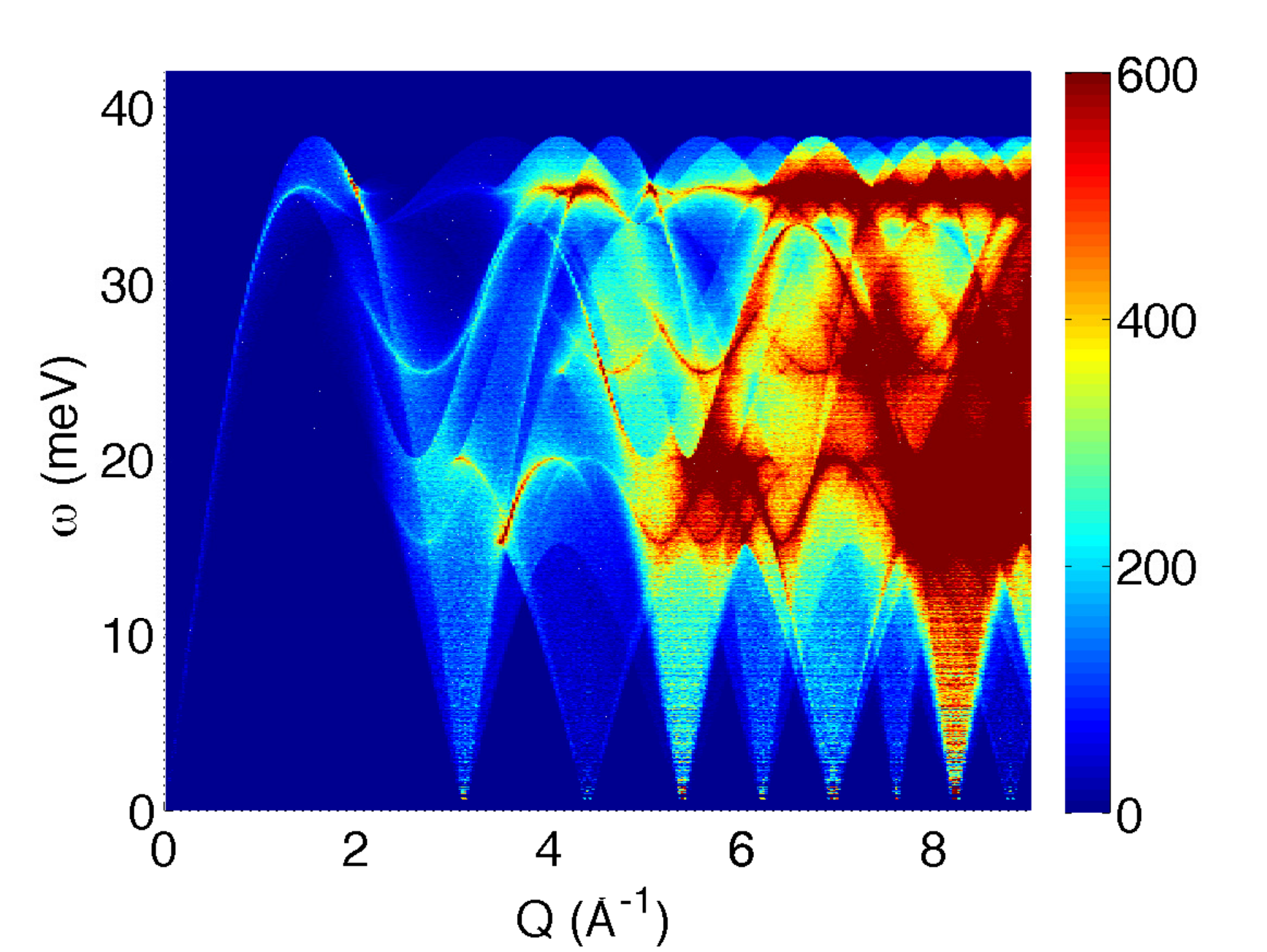} 
\includegraphics[width=4.25cm, height=4.0cm]{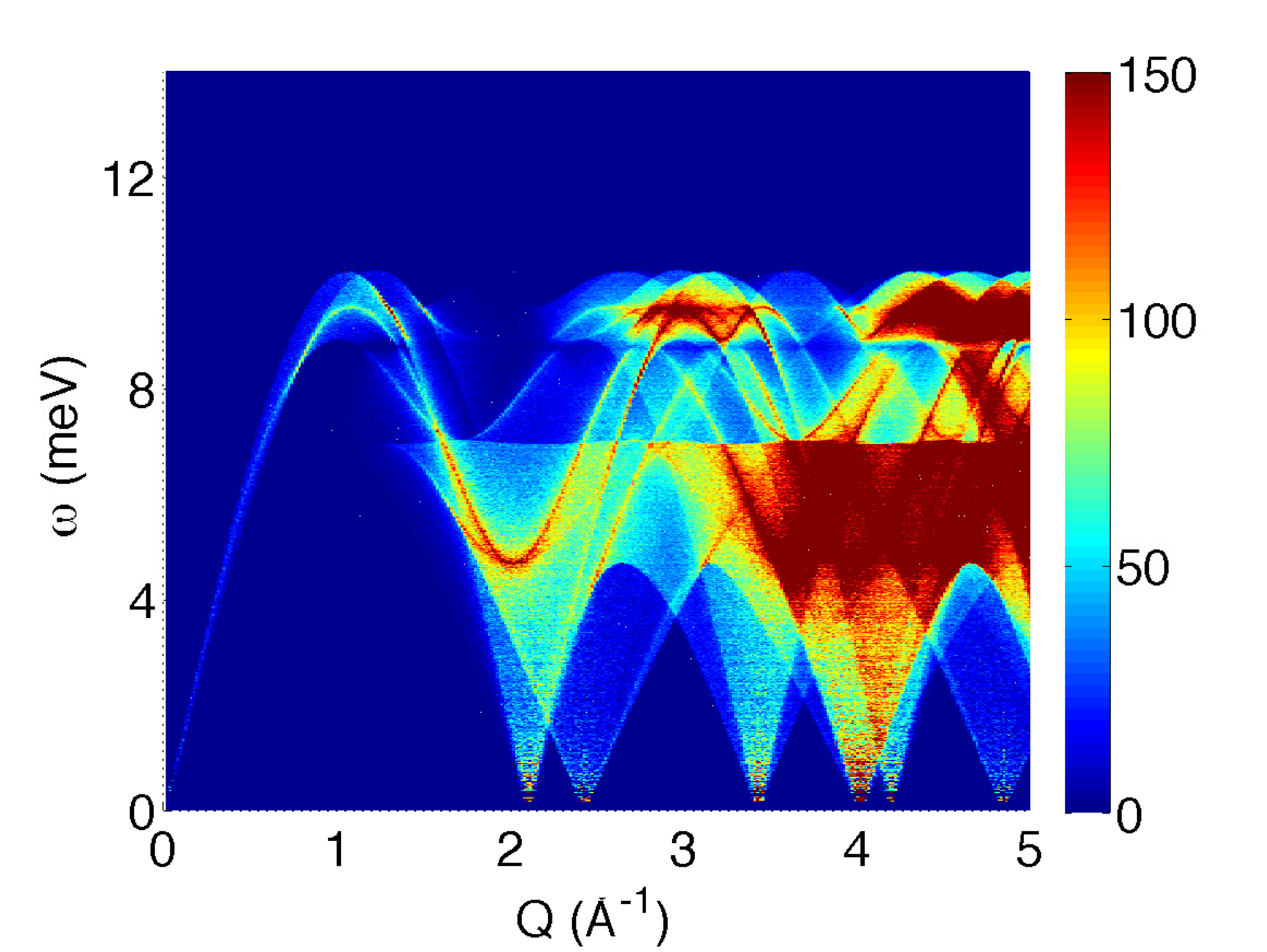} 
\includegraphics[width=4.25cm, height=4.0cm]{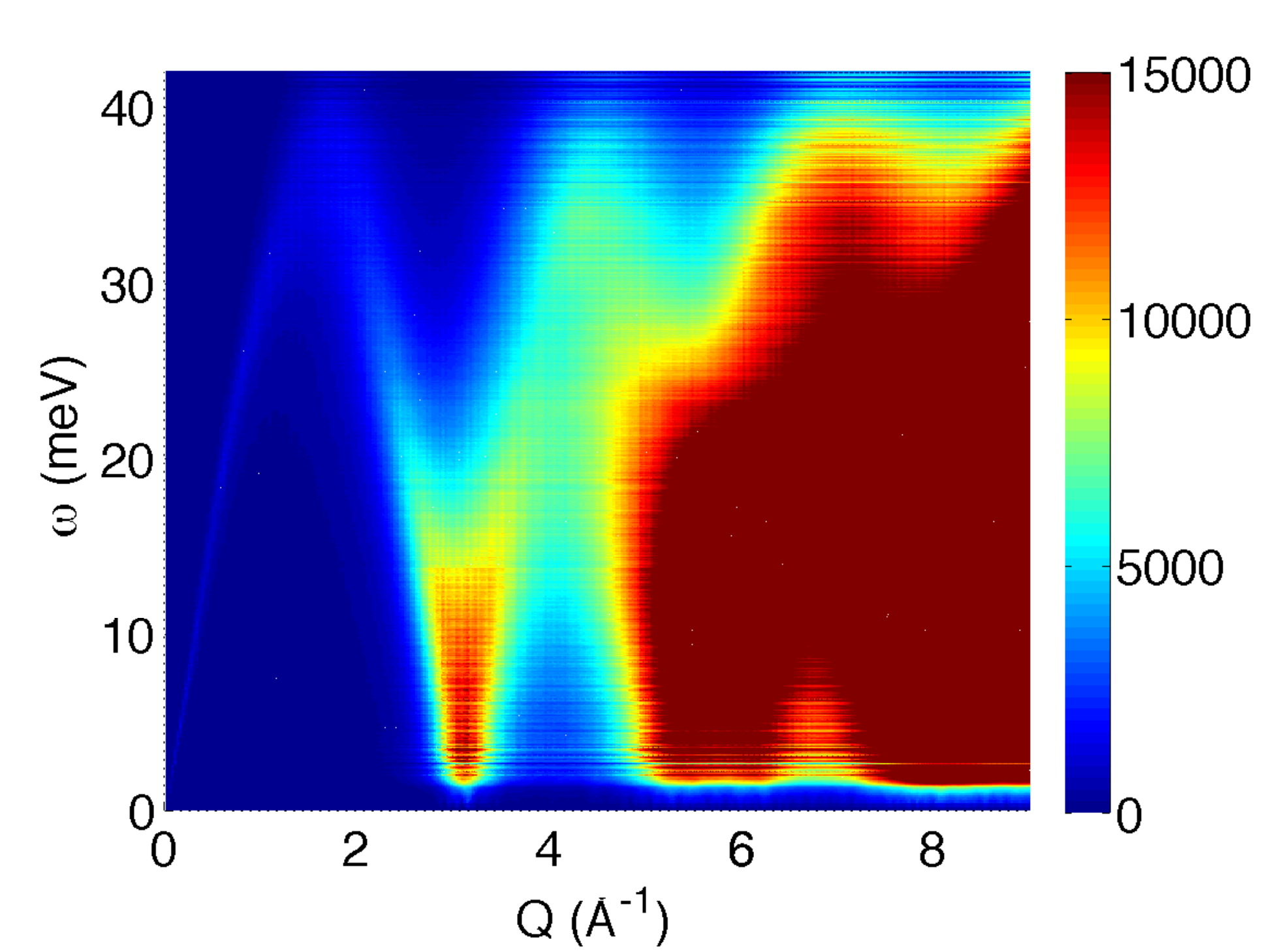} 
\includegraphics[width=4.25cm, height=4.0cm]{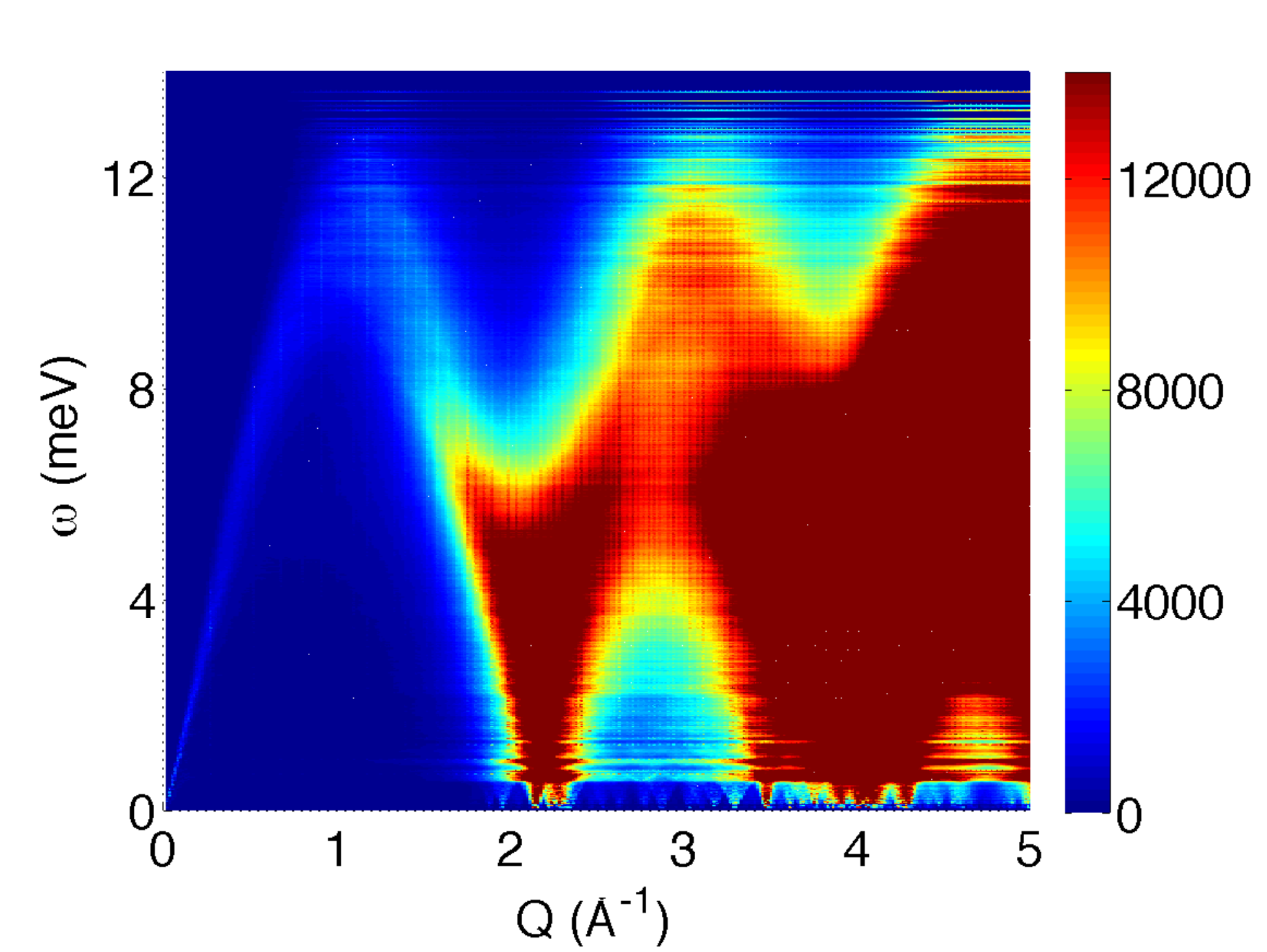} 
\caption{Collective excitation spectra from INS simulations for Fe and Ar (left to right). 
The graphs on the top show the spectra of crystalline structure and the bottom the spectra of the supercooled liquid. 
The colour-scale indicates the high and low scattering intensity regimes.}
\label{fig:INSsims}
\end{center}
\end{figure}

\begin{figure}[h!] 
\begin{center}
\includegraphics[width=4.24cm, height=3.8cm]{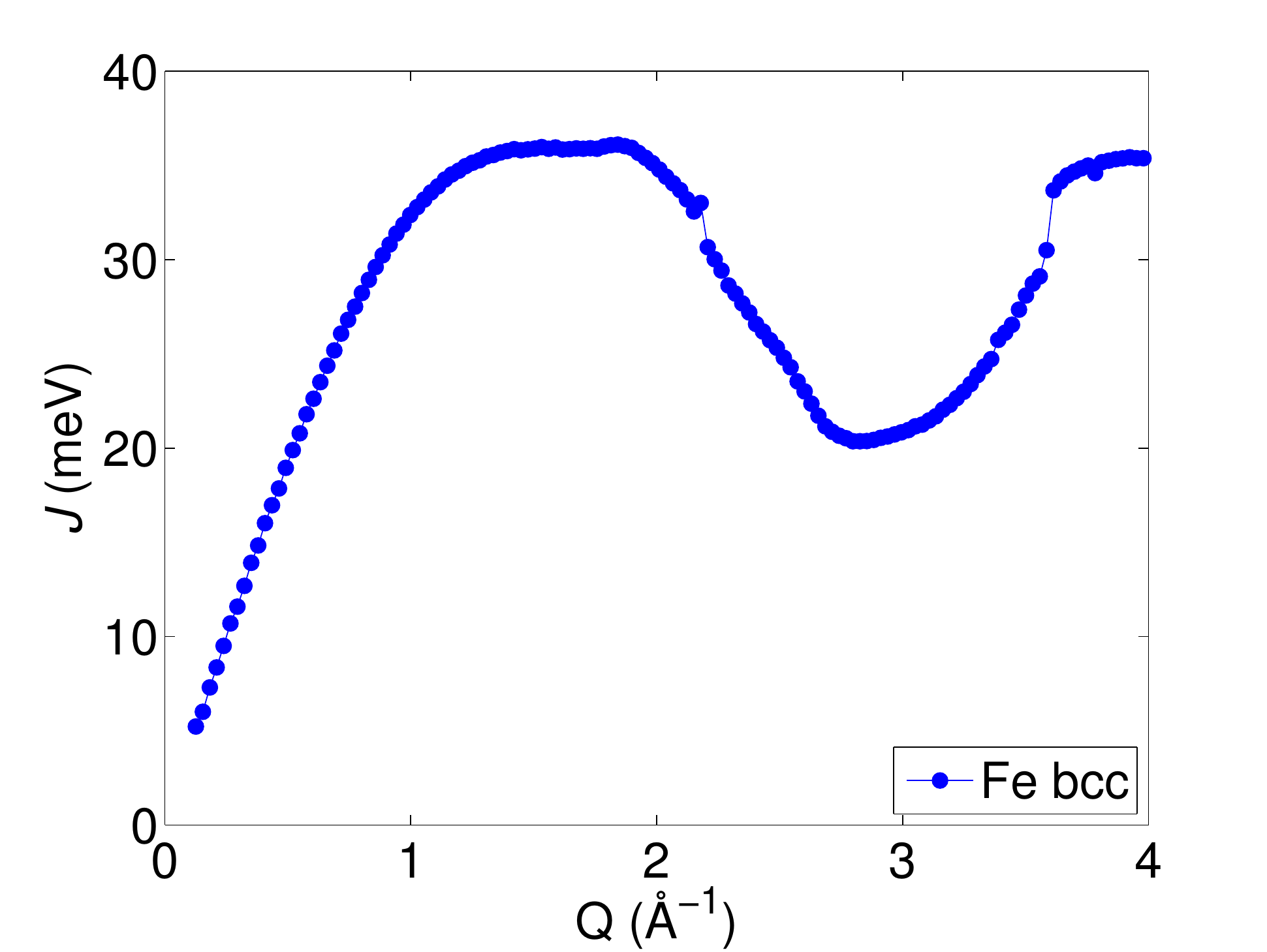} 
\includegraphics[width=4.24cm, height=3.8cm]{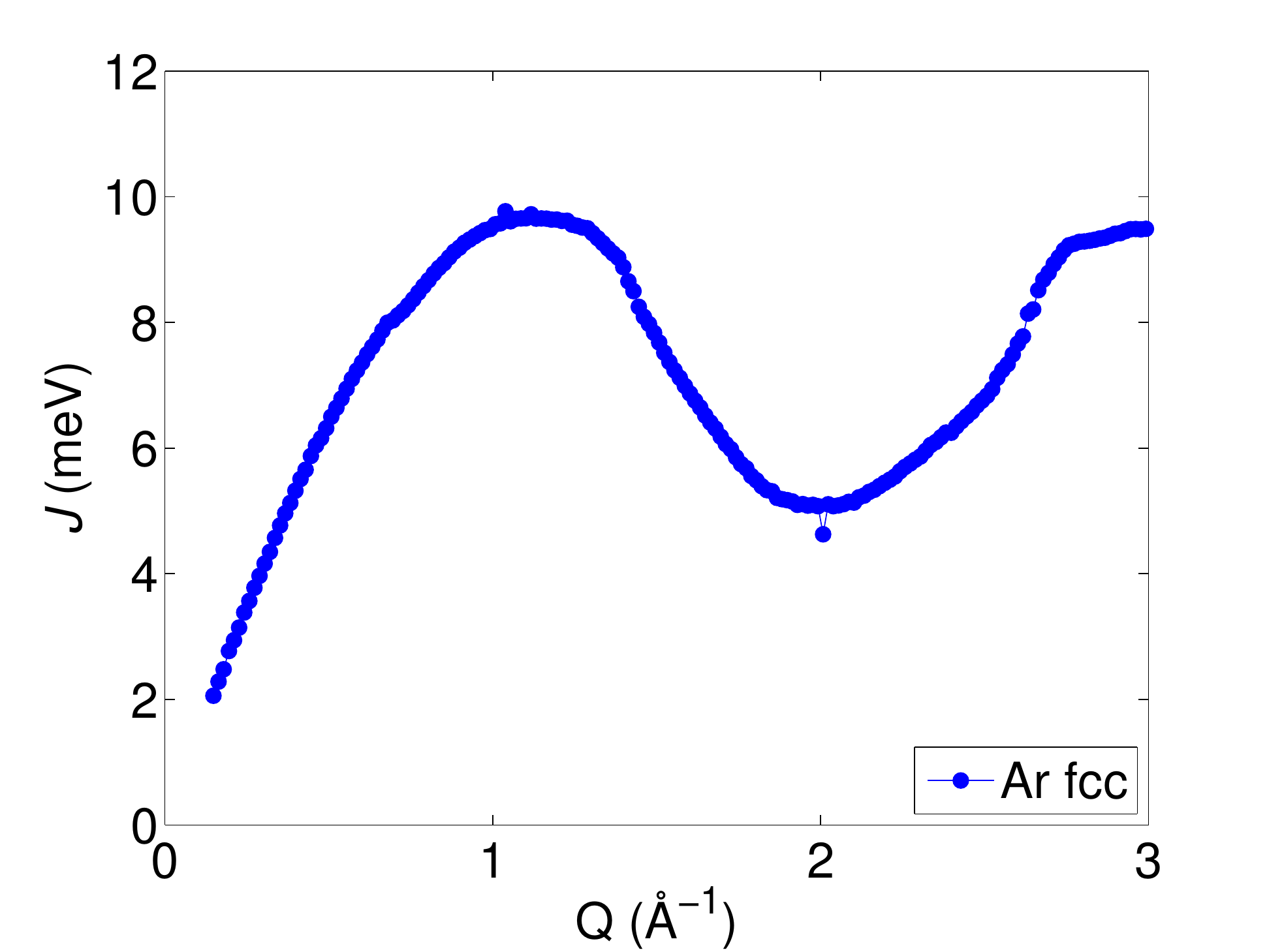} 
\includegraphics[width=4.24cm, height=3.8cm]{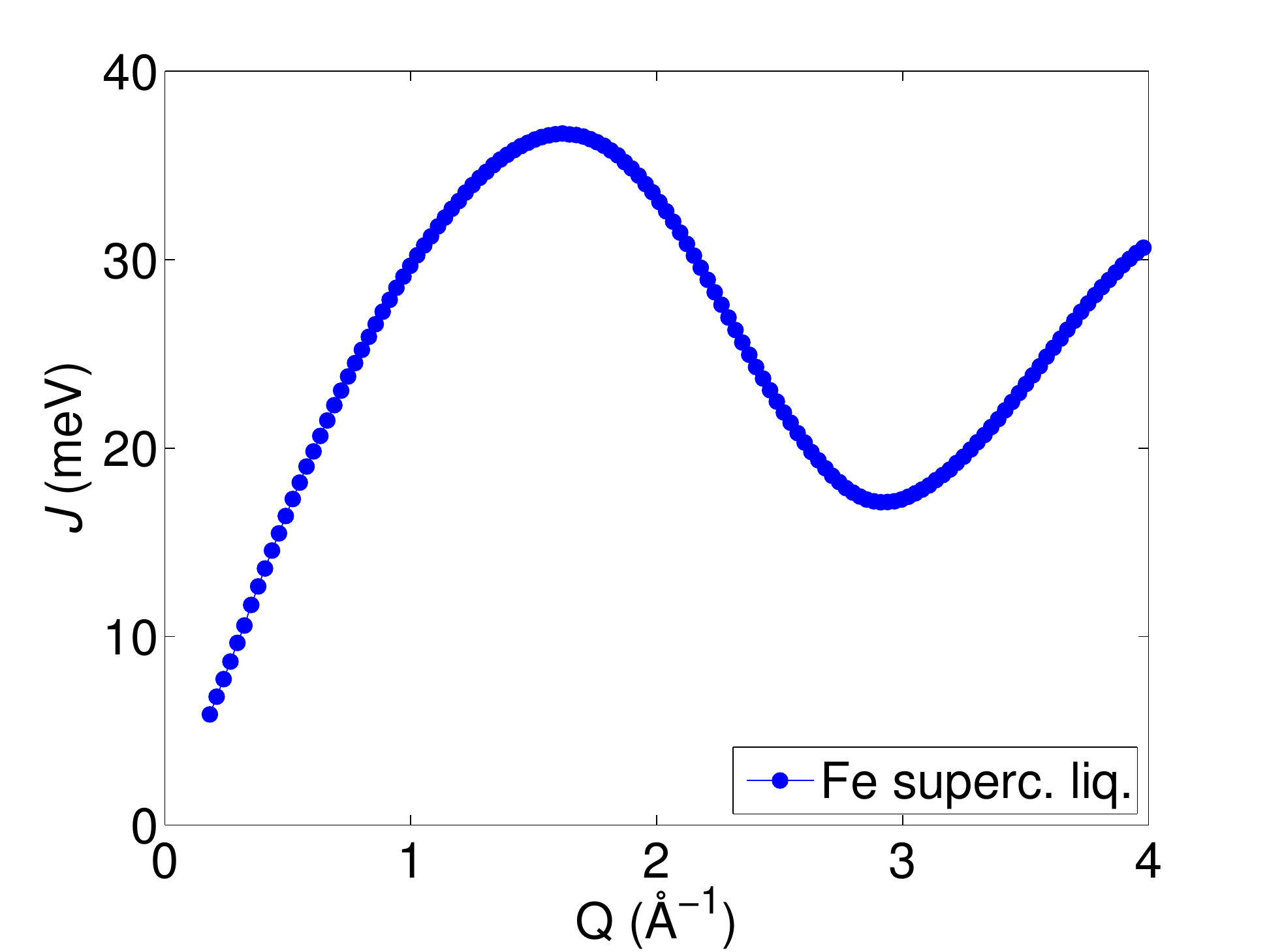} 
\includegraphics[width=4.24cm, height=3.8cm]{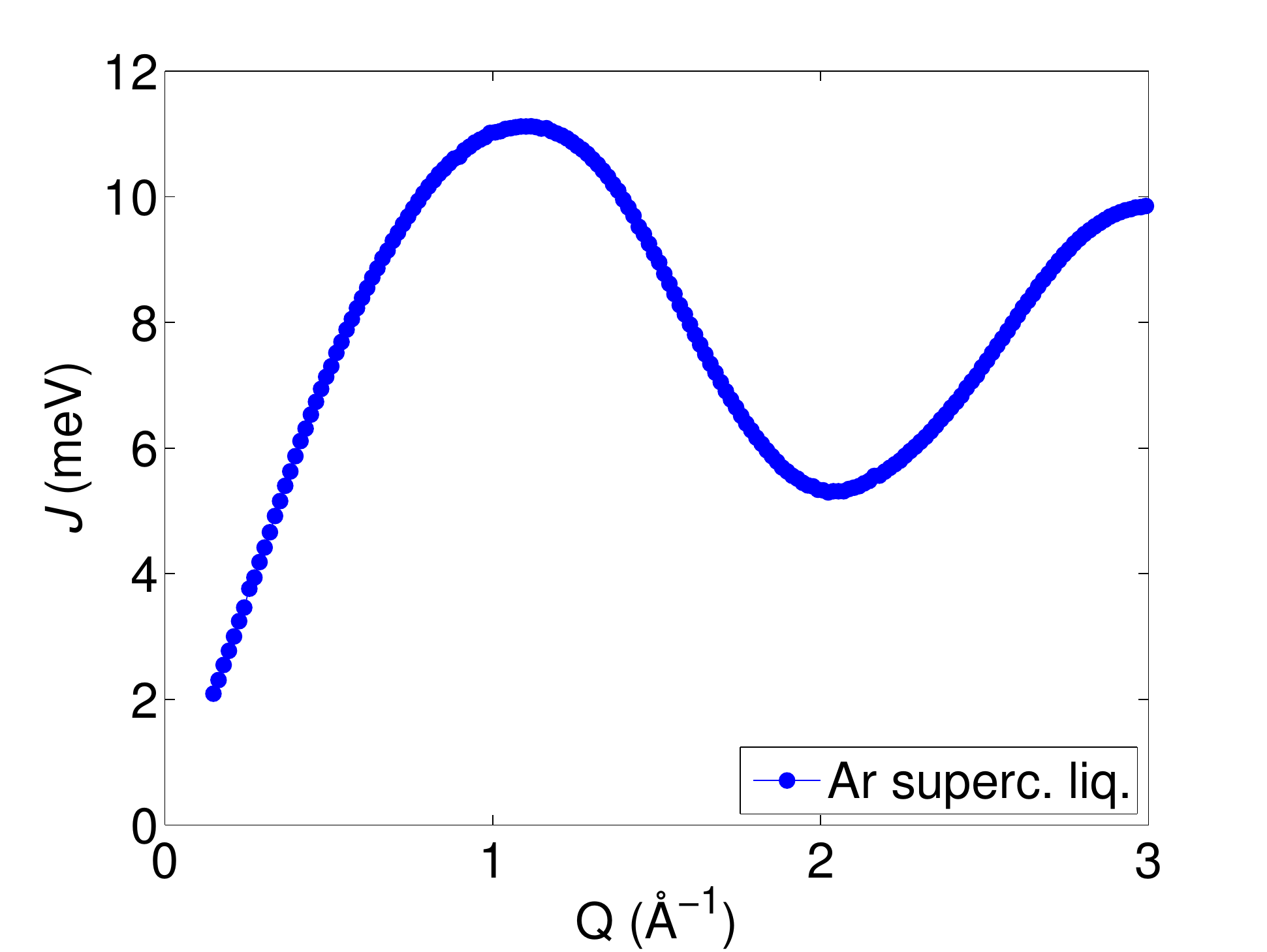} 
\caption{$J(\mathit{Q},\omega)$ of Fe and Ar for the crystalline and supercooled liquid structure.
The phonon-roton minimum is located at similar \textbf{Q} for both phases.}
\label{fig:DHOsims}
\end{center}
\end{figure}

In Fig. \ref{fig:INSsims} we present results from INS simulations of Fe and Ar (left to right plots) while results for K, Ta and W can be found in Appendix A.
The top plots represent the crystalline structure and the bottom plots the supercooled liquid structure.
Similarly to Fig. \ref{fig:INSexp} the plots show the intensity of frequency with respect of the scattering vector in a colour-scale map.
In these plots we have calculated only the energy gain of the acoustic spectra.
For all materials the crystalline structure exhibits distinct phonon dispersion curves while the supercooled liquid structure exhibit collective excitations.
In all plots we can clearly see the linear increase of the phonon frequencies at low \textbf{Q}, followed by the formation of a minimum.\par

In Fig. \ref{fig:DHOsims} we show the $J(\mathit{Q},\omega)$ of Fe and Ar from INS simulations, similarly to the calculations presented in Fig. \ref{fig:DHOexpall}. Results for K, Ta and W are presented in Appendix A.
The top plots show the $J(\mathit{Q},\omega)$ for the crystalline structure while the bottom graphs the supercooled liquid structure.
All curves follow the same pattern starting with a linear increase and followed by a minimum.
Although the minimum \textbf{Q} point for the crystalline structure cannot be accurately defined for all materials, due to overlap of several dispersion curves, it is located at similar \textbf{Q} value as this of the supercooled liquid for each material. 
We present the calculated values for all materials in Table \ref{table:Qmin}, where $\mathrm{Q}^{\mathrm{cr}}_{\mathrm{min}}$ and $\mathrm{Q}^{\mathrm{sl}}_{\mathrm{min}}$ are the minimum \textbf{Q} for the crystalline and supercooled liquid structures respectively.
We calculate the value $2\pi / \mathrm{Q}^{\mathrm{cr}}_{\mathrm{min}} (\mathrm{\AA})$ for each material for the crystalline phase and compare it with the nearest neighbour distance ($d$) calculated from the $g(r)$ in the next section.
For all studied bcc materials the difference of $2\pi / \mathrm{Q}^{\mathrm{cr}}_{\mathrm{min}} (\mathrm{\AA})$ and \textit{d} is less than 10$\%$ and 14$\%$ for the fcc.\par

\begin{table}
\caption{Values of $\mathbf{Q}$ for the crystalline and supercooled structures at the phonon-roton minimum, extracted from from Figs. \ref{fig:DHOsims} and \ref{fig:DHOsimsAppA}. 
The bond length ($d$) is extracted from Figs. \ref{fig:pdfnna} and \ref{fig:pdfnnaAppA}. }\label{table:Qmin} 
\begin{tabular}{l c c c c }
\hline
\hline
Material \ & $\mathrm{Q}^{\mathrm{cr}}_{\mathrm{min}} (\mathrm{\AA^{-1}})$ \ & $\mathrm{Q}^{\mathrm{sl}}_{\mathrm{min}} (\mathrm{\AA^{-1}})$ \ & $2\pi / \mathrm{Q}^{\mathrm{cr}}_{\mathrm{min}} (\mathrm{\AA})$ \ & $d (\mathrm{\AA})$ \\
\hline
Fe & 2.86 & 2.91 & 2.20 & 2.44\\
Ar & 2.04 & 2.02 & 3.08 & 3.59\\
K & 1.54 & 1.38 & 4.08 & 4.49\\
Ta & 2.44 & 2.54 & 2.58 & 2.84\\
W & 2.46 & 2.66 & 2.55 & 2.71\\
rscFe & 1.64 & 1.66 & 3.85 & 4.27\\
\hline
\hline
\end{tabular}
\end{table}

\section{III. Short-range order}
\subsection{A. Population distribution}

For the previously discussed materials we analyse their atomic arrangements for the crystalline, supercooled liquid and liquid structures with two methods: the calculation of the $g(r)$ and the nearest neighbour analysis (NNA).
The NNA is defined as the definite integral of the $g(r)$ for a specific area, from $r_1$ to $r_2$.
\begin{equation} \label{eq:IntPDF}
n_{\mathrm{AB}}(r) = \int^{r_2}_{r_1} 4\pi r^2 \rho_{\mathrm{B}} g_{\mathrm{AB}}(r) \mathrm{dr}
\end{equation}
where for two atom types A and B, $n$ is the number of atoms of type B that can be found in a distance of radius $r$, $\rho_{\mathrm{B}}$ is the density of atoms B.
It is known that for bcc crystals there are 8 first and 6 second neighbour atoms while for fcc structures there are 12 first and 6 second neighbour atoms. \par

\begin{figure}[t]
\begin{center}
\includegraphics[width=4.2cm, height=3.70cm]{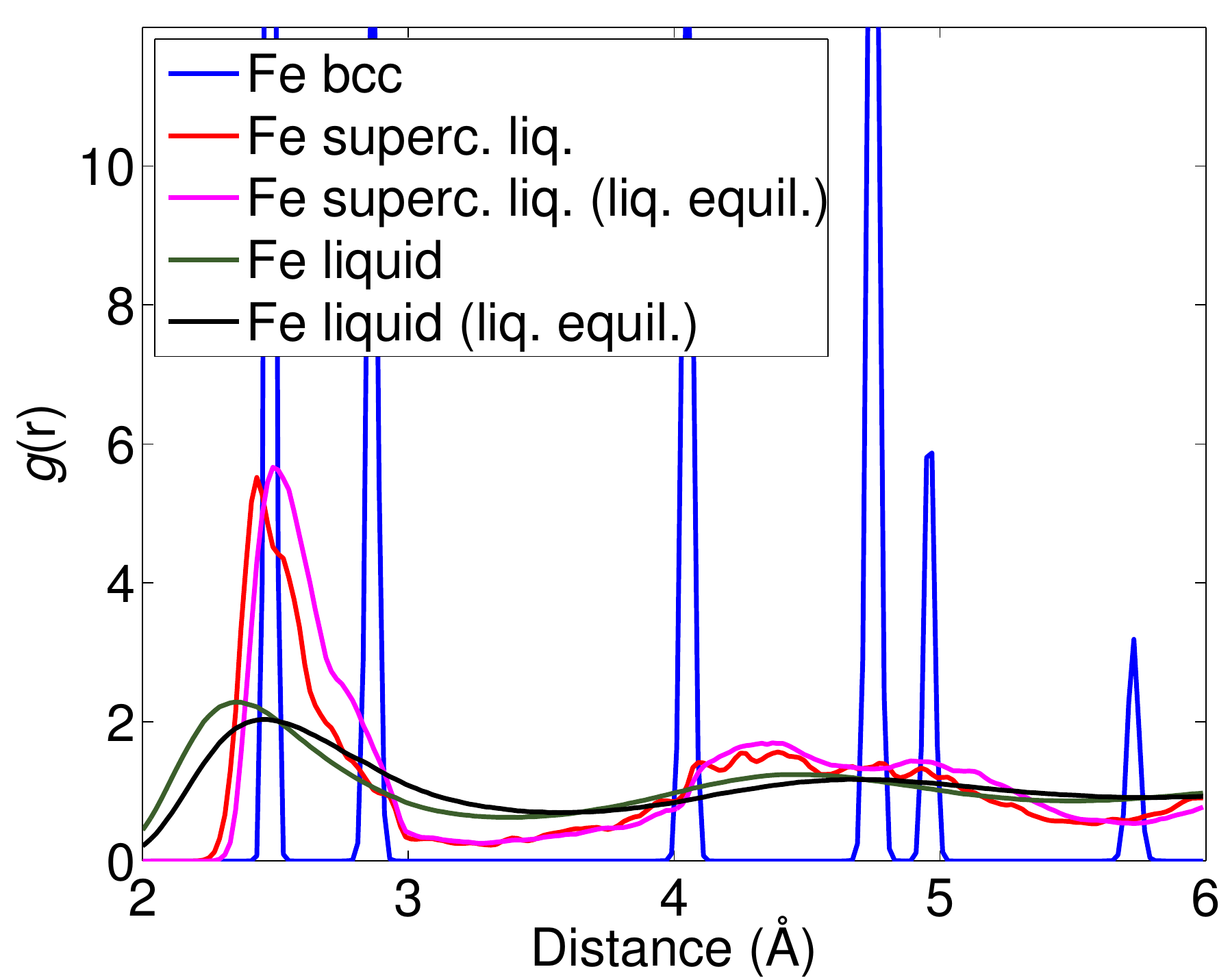} 
\includegraphics[width=4.2cm, height=3.70cm]{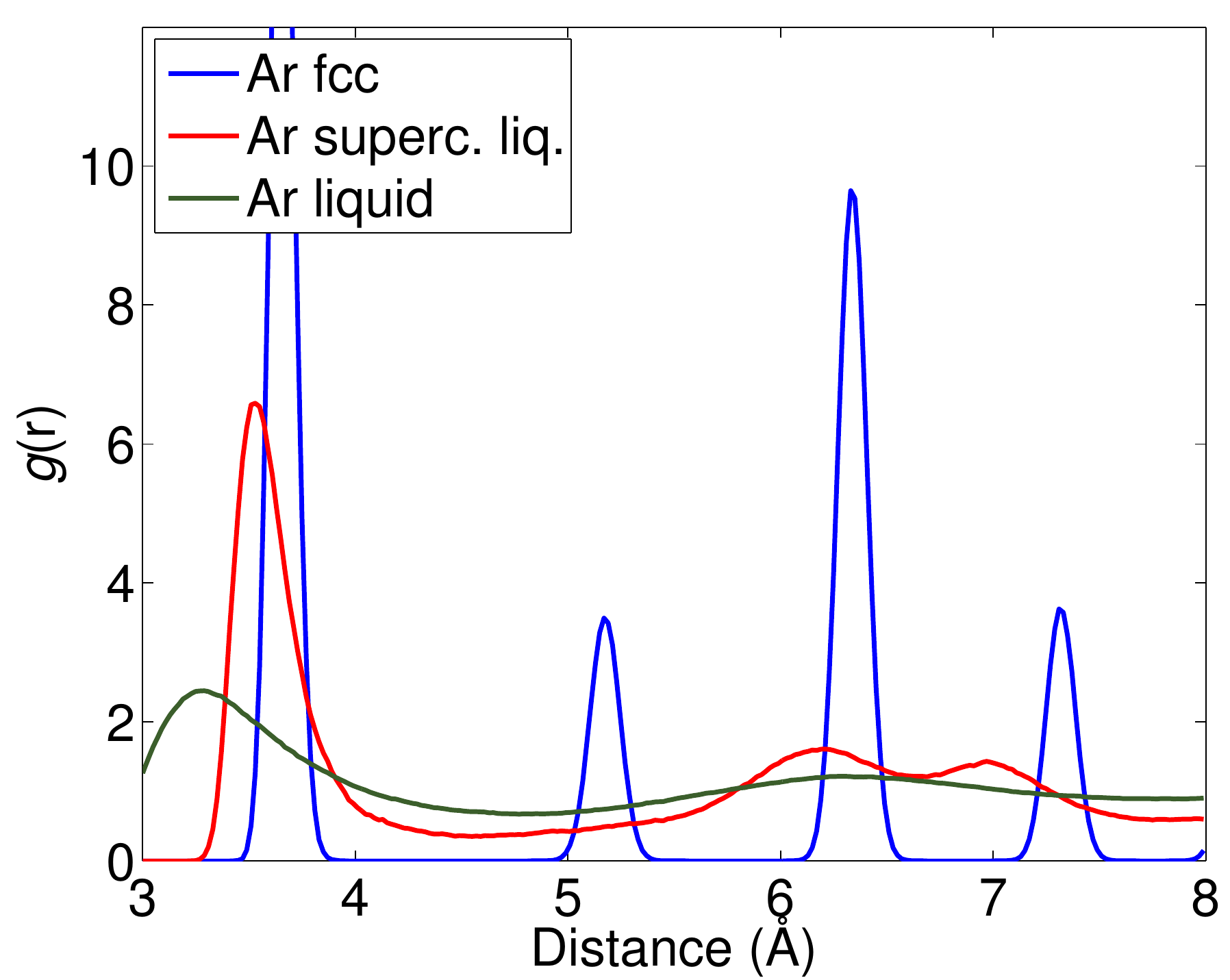} 
\includegraphics[width=4.2cm, height=3.70cm]{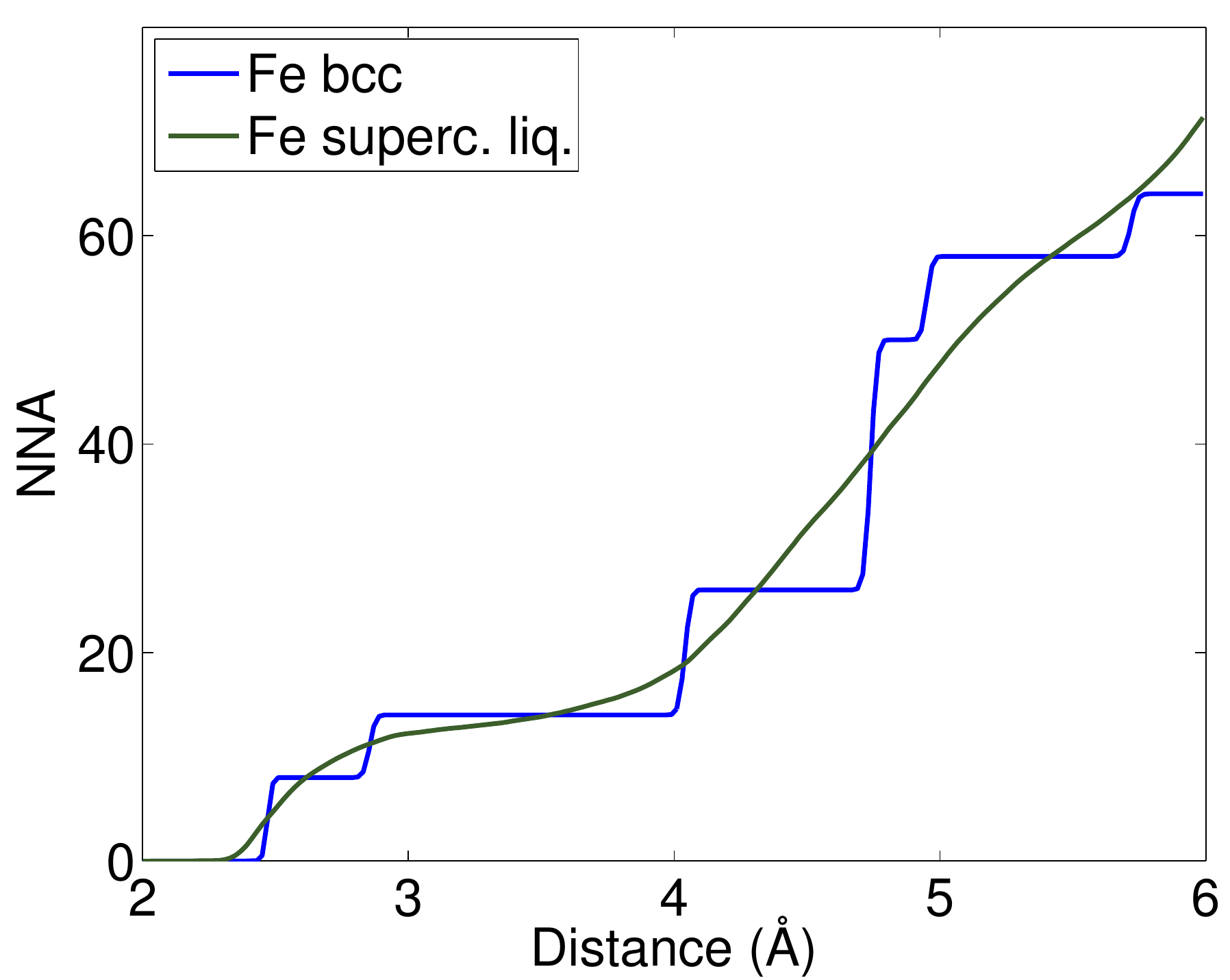} 
\includegraphics[width=4.2cm, height=3.70cm]{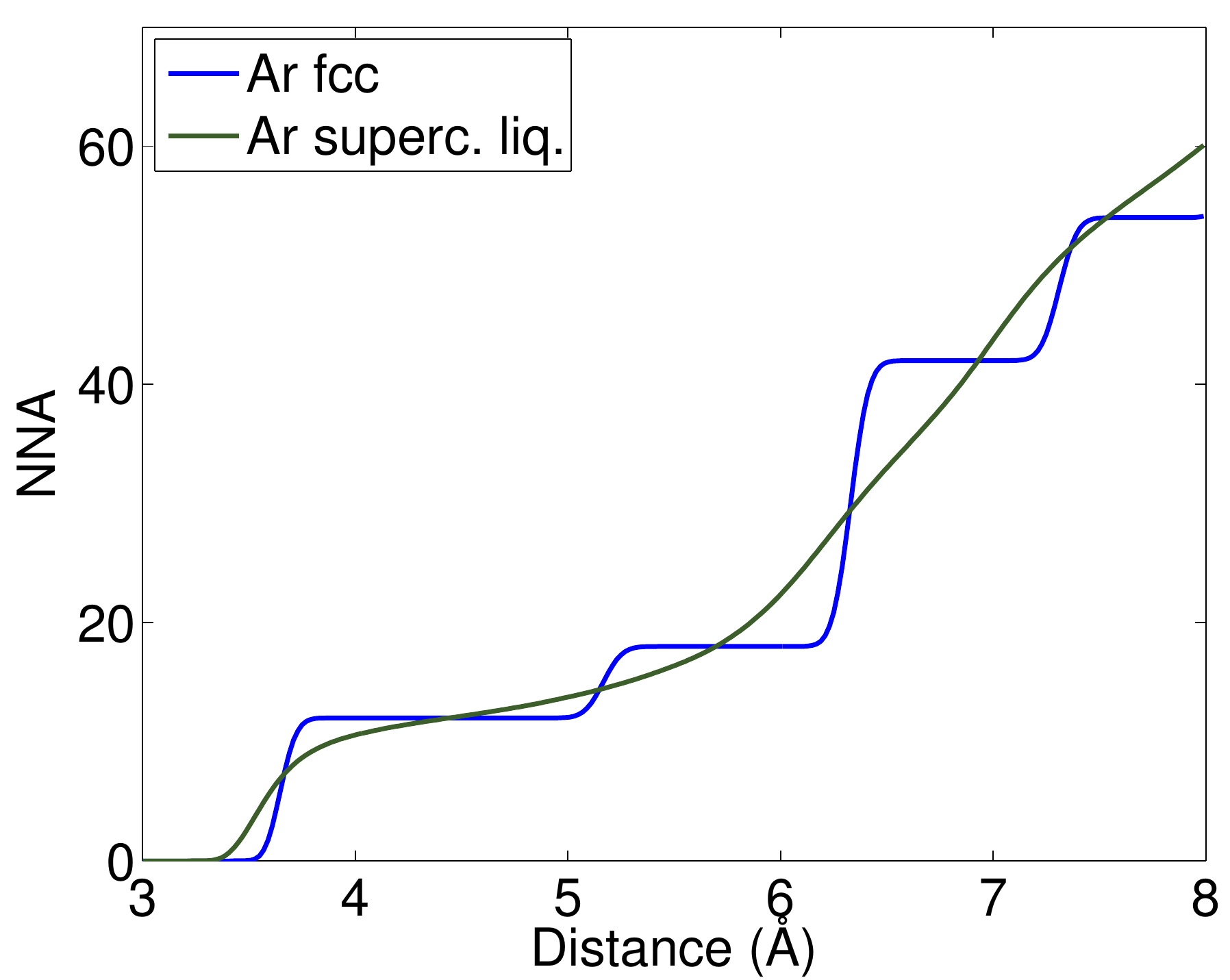} 
\caption{$g(r)$ and NNA results for crystalline, supercooled liquid and liquid Fe and Ar from MD simulations.
The $g(r)$  for Fe includes results for the supercooled liquid and liquid structure initially prepared with two different methods.}
\label{fig:pdfnna}
\end{center}
\end{figure}

In Fig. \ref{fig:pdfnna} we present results from calculations of the $g(r)$ and NNA for Fe and Ar.
As it is expected from the theory the $g(r)$ for the crystalline structure is well defined with well defined distances for the neighbour atoms while for the other two structures there is continuity in the curve, showing a not well-defined structure.
For the crystalline structure the NNA is in good match with the $g(r)$, exhibiting sharp increase at distances where the $g(r)$ curve exhibits a peak and straight plateaus where the $g(r)$ curve is zero.
Despite the fact that for the both supercooled liquid structure the NNA curve is smoother with less well-defined plateaus, as it is for the crystalline structure, it exhibits two main features. 
First, the supercooled liquid curve exhibits a different behaviour which depends on the initial crystalline phase of the material.
This can be understood better if we compare the behaviour of the supercooled liquid NNA curves for Fe and Ar.
Second, the number of atoms at the short-range limit (a distance of about four nearest neighbour atoms) for both structures is very similar.
This can be seen for both Fe and Ar.
This shows a consistency in the local atomic density regardless of the structural phase.
Similar analysis for results from population distribution analysis for K, Ta and W can be found in Fig. \ref{fig:pdfnnaAppA} in Appendix A. \par

For the preparation of the supercooled liquid and liquid structures we applied a different equilibration method to the previously described one. 
The new method, noted as "(liq. equil.)" in Fig. \ref{fig:pdfnna}, follows the same procedure as in the standard equilibration but at temperature above the melting point instead of room temperature.
In this method the system does not have an initial well defined structure that could possibly affect the local atomic arrangement.
Despite the different equilibration method the $g(r)$ for the supercooled liquid and liquid structures exhibit very similar results and prove that our simulations are independent on the preparation method.\par

\subsection{B. Bond angle distribution}

To investigate further the short-range order of the three phases we analyse the local bonding.
Similar analysis has been used in the past for a number of cases comparing the supercooled liquid and liquid structures \cite{Steinhardt1983, Tomida1995, Price2010} but direct comparison with the crystalline phase is limited in literature. 
We perform classical MD simulations comparing the crystalline and the supercooled liquid structures, showing evidence of local order dependence of the two phases.
For the previously mentioned materials we measure the angle between two consecutive neighbour atoms and compare.
The cut-off distance we use for the angle calculation for both structures is defined from the $g(r)$ of the crystal and it corresponds to the first neighbour distance.\par

\begin{figure}[t]
\begin{center}
\includegraphics[width=4.2cm, height=3.90cm]{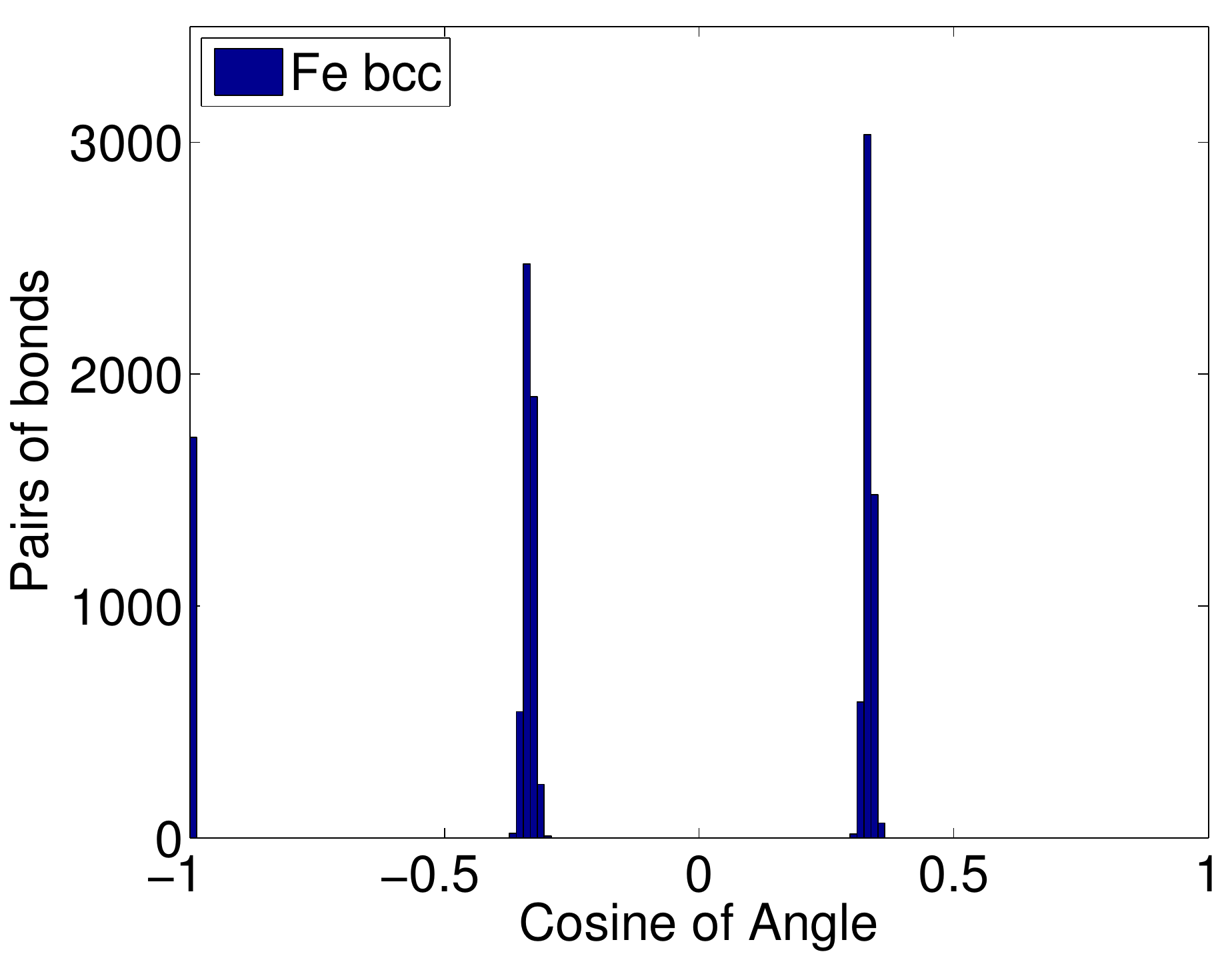} 
\includegraphics[width=4.2cm, height=3.90cm]{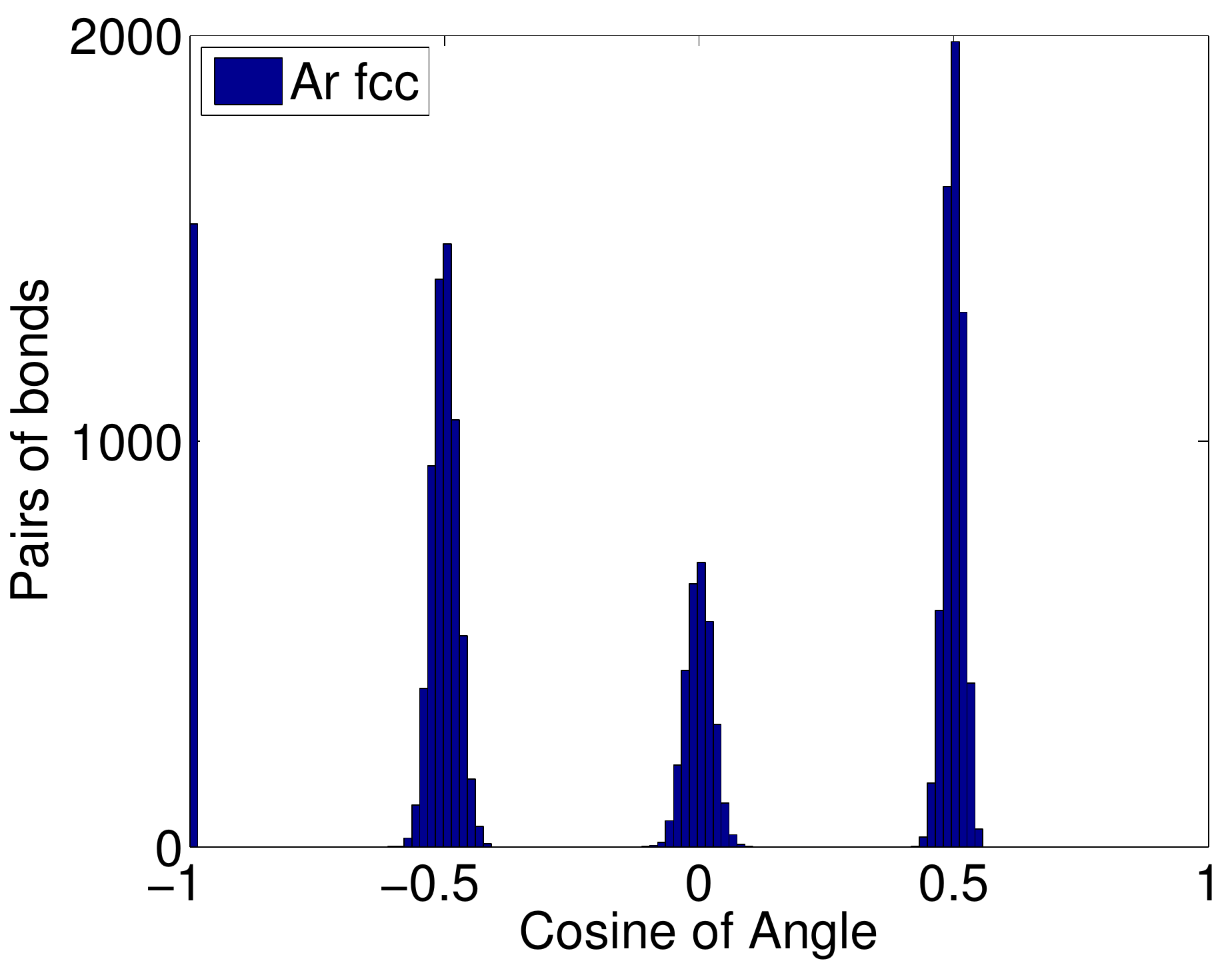} 

\includegraphics[width=4.2cm, height=3.90cm]{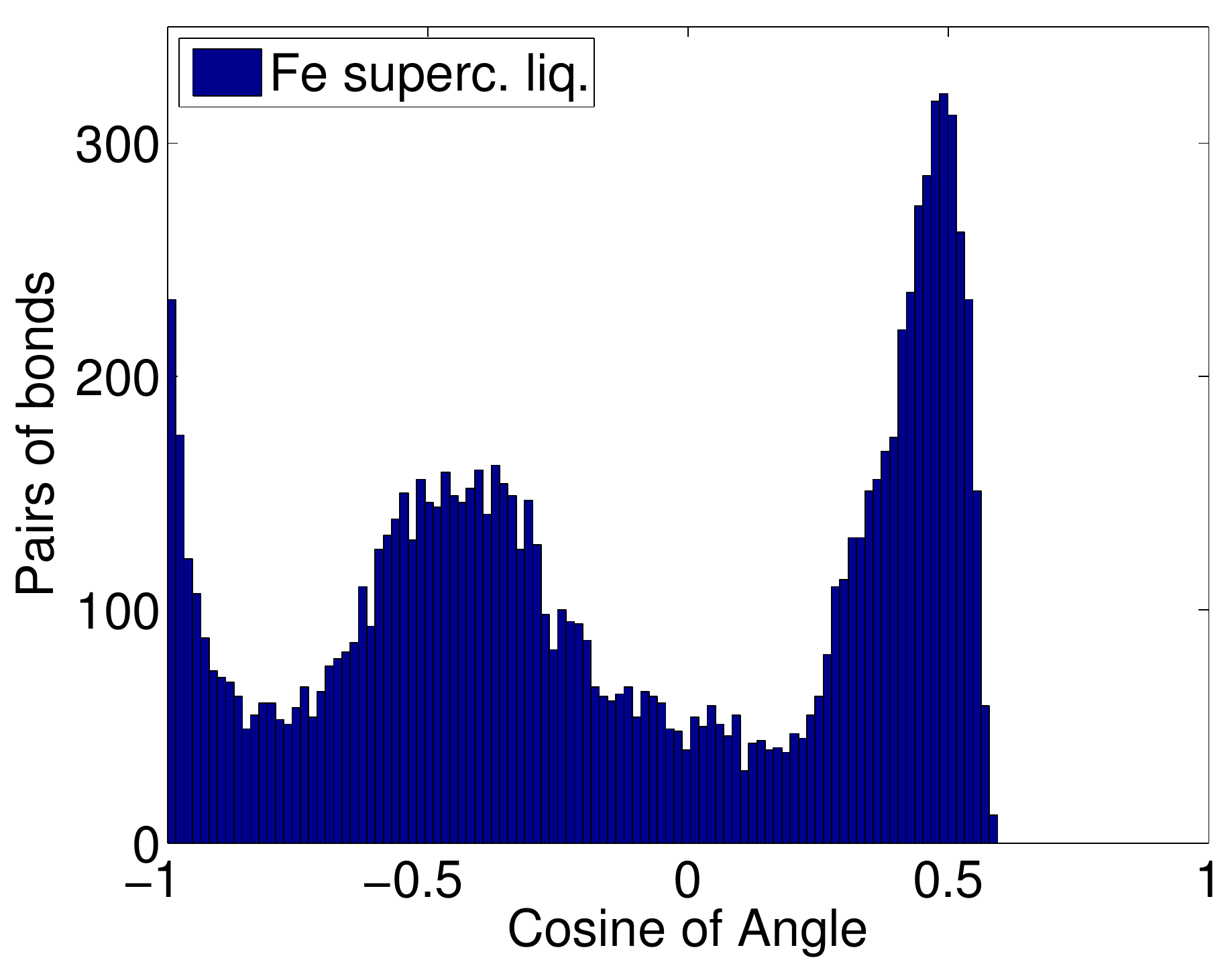}
\includegraphics[width=4.2cm, height=3.90cm]{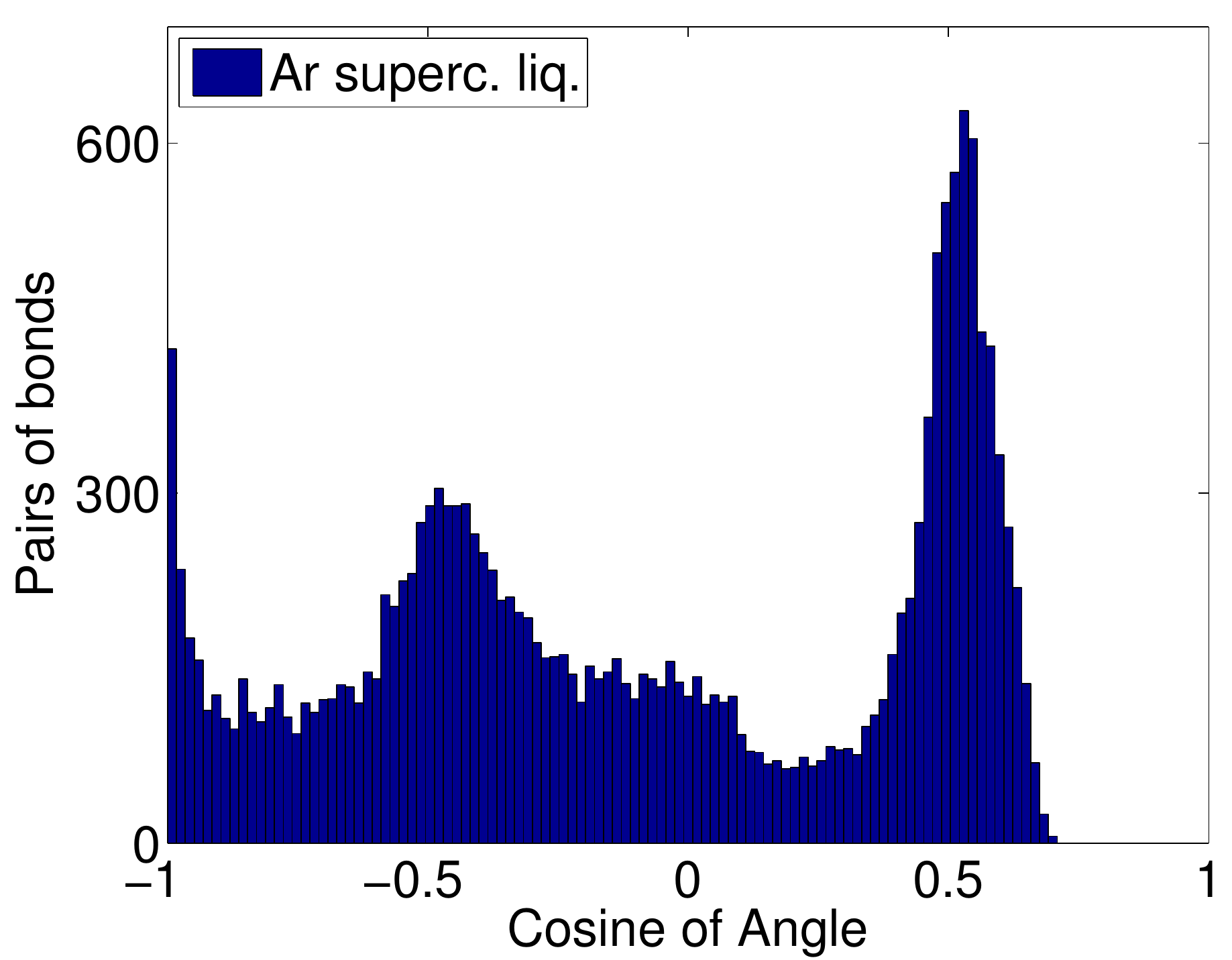}

\includegraphics[width=4.2cm, height=3.90cm]{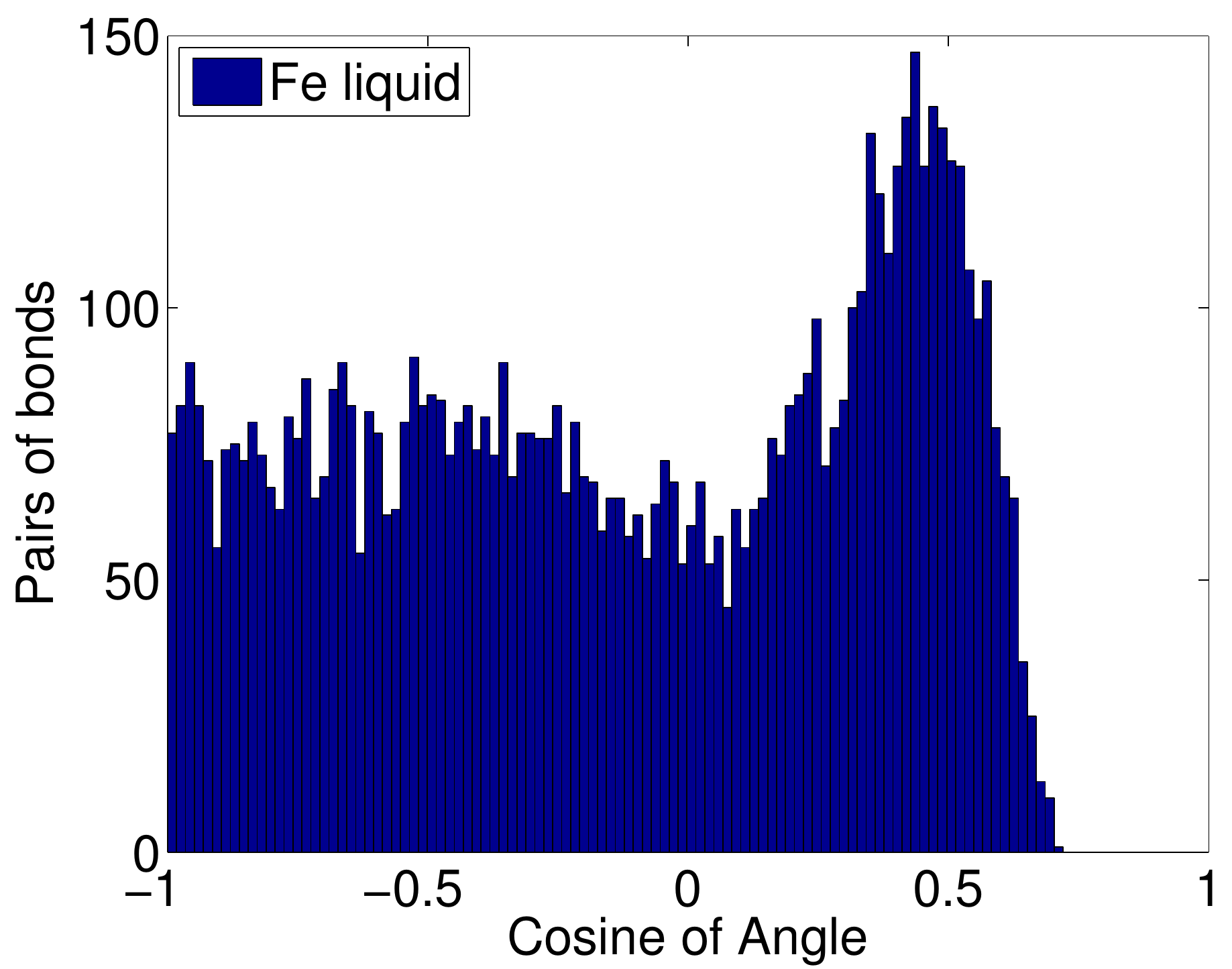}
\includegraphics[width=4.2cm, height=3.90cm]{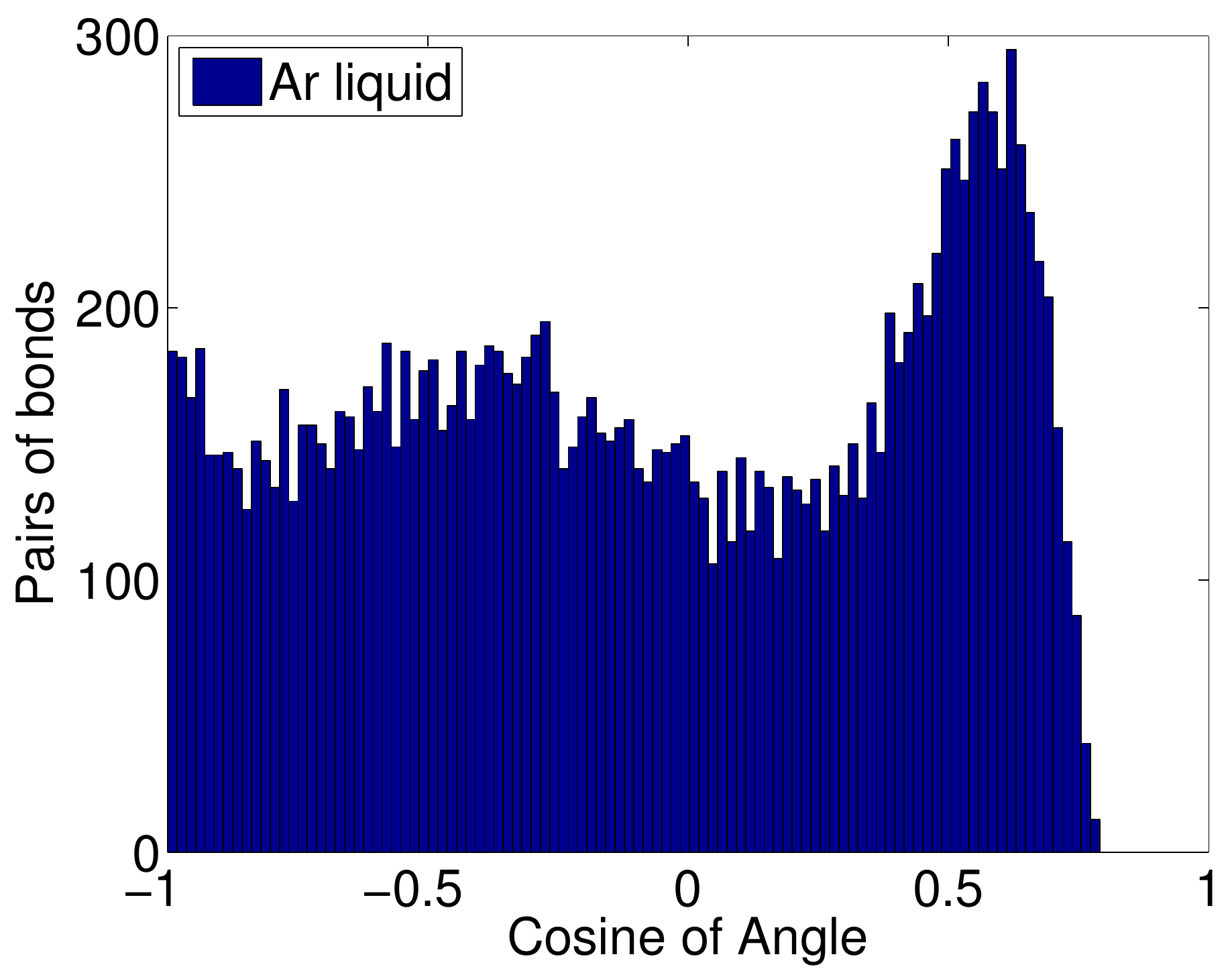}
\caption{Nearest neighbour angle distribution of Fe and Ar at the crystalline, amorphous and liquid phases calculated from MD simulations.
The supercooled liquid maintains local order features from both the liquid and crystalline phases showing evidence of dependence on the crystalline structure.}
\label{fig:angles}
\end{center}
\end{figure}

In Fig. \ref{fig:angles} we present histograms  for Fe and Ar projecting the amount of pairs of bonds as function of the cosine of the degree of the angle for each pair of bond for the crystalline, supercooled liquid and liquid structures.
The histograms for K, Ta and W can be found in Fig. \ref{fig:anglesAppA} in Appendix A.
For all histograms the bin size is 1.8$^\circ$.
For bcc crystals there are three characteristic angles between two nearest neighbour bonds of 70.53$^\circ$, 109.47$^\circ$ and 180.00$^\circ$.
For a fcc crystal there are four characteristic angles of 60.00$^\circ$, 90.00$^\circ$, 120.00$^\circ$ and 180.00$^\circ$.
Due to atomic vibration during MD simulations the atoms are displaced from their equilibrium position even at 10 K where the kinetic energy is low.
As a result the calculated angles have a small variation from the theoretical values.
The calculation of the cosine of the angles summarises all the similar angles, provides results for small groups of angles and gives a better understanding of the main angle the atoms move around. \par

As we can see in Fig. \ref{fig:angles}, both materials in crystalline phase exhibit the characteristic angles of the bcc and fcc phase accordingly.
In the amorphous phase both systems are expected to be totally disordered since the system was initially liquefied before instantly frozen at 10 K.
Instead we can see that for both materials the supercooled liquid structure maintains the major crystalline angles, showing a remaining local order that depends on the crystalline structure.
In the liquid phase the degree of freedom is higher and the atoms are free to flow in the system.
The angle distribution is even wider than this of the amorphous phase but we can observe a similar trend in the formation of the main peaks. 
Evidence of short-range order dependence of the supercooled liquid and liquid structures on the initial crystalline structure has been observed in a number of cases, such as in Zr \cite{Jaske2003} and Cu \cite{Ganesh2006}, Fe and W \cite{Ganesh2008} from high quality \textit{ab initio} MD simulations and in Cu, Ni and Au from classical MD simulations \cite{Murin2014}.
Our results agree with previous findings in literature and demonstrate  the good predictive power of our model. \par

\section{IV. The origin of phonon-roton minima}
 
The term roton minimum was initially introduced by Landau \cite{Landau1947} to describe the elementary excitations in superfluid $^4$He.
He suggested that the phonon excitation should not be monotonic but it starts with a longitudinal phonon branch with linear initial gradient which reaches a maximum value and then turns towards a deep minimum on increasing wave vector.
At the minimum the phonon energy is equal to the absolute value of the wave vector times the critical flow velocity.
Feynman \cite{Feynman1954} and later Feynman and Cohen \cite{Feynman1956} suggested that the roton is a quantum-mechanical analog of a microscopic vortex ring, of diameter roughly equal to the atomic spacing.
It is known nowadays this model gives only qualitative agreement with experiments.\par

Glyde and Griffin \cite{Glyde1990} developed Landau's idea to relate rotons with the existence of a Bose condensate, and following Miller, Pines, and Nozi\`{e}res \cite{Miller1962} and Burke, Major and Chester \cite{Burke1967} they presented the roton as a renormalised single-particle excitation 
mixed with the density fluctuation spectrum for $T < T_{\lambda}$, where $T_{\lambda}$ is the superfluid transition temperature. 
Therefore one might expect distinct changes in the excitation spectrum on cooling through $T_{\lambda}$.
Different behaviour of the neutron scattering in the phonon and roton region motivated Glyde and Griffin to propose that the phonon--maxon--roton excitation curve arises from two quite distinct processes, a zero sound mode at small scattering vector and a single particle mode beyond the maxon scattering vector.
This has been used extensively to analyse neutron scattering results for liquid $^4$He \cite{Andersen1994, Gibbs1996, Tozzini1999}, but there are cases with qualitative disagreement with this model \cite{Svensson1996}.\par 

Two interesting theories on the origin of roton minimum and the physical mechanism of its formation was pointed out by Schneider and Enz \cite{Schneider1971} and Celli and Ruvalds \cite{Celli1972}.
In Ref. \onlinecite{Celli1972} the authors mention that the roton minimum is not characteristic of $^4$He only but is observed in non-superfluid materials as well, suggesting that any theory on rotons should apply to phonons and maxons as well. In Ref. \onlinecite{Schneider1971}, the authors refer to the roton as a soft mode, something that later has been proposed as the cause of dynamic instability for the creation of a new phase of calcite \cite{Dove1992}.
Based on these ideas, Galli, Cecchetti and Reatto  \cite{Galli1996} and later Nozi\`{e}res \cite{Nozieres2004} stated that the origin of the roton minimum arises from strong correlations dominating the system and that the roton is not behaving exactly as a single particle excitation.
Interestingly  enough, Nozi\`{e}res discusses superfluid $^4$He and the roton minimum by starting from the solid phase rather than the dilute gas as before.
In Ref. \onlinecite{Galli1996} the authors propose that the roton is similar to a single particle excitation but for wave vector larger than the roton minimum regime where the excitation is quite distinct from a single particle excitation due to interference effects between atoms. \par

Following Nozi\`{e}res's ideas, Kalman \textit{et al.} \cite{Kalman2010} studied the roton minimum and strong correlations in classical systems.
They calculated the phonon dispersion curves of a classical system from classical MD simulations, focusing on the longitudinal phonon dispersion curve and compared them with results from quantum simulations.
Their results show that there is no major difference in the dispersion curves from quantum and classical systems and they conclude that the roton minimum is a classical phenomenon.
They argue that dispersion curves explained through strong correlations in a classical system cannot be described by vortices or single-particle excitations.
Based on that and on results from collective excitation spectra they discuss roton as a universal effect. \par

Before Kalman \textit{et al.} \cite{Kalman2010} comment on roton's universality, Tozzini and Tosi \cite{Tozzini1999} studied the phonon dispersion curves of crystalline He.
The authors commented on the extent to which phonons and rotons of the superfluid $^4$He reflect the longitudinal and transverse phonons of the corresponding crystal.
From Tozzini's and Tosi's results, as well as from references in their paper (\textit{e.g.} Ref. \onlinecite{GlydeBook1994}) we understand that the roton minimum, or more general the phonon-roton minimum, is a classical phenomenon and it exists for both the crystalline and liquid phases of a material, not only for superfluid materials but for a wide variety of strongly coupled materials. \par

Our own results for barium, coupled with recent results on sodium \cite{Monaco2010}, show clearly that the bcc metals show dispersion curves that are similar in both the polycrystalline and liquid phases. 
Our simulations of other materials presented here show that this is a general feature. 
The origin of the minimum in the metals can easily be understood with reference to the longitudinal acoustic phonon dispersion curves for wave vectors along directions parallel to the nearest-neighbour bonds. 
In the case of bcc metals this is for wave vectors along $\{111\}$, and in the cubic-close-packed metals this is for wave vectors along $\{110\}$. In the latter case the minimum occurs at the Brillouin zone boundary, but in the case of the bcc metals the minimum occurs at a wave vector two-thirds of the way towards the Brillouin zone boundary. 
In both cases, the minimum corresponds to the wave vector at which the relevant phonon phase factors mean that the nearest-neighbour atoms lying along the direction of the wave vector move together. The lack of compression of the bonds in such a case is reflected in the calculation of a significant minimum in the phonon dispersion curves.

In short, the minimum in the liquids of bcc or ccp metals occurs for scattering vectors $Q = 2 \pi / d$, where $d$ is a bond length. 
In any disordered state (liquid, glass of polycrystalline solid) there will be some bonds lying close to the direction of the scattering vector $\mathbf{Q}$, which plays the role of the phonon wave vector. 
The frequencies of the longitudinal phonon branches are primarily determined by the strength of the compressional force constant, but when $Q = 2 \pi / d$ the phase factors lead to cancellation of the effect of compression in the dynamical equations. 
Since this is the exact explanation for the origin of the minimum in the polycrystalline materials, and since the maps of scattering intensity we have calculated for the supercooled liquids match closely to those for the polycrystalline materials, we can surmise that the same basic mechanism explains the minimum in the liquid state. 
In the case of the bcc metals, the position of the minimum does not correspond to a pseudo-Brillouin-zone-boundary wave vector.

\section{V. Conclusion}

In this work we presented results from INS experiments and simulations of a range of simple materials, comparing their phonon spectra in their polycrystalline, amorphous and liquid phases.
Analysis of the atomic structures of the supercooled liquid phases show that the liquid states preserve similarities in their local structure with the short-range structures of the corresponding crystalline phases. 
From our experiments and simulations we also show that the phonon spectra of the liquid and polycrystalline states are similar across a wide range of values of the scattering vector, including the presence of the significant minima in the dispersion curves. 
We argue that this arises at the scattering commensurate with the nearest-neighbour distance, and corresponds to correlated motions in which neighbouring atoms whose bond lies approximately parallel to the wave vector move in phase and therefore have no compression of the bond in the vibration. 
This is consistent with the previous suggestion of Tozzini and Tosi \cite{Tozzini1999} and Kalman \textit{et al.} \cite{Kalman2010} that the roton minimum in simple monatomic liquids is a classical phenomenon.

\section{VII. Acknowledgements}
This work has been partially supported by the Science and Technology Facilities Council.
We are grateful to Dr Jon Taylor for his support during the INS experiments.

\section{Appendix A: Additional materials}

\begin{figure*}[h!]
\begin{center}
\includegraphics[width=5.5cm]{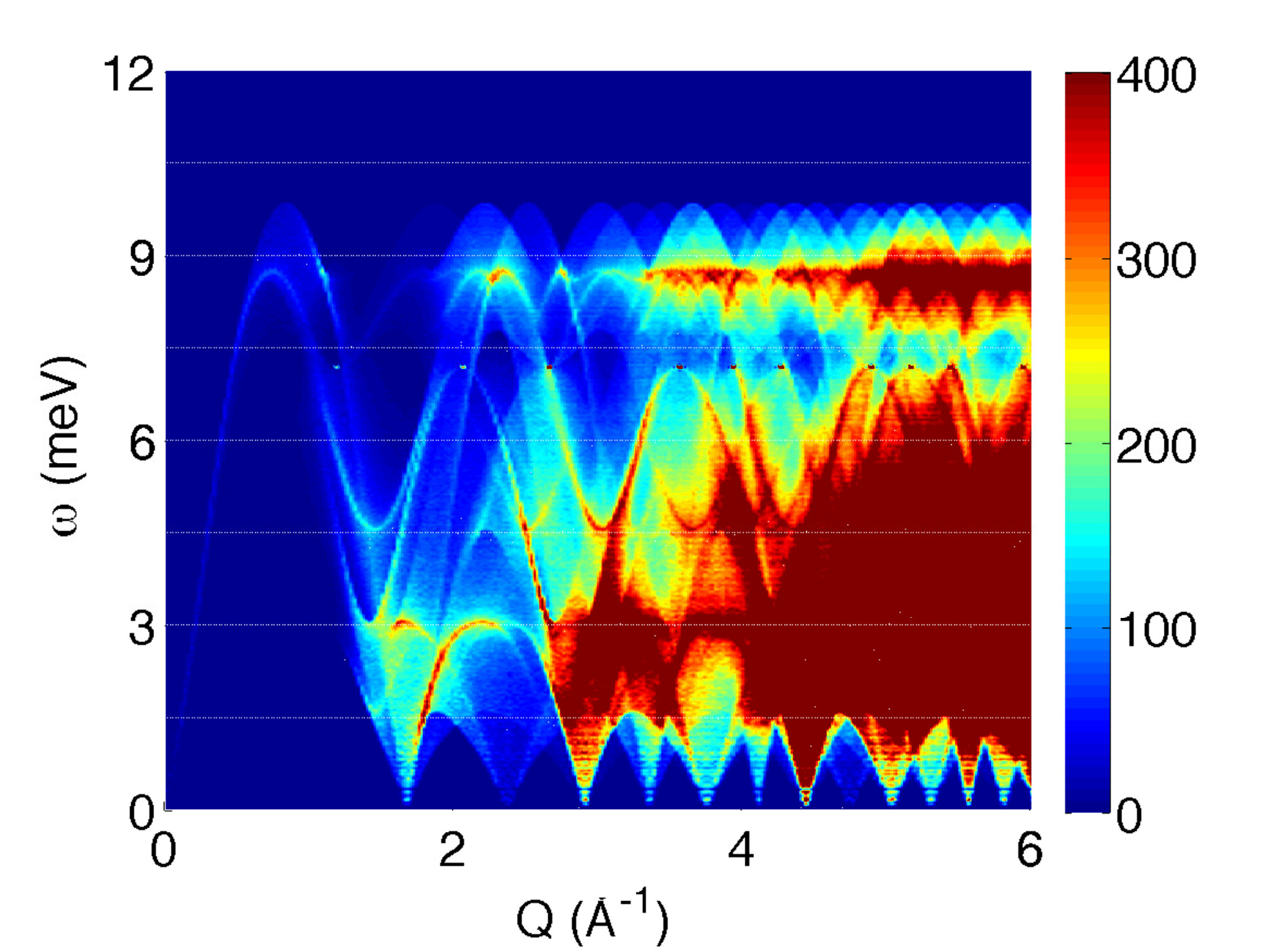} 
\includegraphics[width=5.5cm]{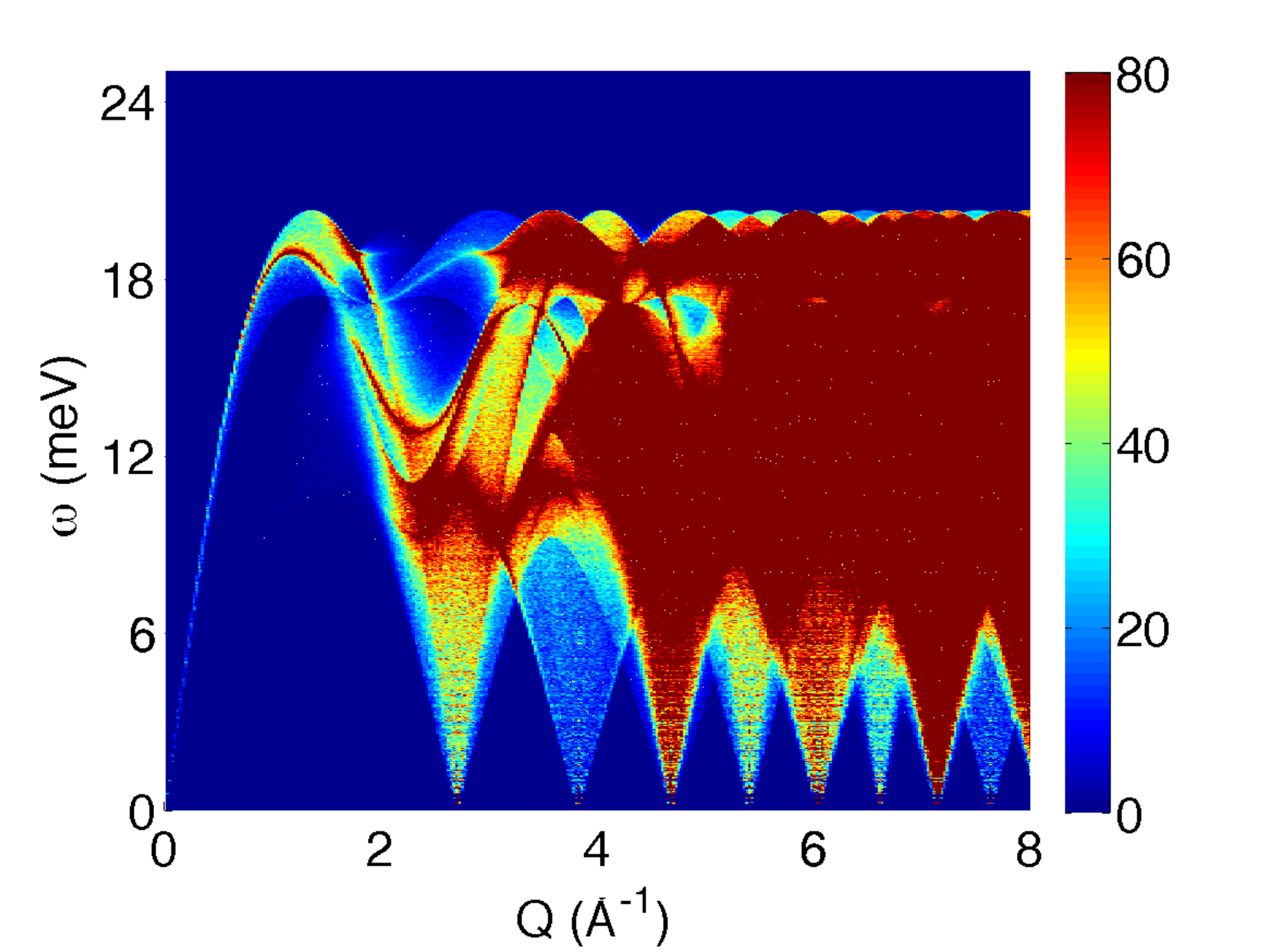} 
\includegraphics[width=5.5cm]{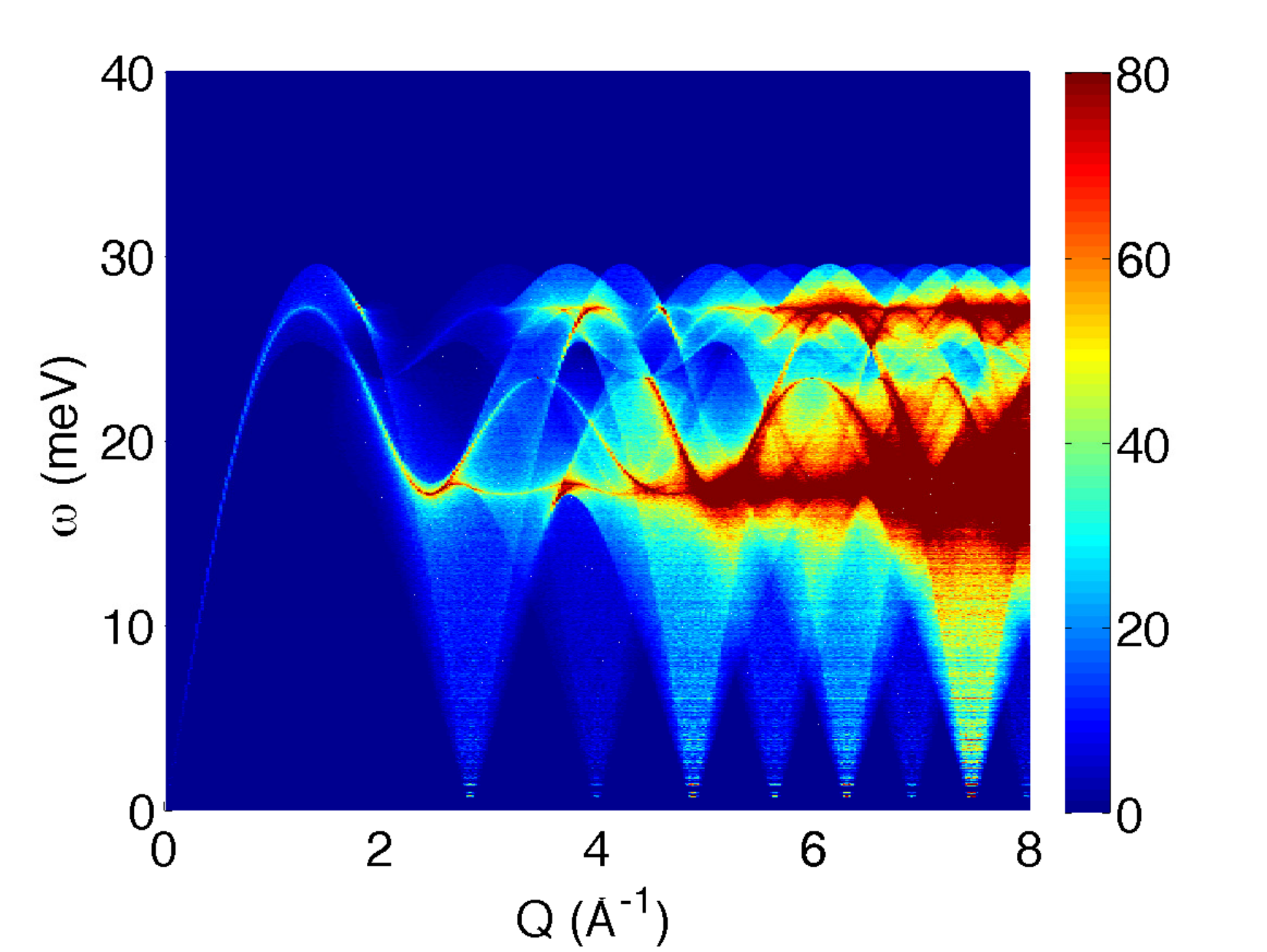} 

\includegraphics[width=5.5cm]{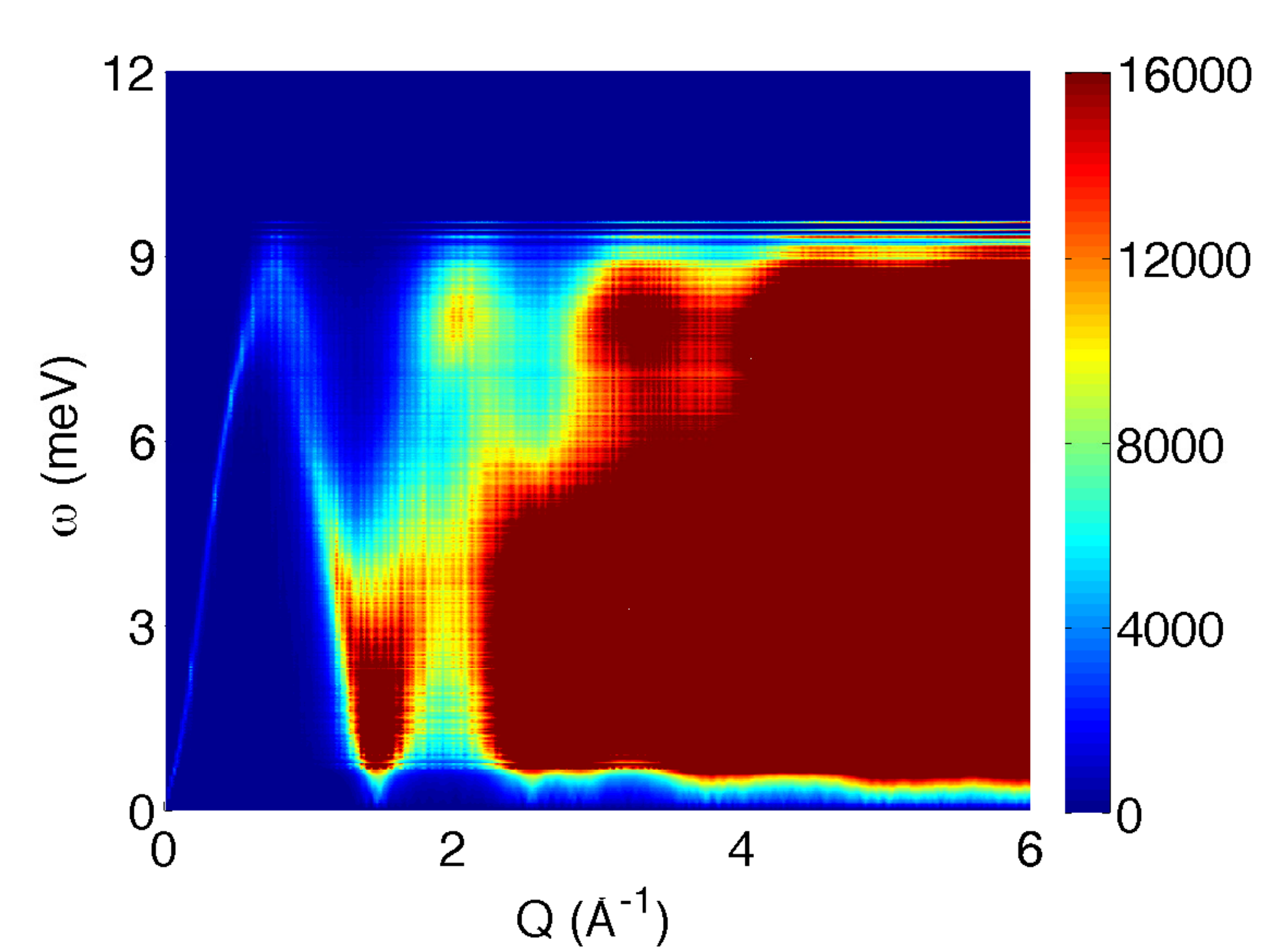} 
\includegraphics[width=5.5cm]{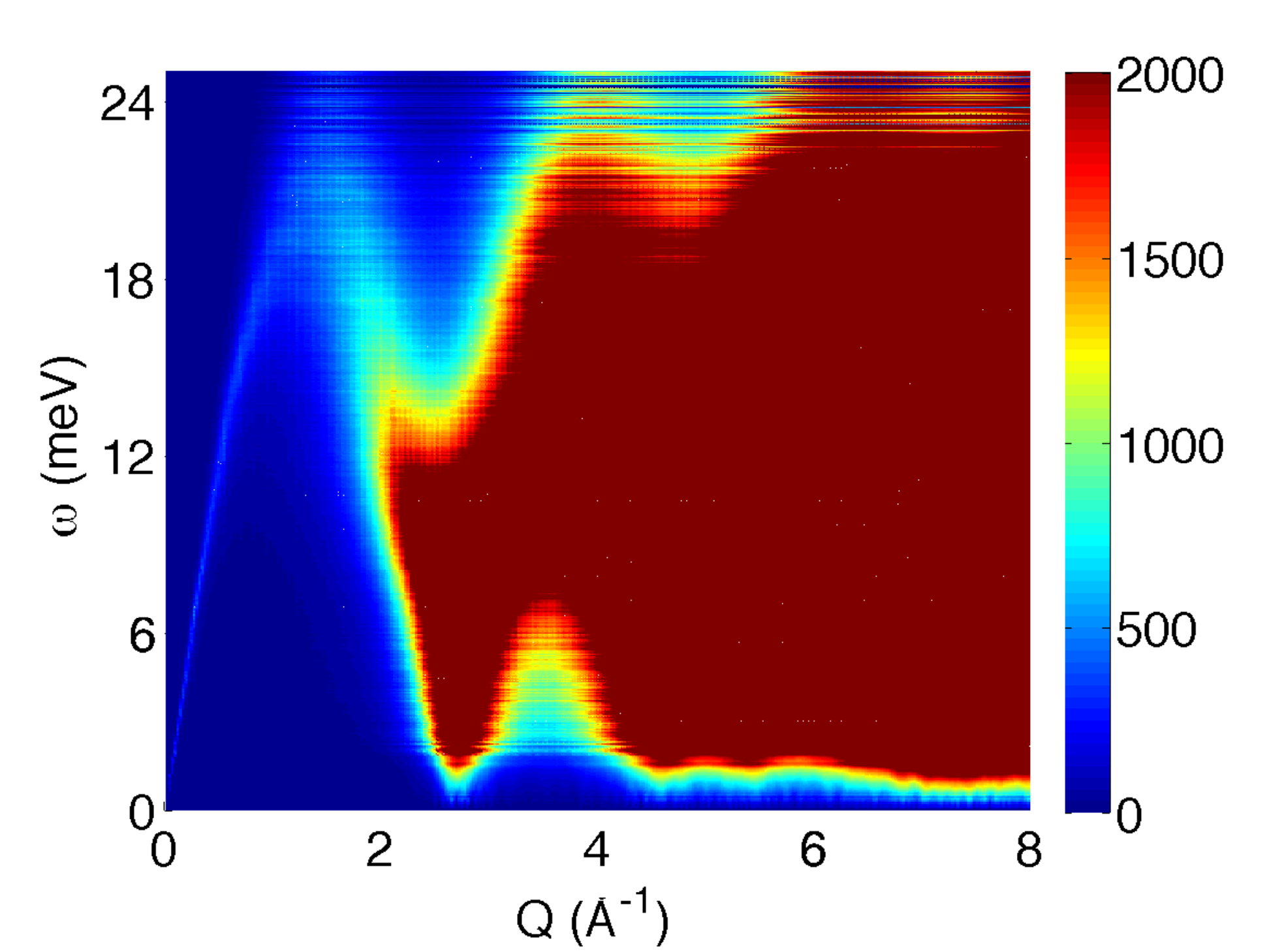} 
\includegraphics[width=5.5cm]{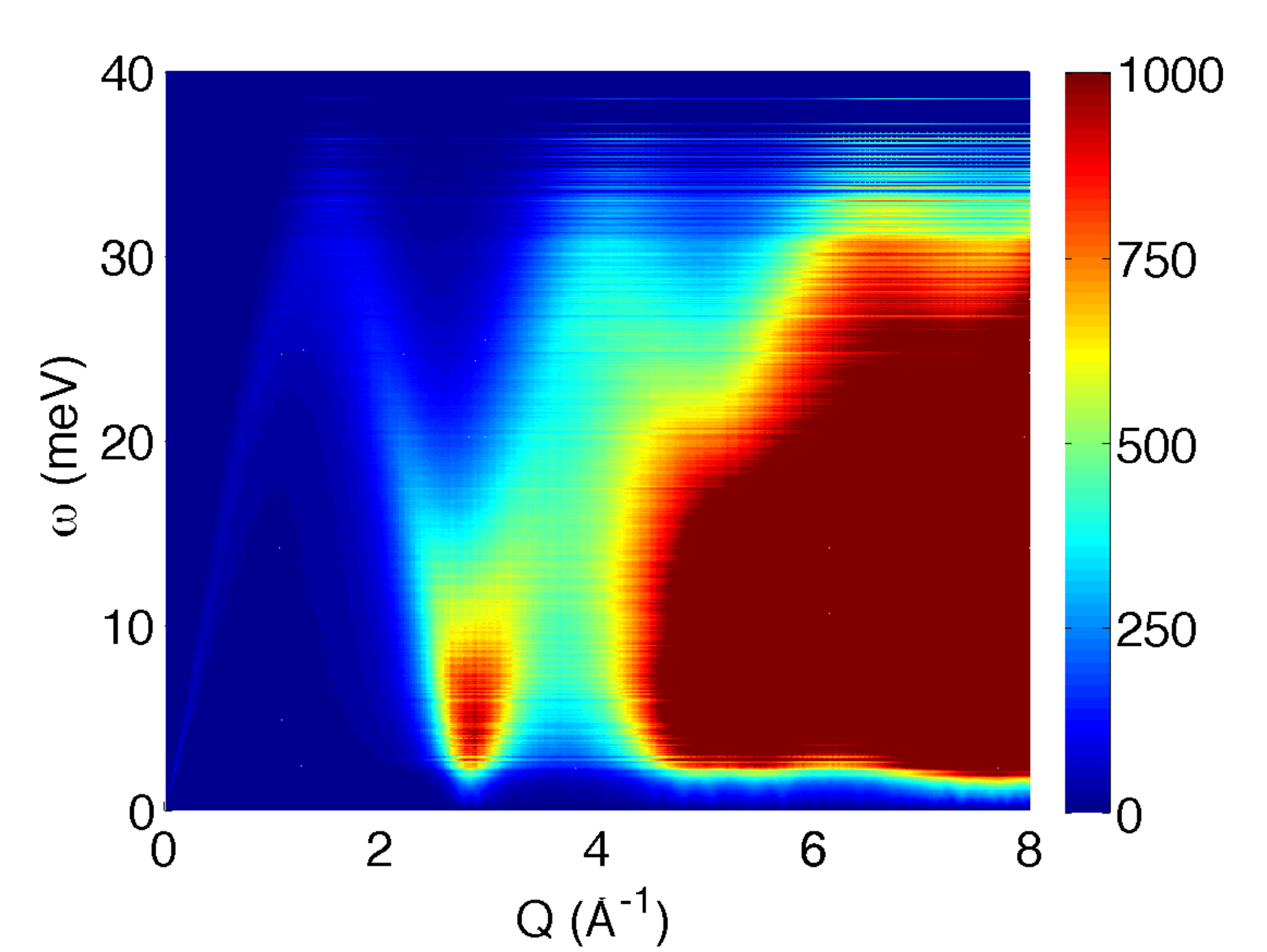} 
\caption{Collective excitation spectra from INS simulations for K, Ta and W (left to right). 
The graphs on the top show the spectra of the crystalline structure and the bottom the spectra of the supercooled liquid structure. 
The colour-scale indicates the high and low scattering intensity regimes.}
\label{fig:INSsimsAppA}
\end{center}
\end{figure*}

\begin{figure*}
\begin{center}
\includegraphics[width=5.5cm]{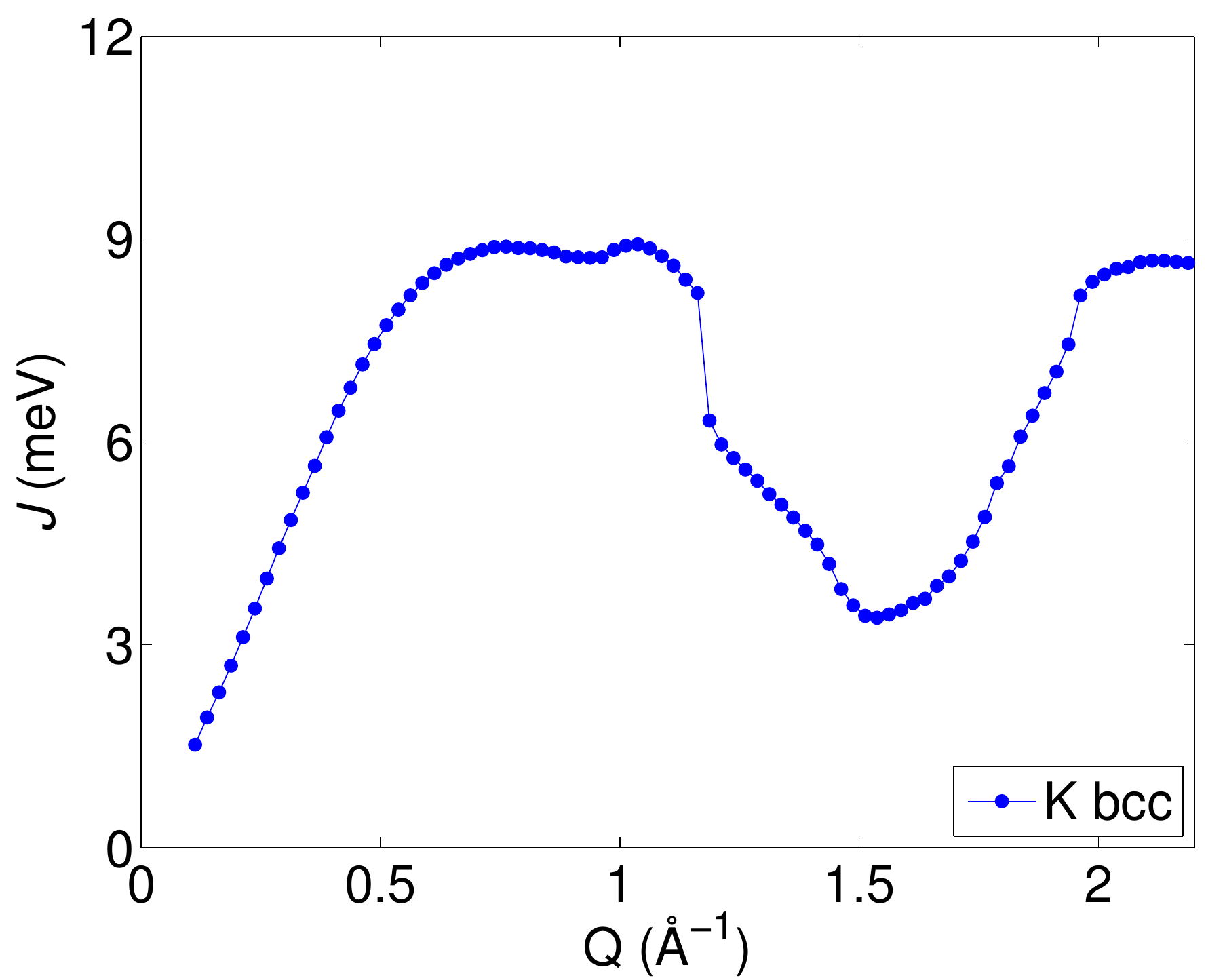}  
\includegraphics[width=5.5cm]{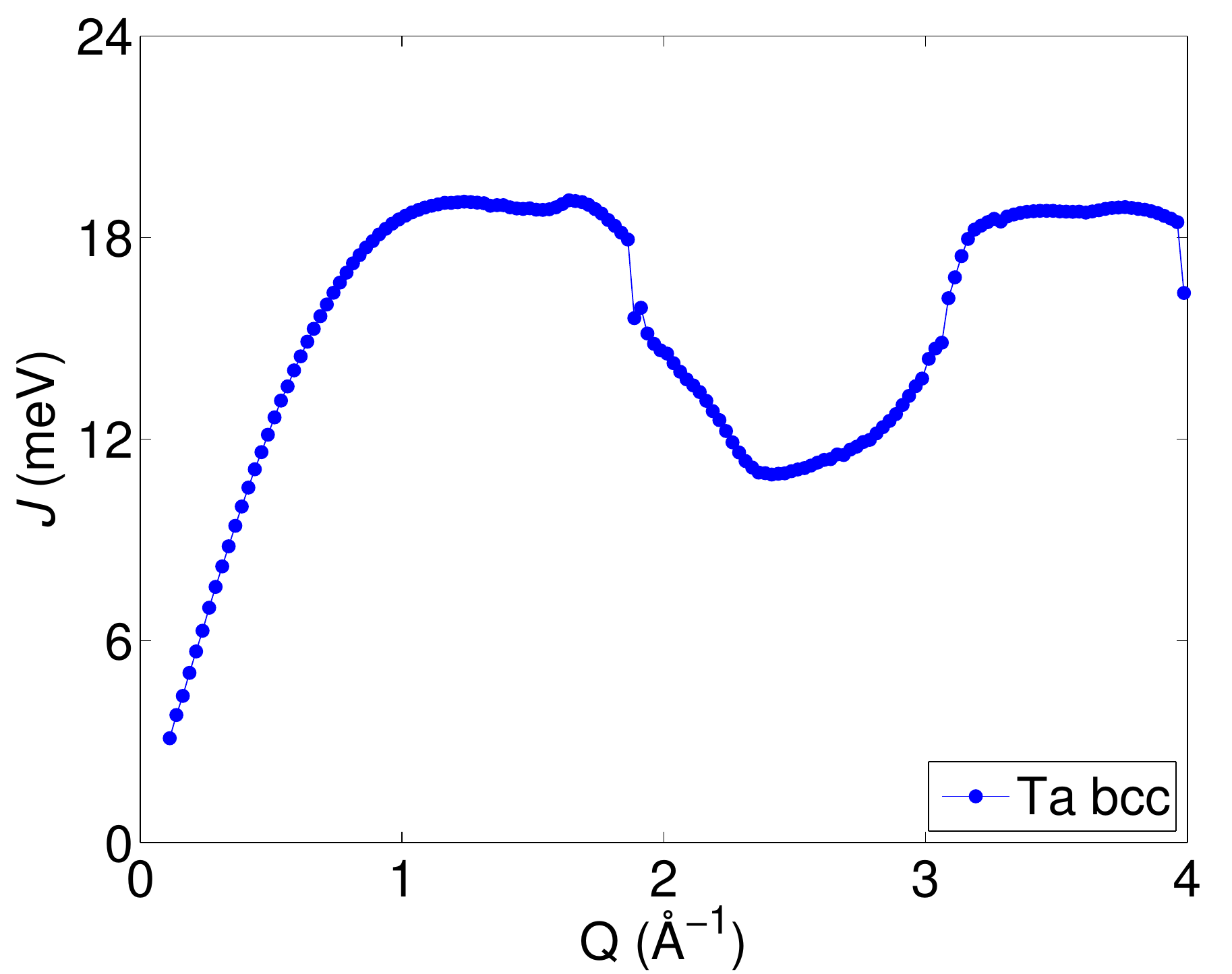} 
\includegraphics[width=5.5cm]{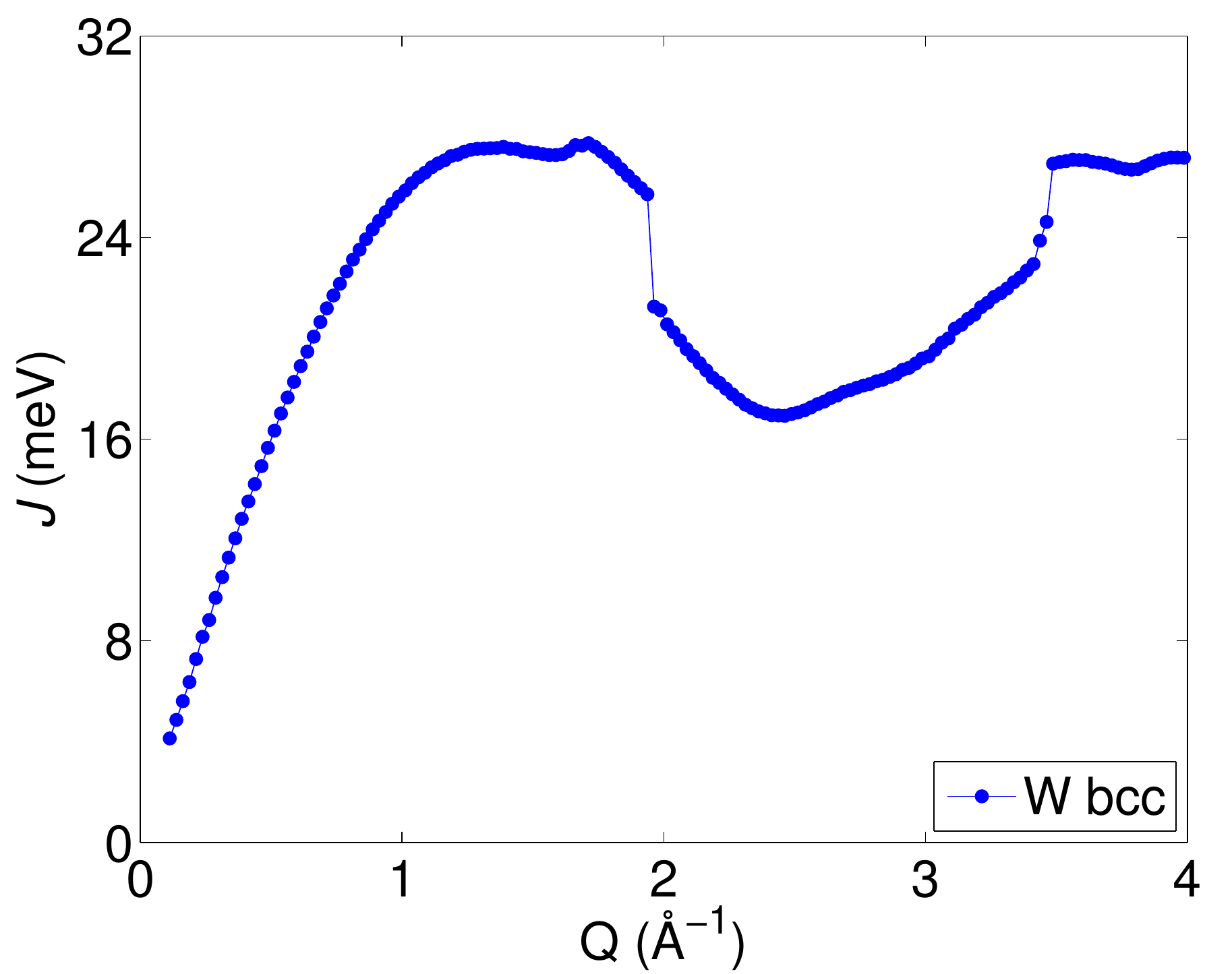}

\includegraphics[width=5.5cm]{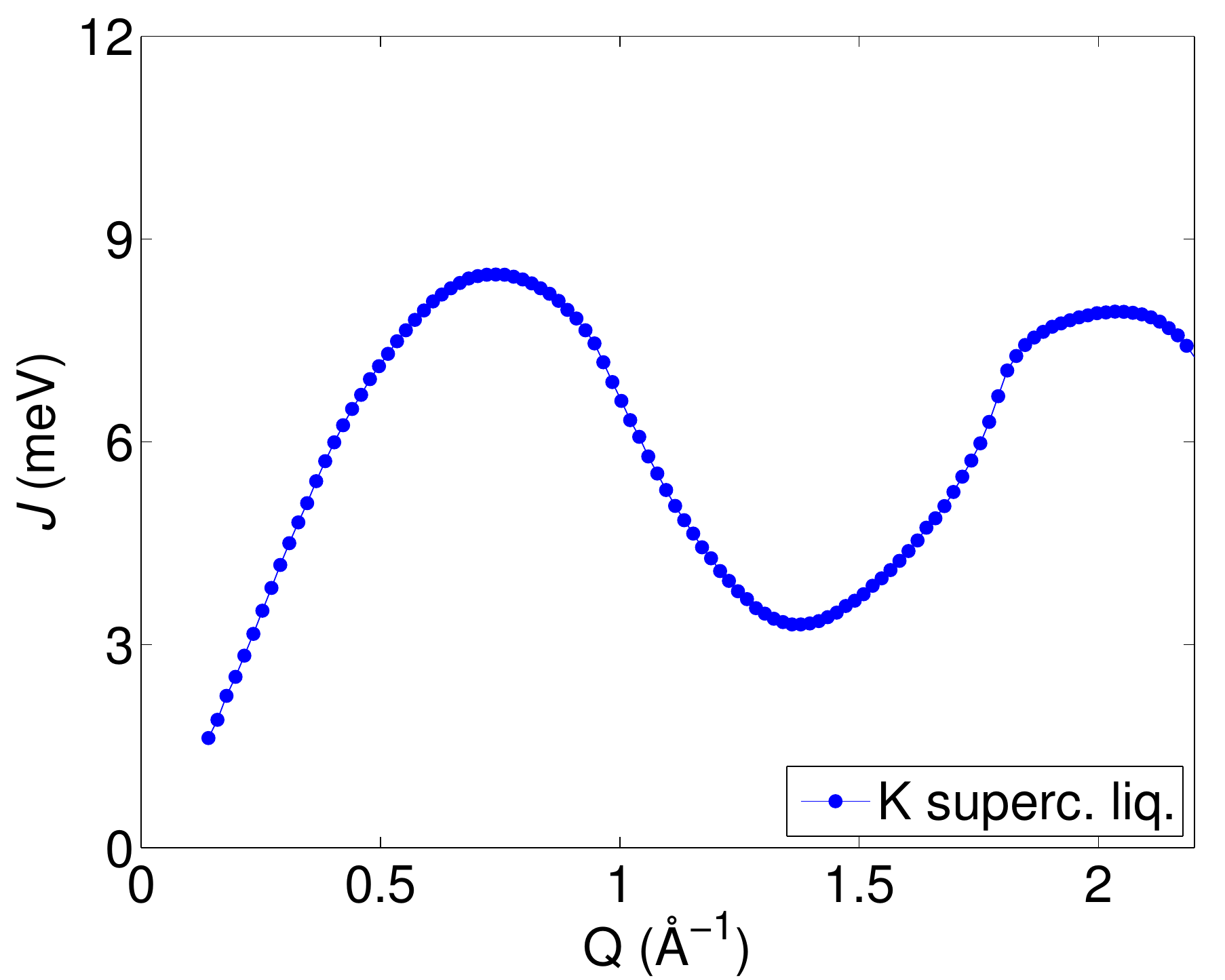}
\includegraphics[width=5.5cm]{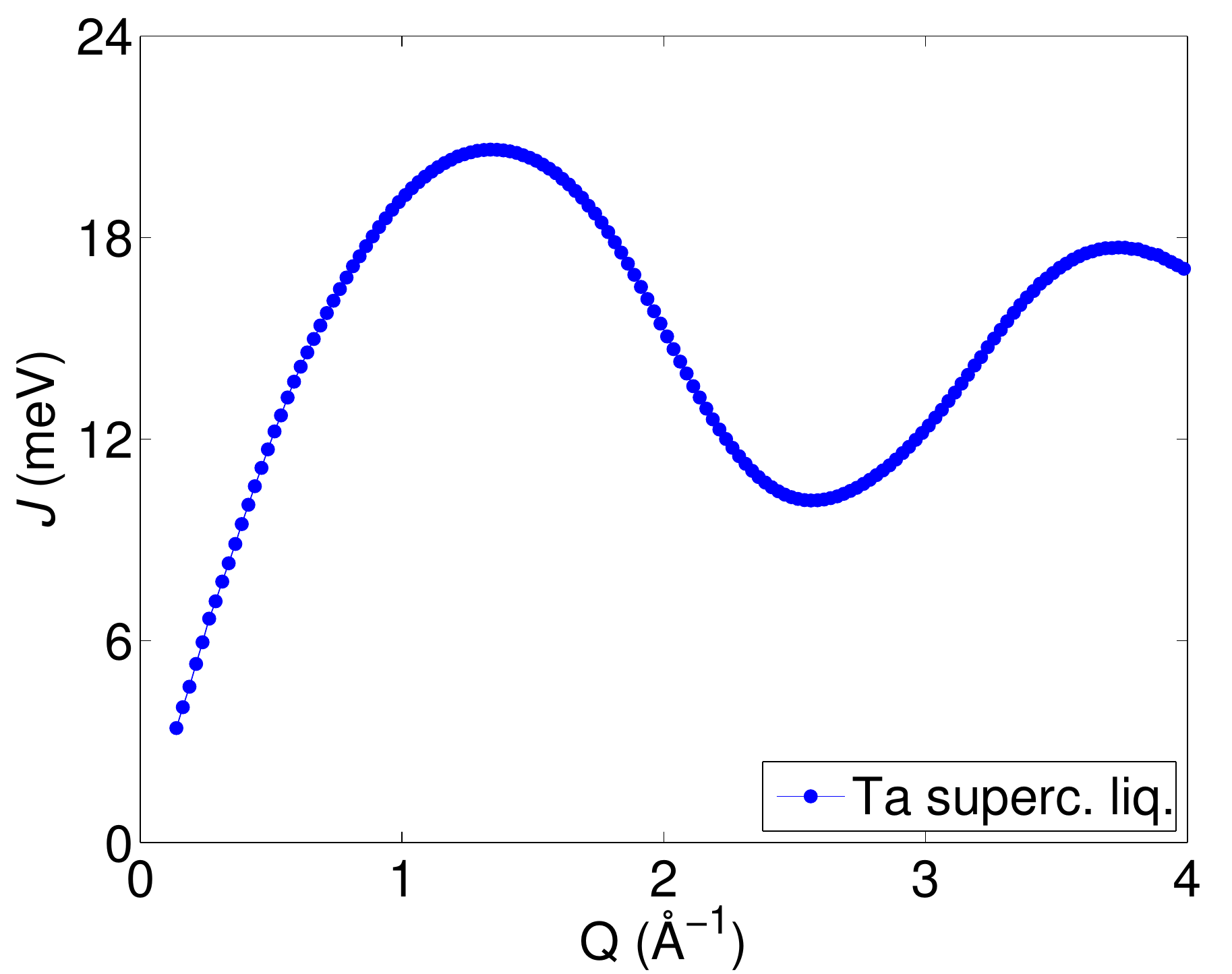}  
\includegraphics[width=5.5cm]{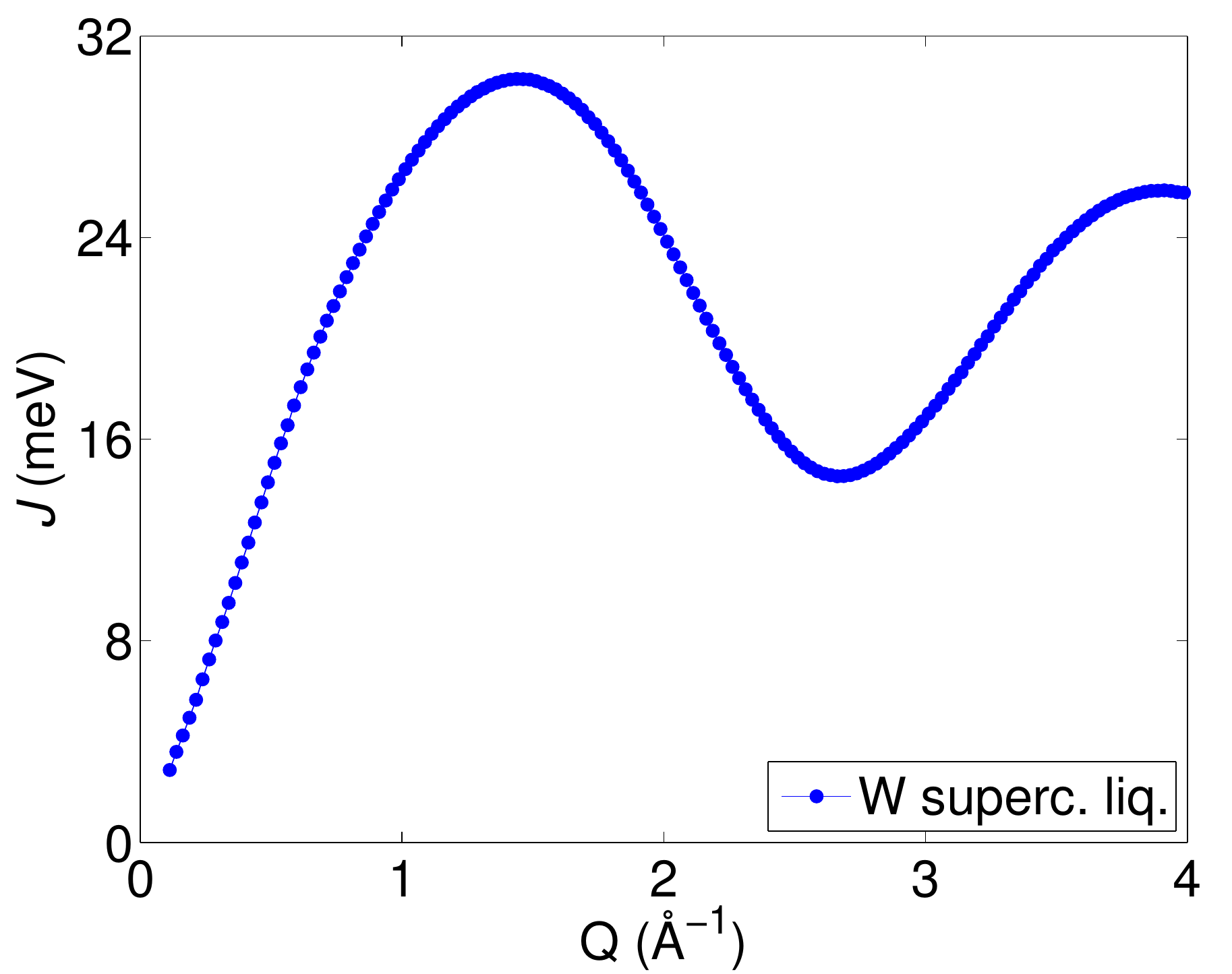} 
\caption{$J(\mathit{Q},\omega)$ of K, Ta and W for the crystalline and supercooled liquid structure.
For all materials the phonon-roton minima are located at similar \textbf{Q} for both crystal and supercooled liquid.}
\label{fig:DHOsimsAppA}
\end{center}
\end{figure*}

\begin{figure*}
\begin{center}
\includegraphics[width=5.5cm]{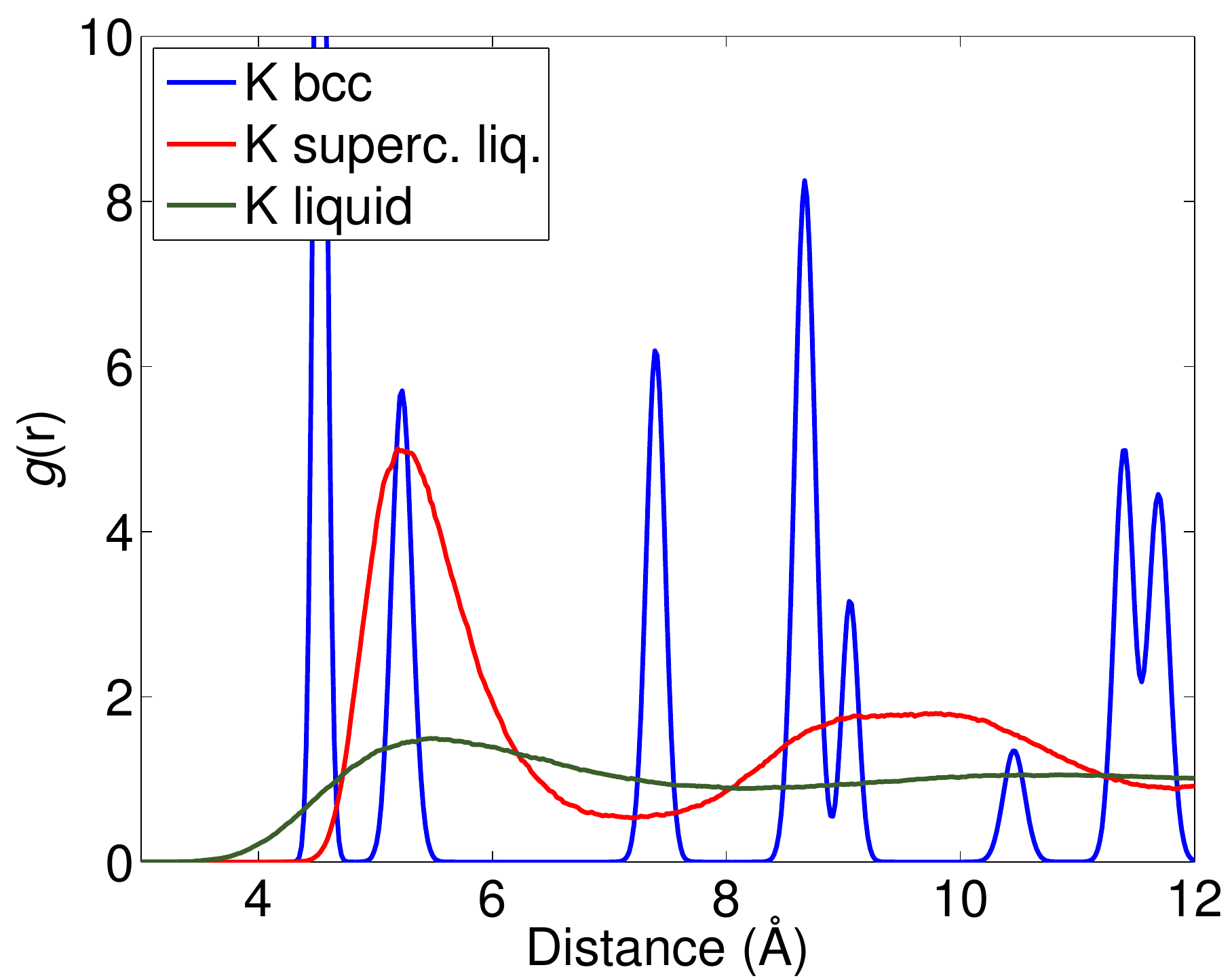} 
\includegraphics[width=5.5cm]{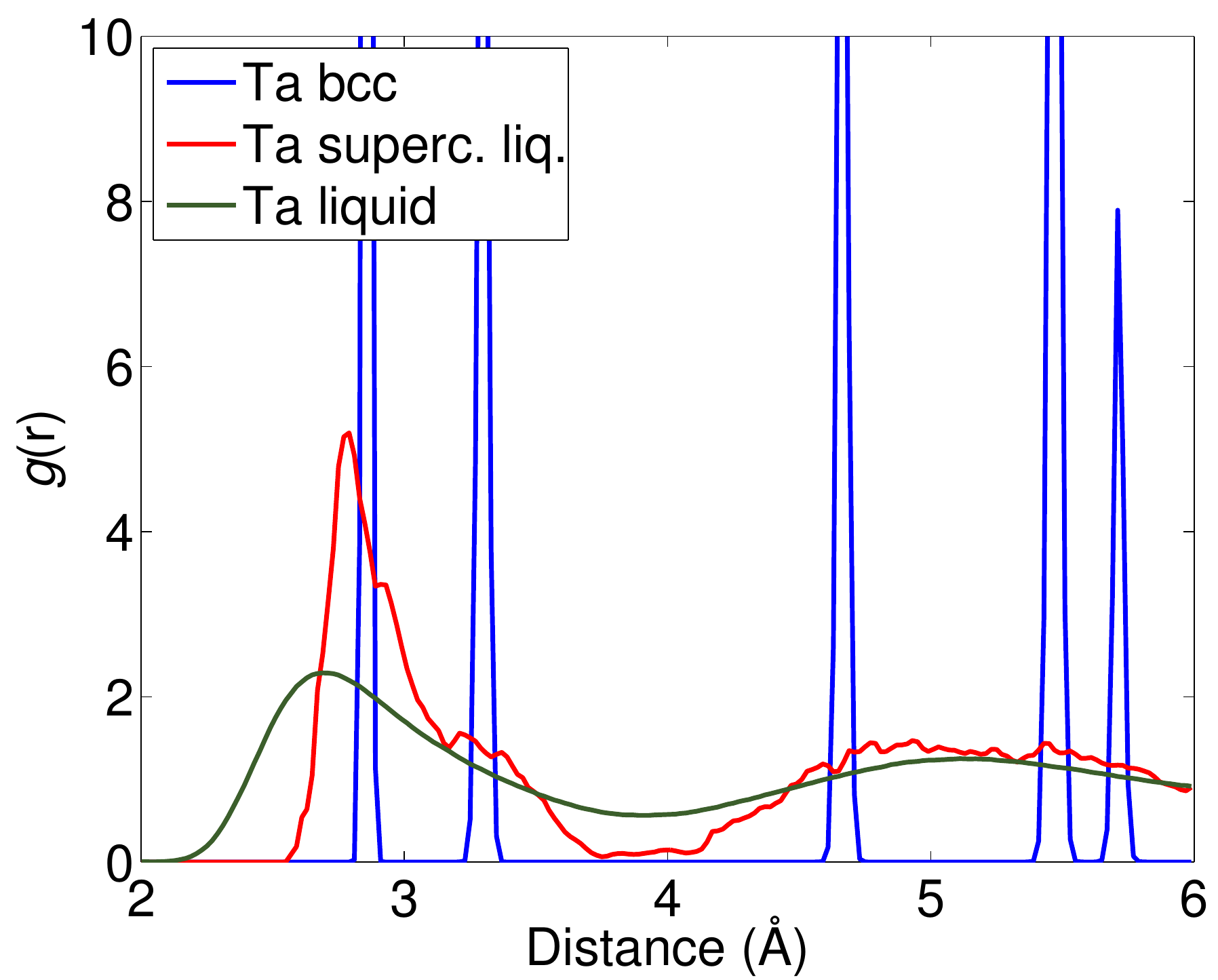} 
\includegraphics[width=5.5cm]{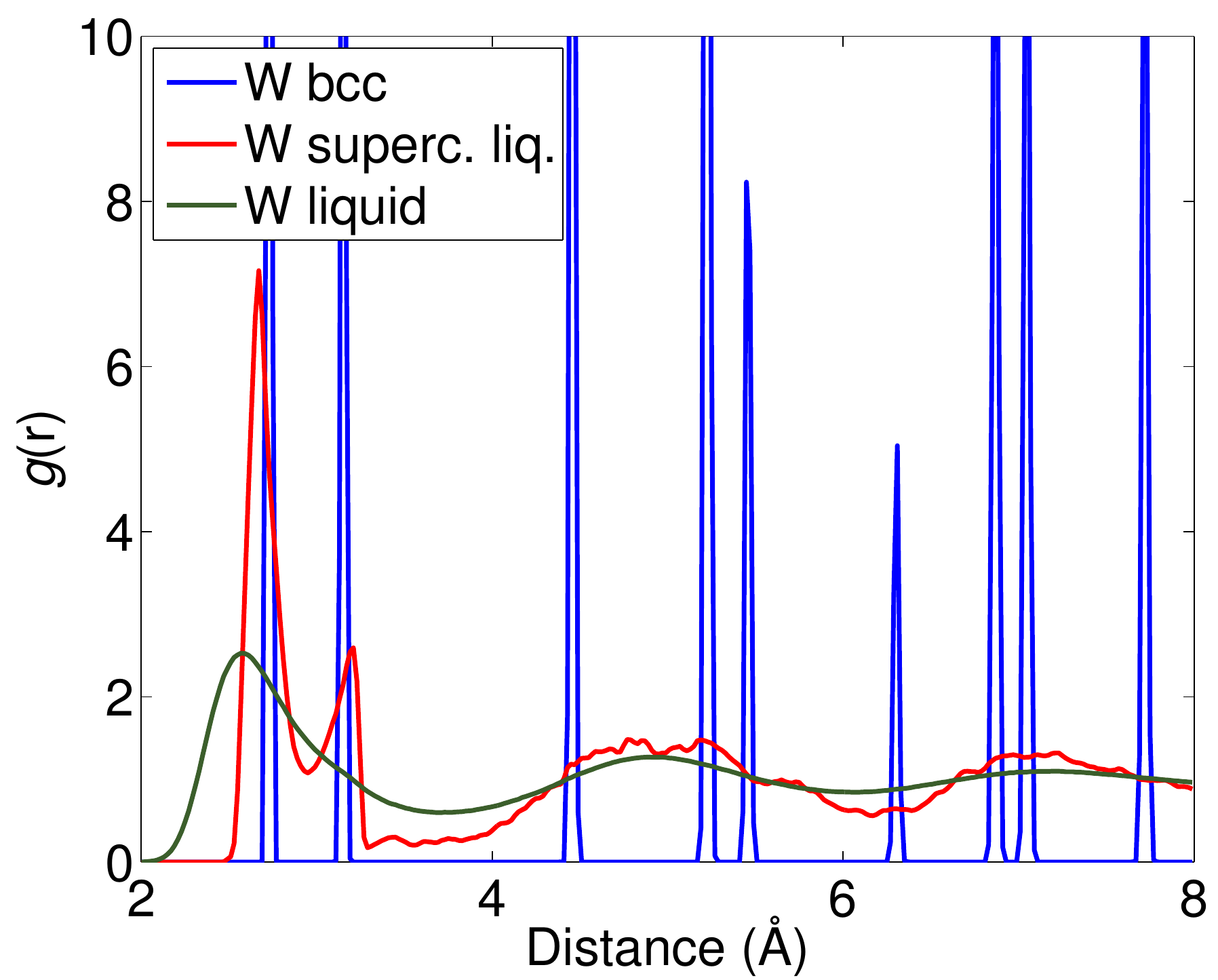}
\includegraphics[width=5.5cm]{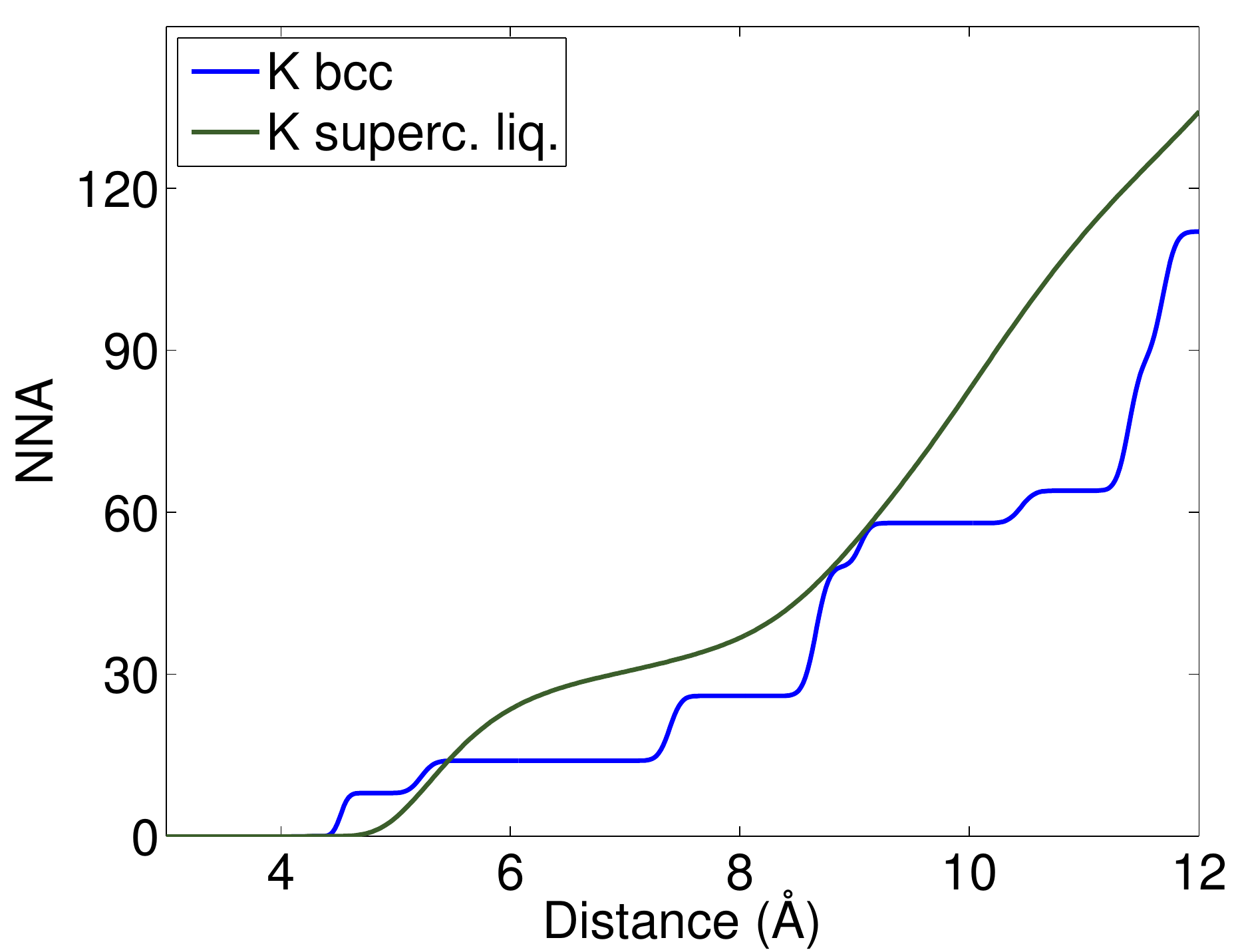}
\includegraphics[width=5.5cm]{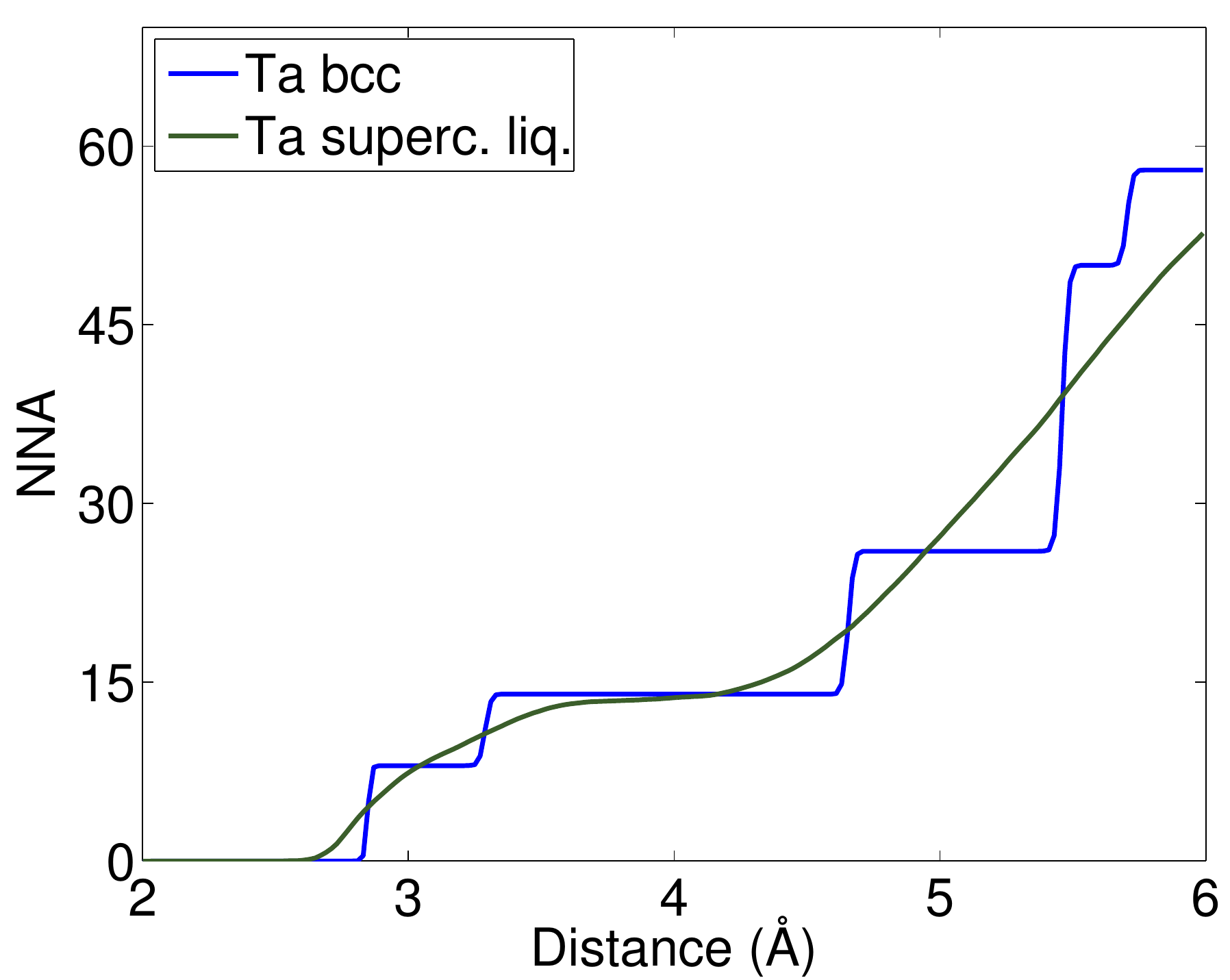}
\includegraphics[width=5.5cm]{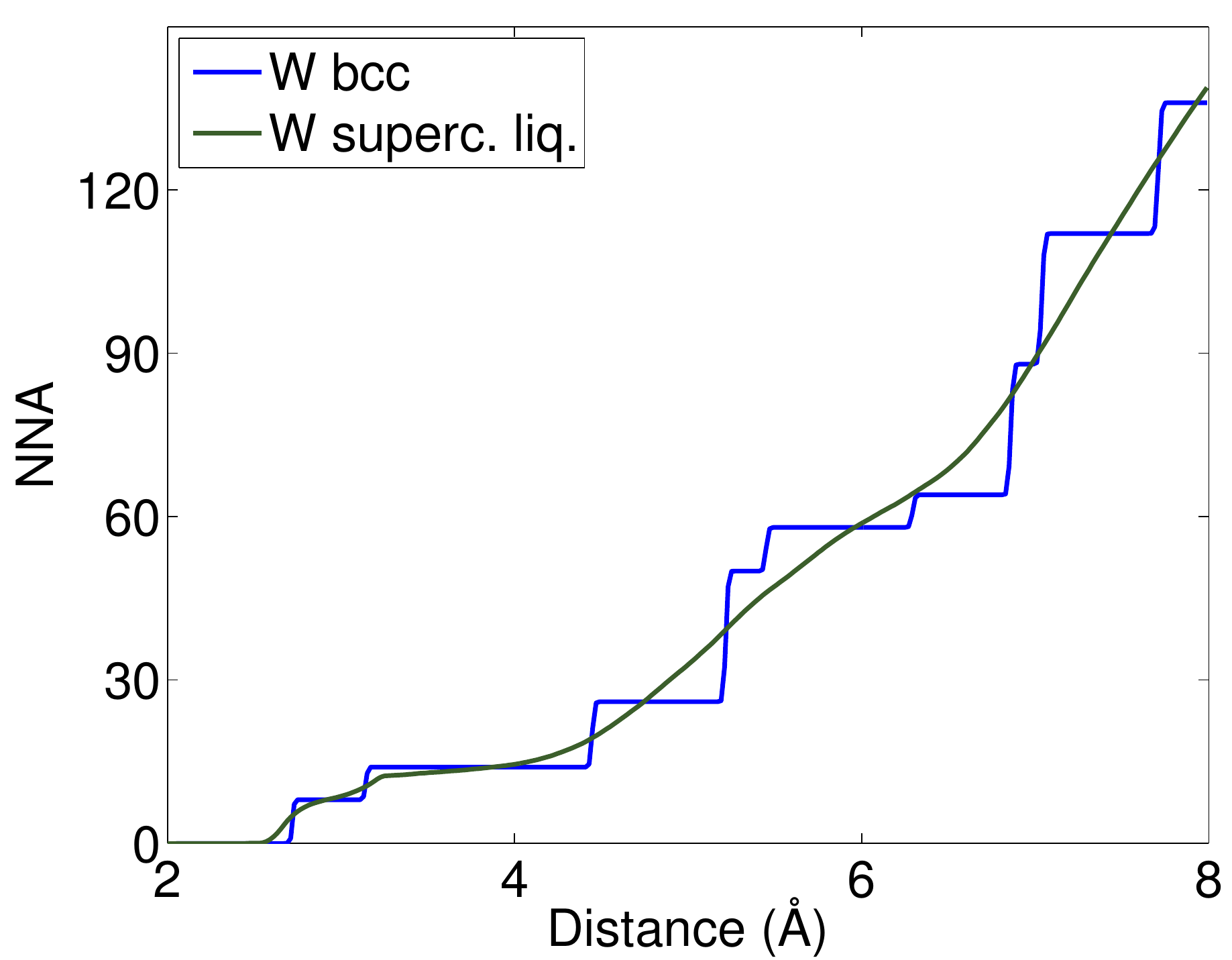}
\caption{$g(r)$ and NNA for K, Ta and W from MD simulations.
For all materials the NNA curve of the supercooled liquid follows the trail of the crystalline curve.}
\label{fig:pdfnnaAppA}
\end{center}
\end{figure*}

\begin{figure*}
\begin{center}
\includegraphics[width=5.5cm]{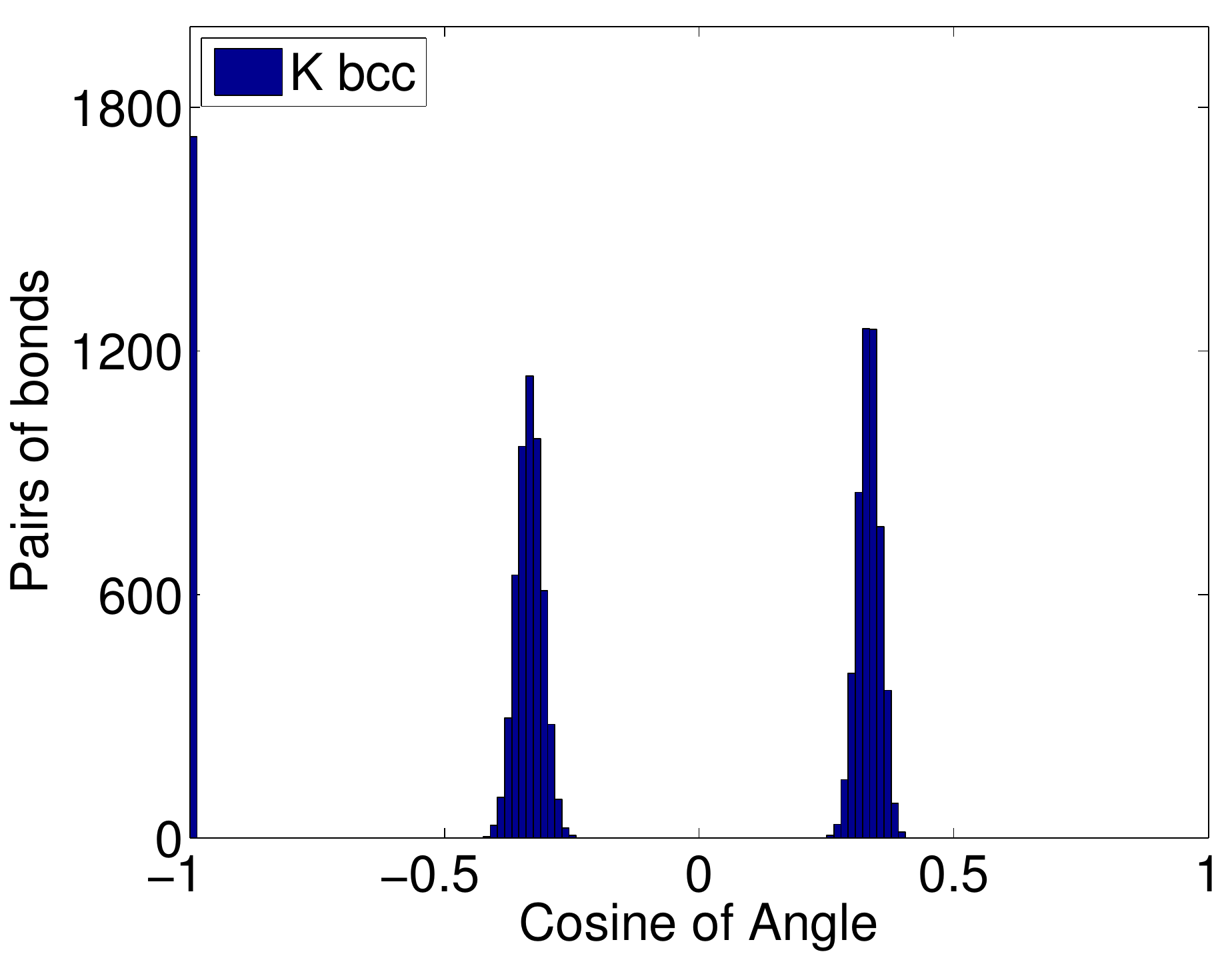} 
\includegraphics[width=5.5cm]{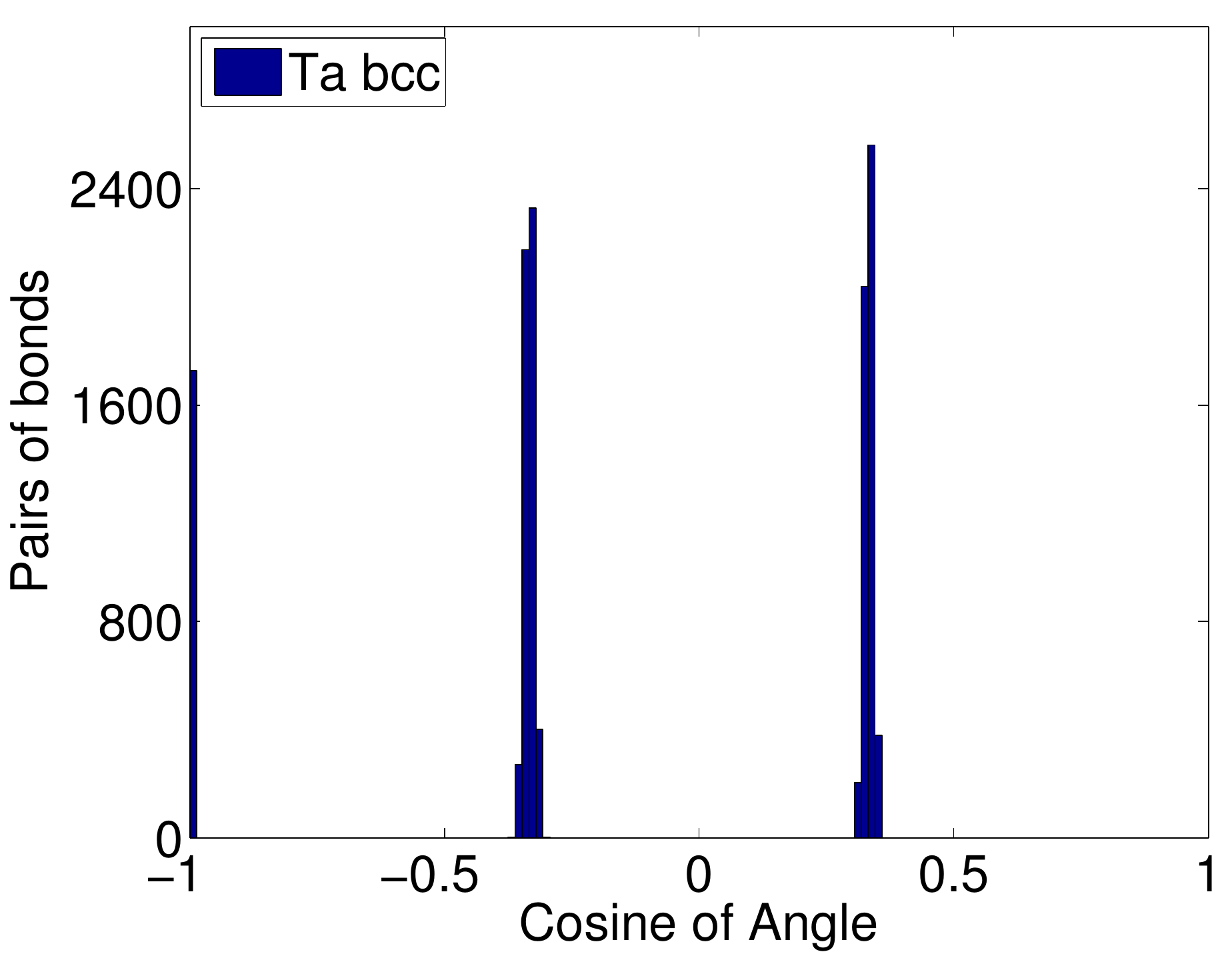} 
\includegraphics[width=5.5cm]{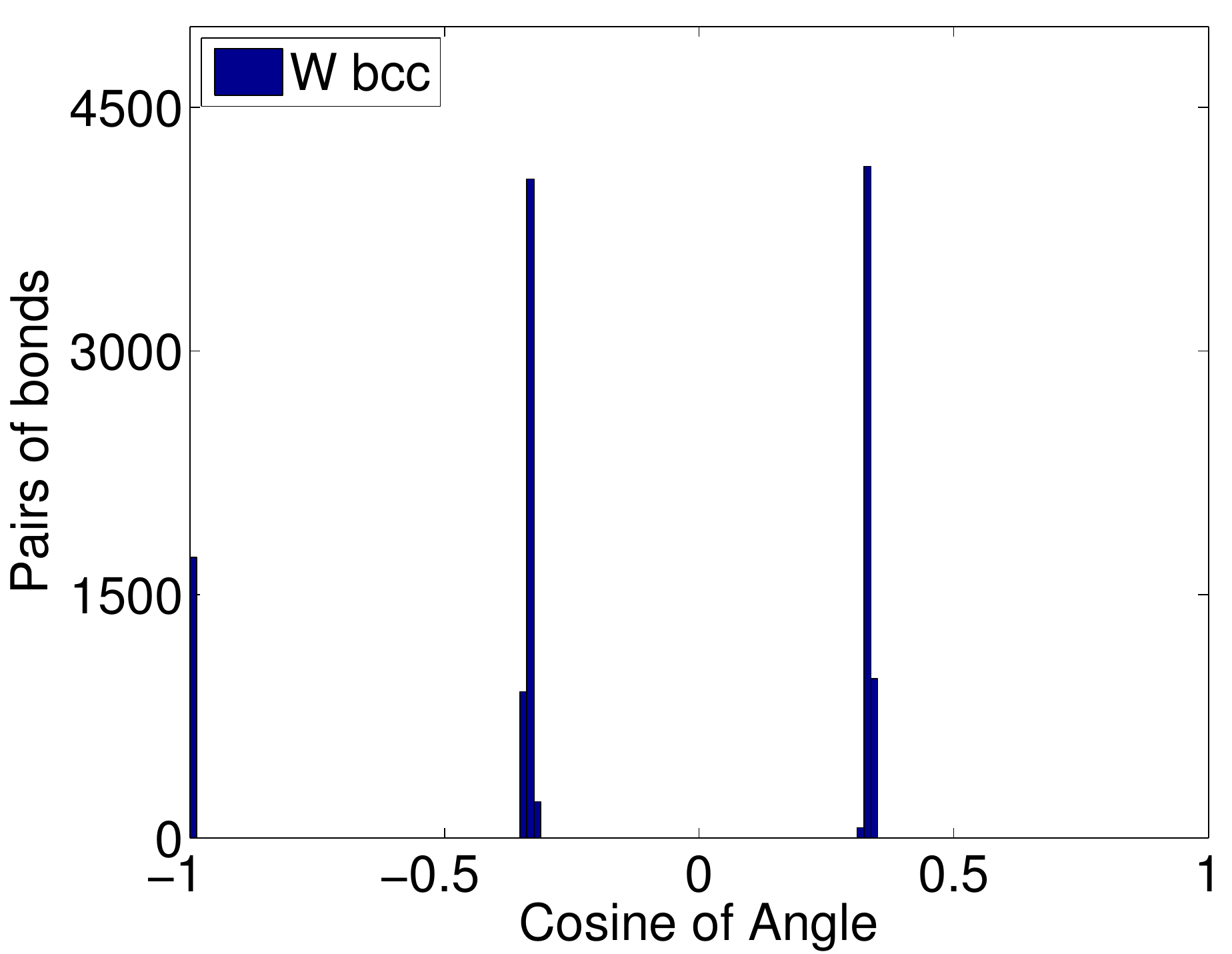} \\

\includegraphics[width=5.5cm]{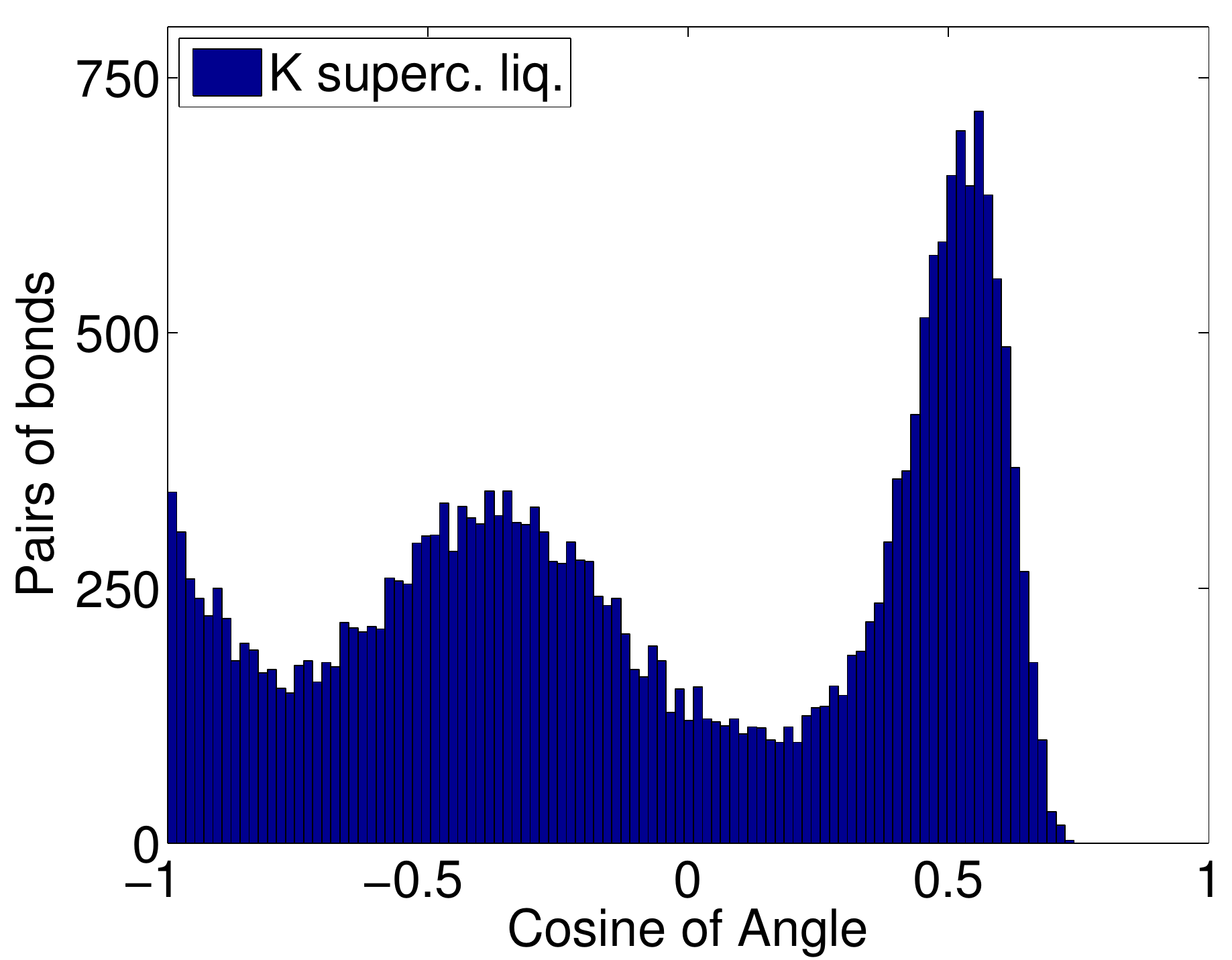}
\includegraphics[width=5.5cm]{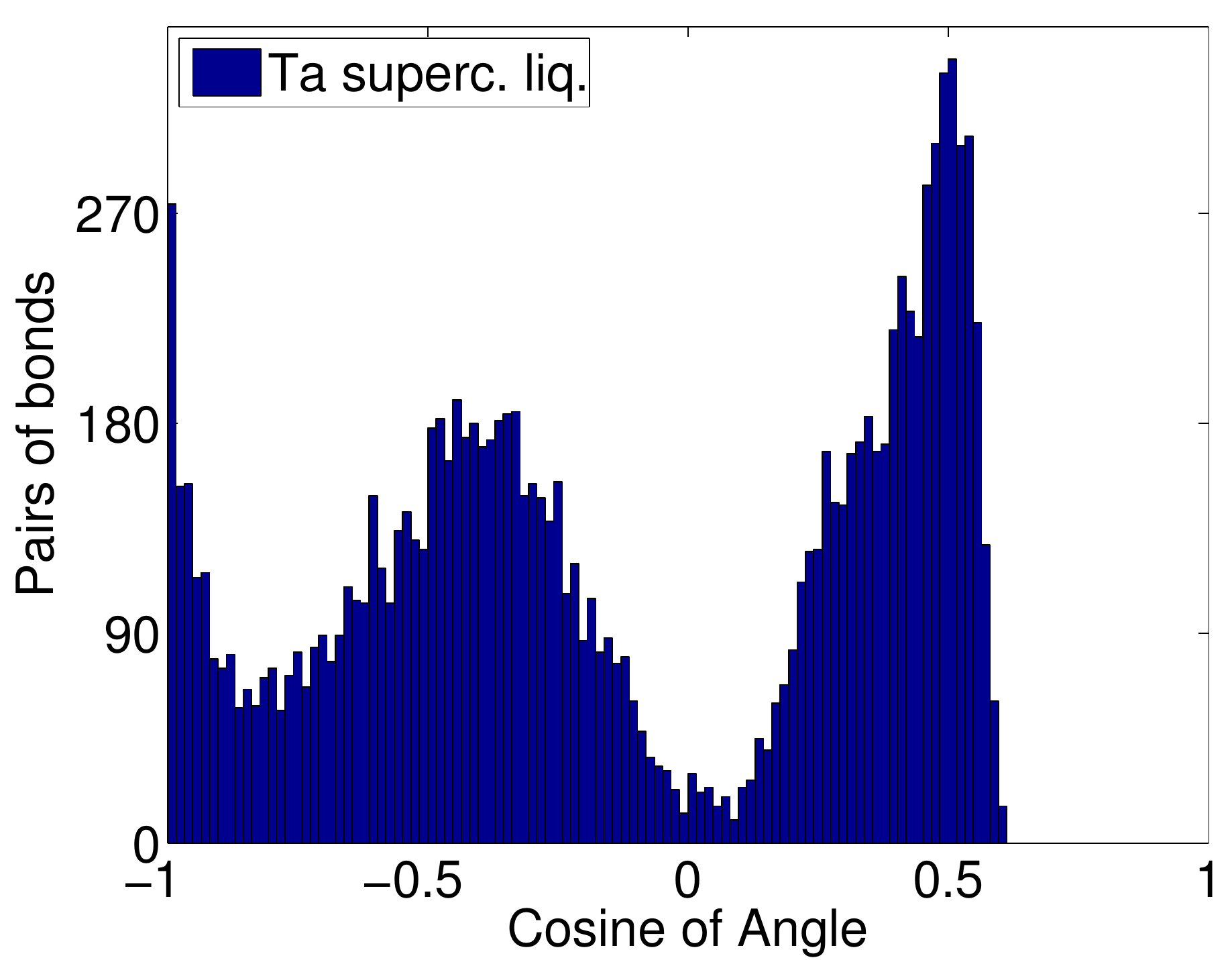}
\includegraphics[width=5.5cm]{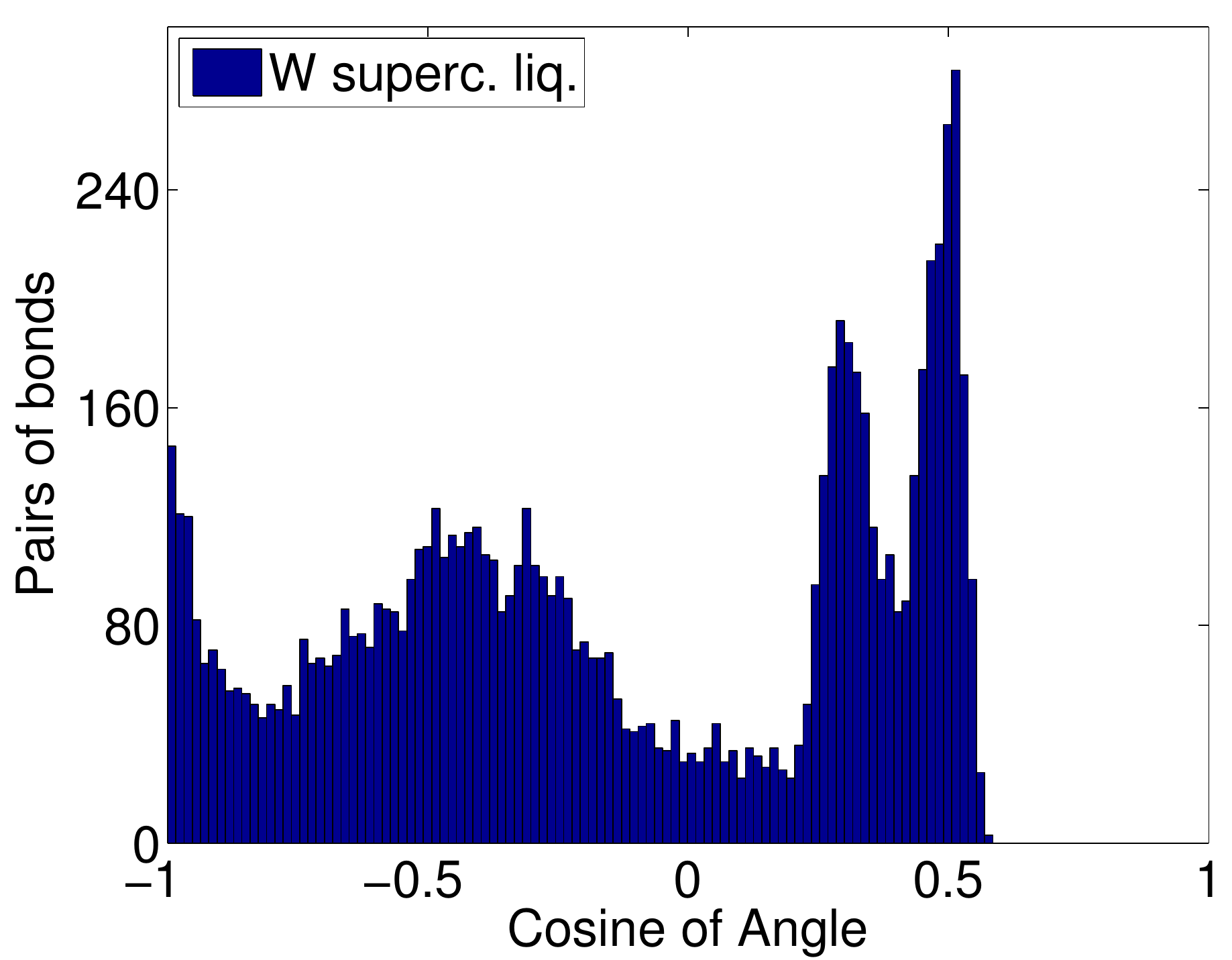}\\

\includegraphics[width=5.5cm]{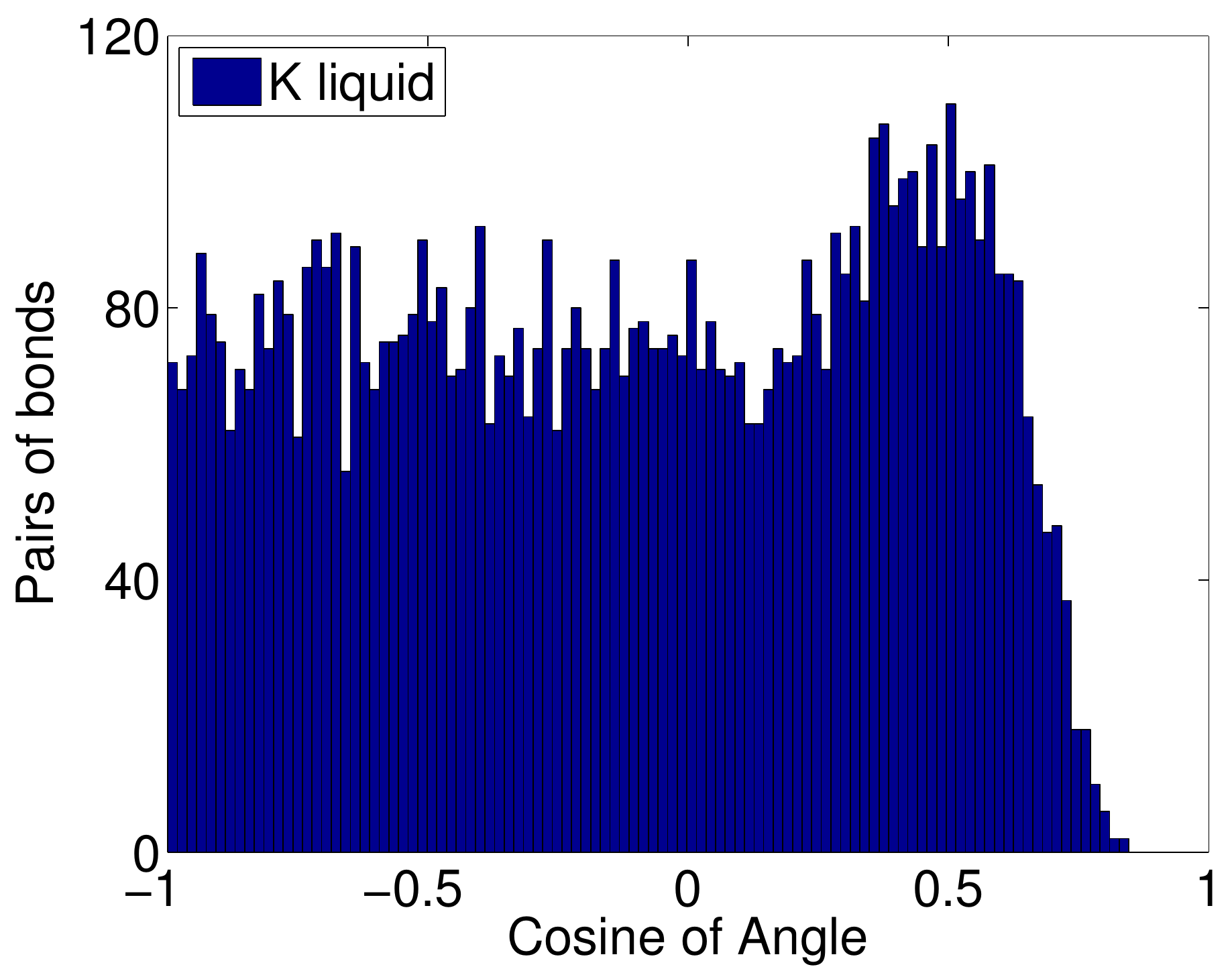}
\includegraphics[width=5.5cm]{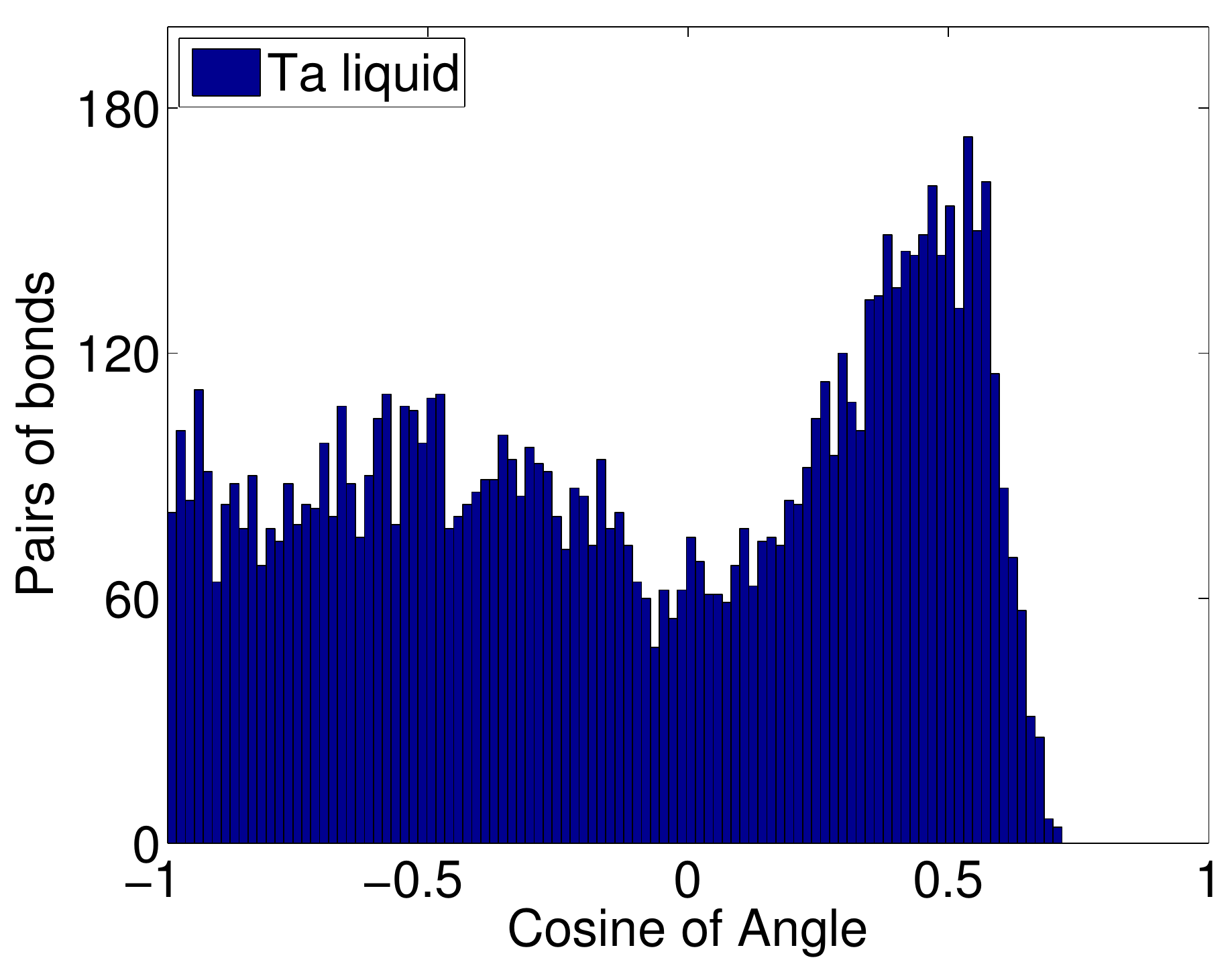}
\includegraphics[width=5.5cm]{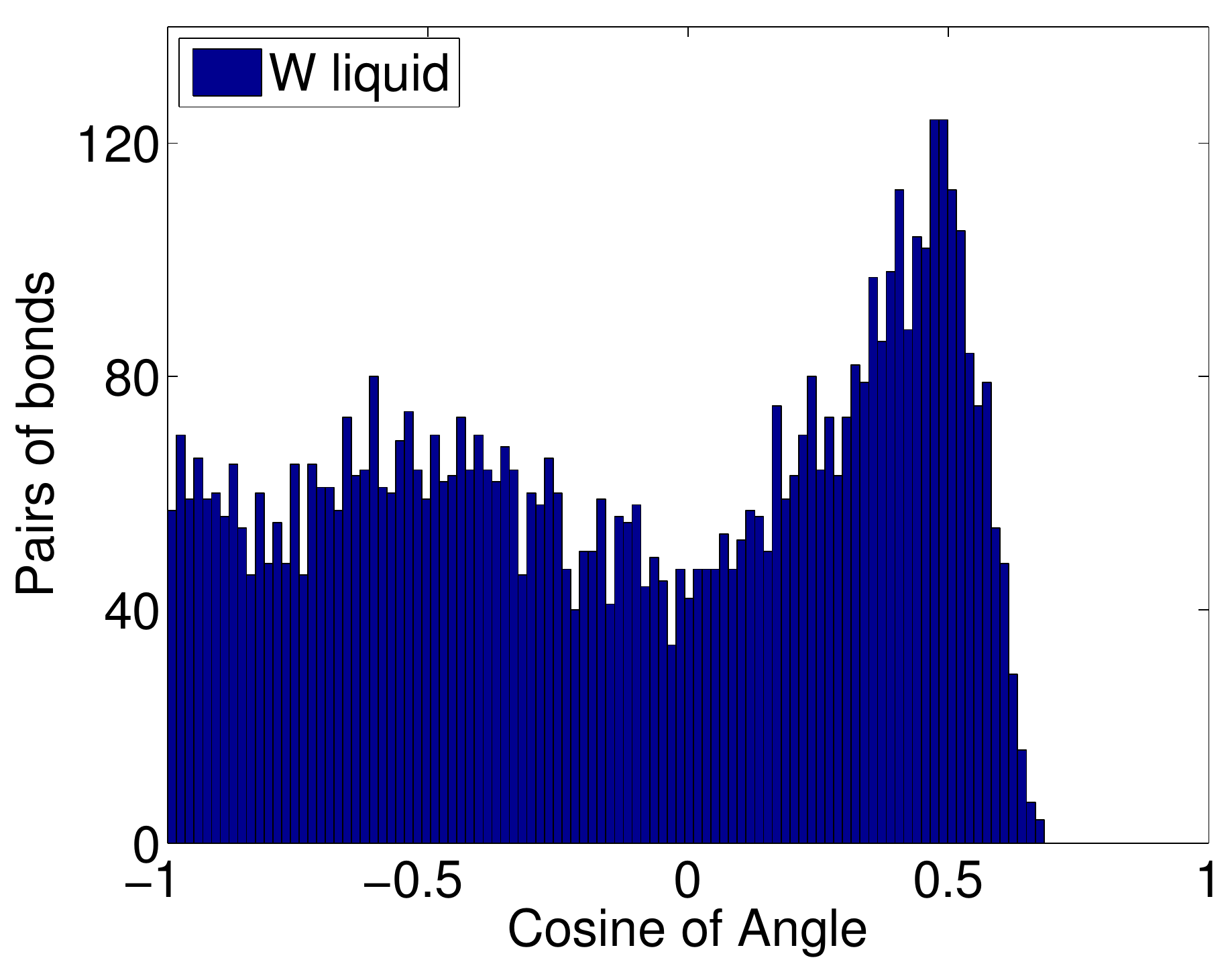}

\caption{Nearest neighbour angle distribution of K, Ta and W at the crystalline, amorphous and liquid phases calculated from MD simulations.
The supercooled liquid maintains local order features from both the liquid and crystalline phases showing evidence of dependence on the crystalline structure.}
\label{fig:anglesAppA}
\end{center}
\end{figure*}

Here we present additional figures for K, Ta and W.
All results are from MD simulations, calculated following the same methods and procedure as described earlier.
The following figures have the same sequence as the previously presented figures for Fe and Ar.
In Fig. \ref{fig:INSsimsAppA} and Fig. \ref{fig:DHOsimsAppA} we present results from INS simulations and the $J(Q,\omega)$ respectively for the crystalline and supercooled structures of the three materials.
In Fig. \ref{fig:pdfnnaAppA} and Fig. \ref{fig:anglesAppA} we present results from population and bond angle distribution analysis respectively, for the crystalline, amorphous and liquid phases of the three materials.
All materials exhibit similar behaviour, typical for a monatomic bcc system and they have qualitative similarities to Fe, hence similar discussion and conclusions apply. \par

\clearpage

\bibliographystyle{unsrt} 


\end{document}